\documentclass[11pt]{article}
\usepackage{float}
\usepackage{comment}
\usepackage{caption}
\usepackage{cancel}
\usepackage{graphicx}
\usepackage{tikz} 
\usetikzlibrary{calc} 
\usetikzlibrary{decorations.pathmorphing,patterns,arrows.meta} 
\pdfoutput=1
\usepackage{amsmath,epsf,amssymb,latexsym,array,color}
\allowdisplaybreaks
\usepackage{graphicx,wasysym}
\usepackage{cancel}
\usepackage[title]{appendix}
\usepackage{hyperref}
\usepackage{orcidlink}
\usepackage{tikz}
\usepackage{xcolor}
\usepackage{adjustbox}
\usetikzlibrary{arrows.meta, positioning, calc}

\definecolor{InputBlue}{HTML}{E8F0FE}
\definecolor{BranchGreen}{HTML}{E6F4EA}
\definecolor{AnsatzOrange}{HTML}{FFF4E5}
\definecolor{EvalRed}{HTML}{FCE8E6}
\definecolor{CombinePurple}{HTML}{F3E8FD}
\definecolor{ReduceTeal}{HTML}{E0F2F1}
\definecolor{FinalYellow}{HTML}{FFF9C4}

\definecolor{xcol}{RGB}{170,55,55}
\definecolor{ycol}{RGB}{45,95,185}
\definecolor{zcol}{RGB}{0,125,110}

\definecolor{propcol}{RGB}{52,92,170}
\definecolor{treecol}{RGB}{180,85,40}
\definecolor{loopcol}{RGB}{0,125,110}

\providecommand{\X}{}
\providecommand{\Y}{}
\providecommand{\Z}{}

\renewcommand{\X}{\ensuremath{\textcolor{xcol}{x}}}
\renewcommand{\Y}{\ensuremath{\textcolor{ycol}{y}}}
\renewcommand{\Z}{\ensuremath{\textcolor{zcol}{z}}}

\newcommand{\circled}[1]{%
  \tikz[baseline=(char.base)]{
    \node[shape=circle,draw,inner sep=1pt] (char) {#1};
  }%
}

\def\be{\begin{equation}}
\def\ee{\end{equation}}
\def\ba{\begin{array}}
\def\ea{\end{array}}
\newcommand{\beq}{\begin{equation}}
\newcommand{\eeq}[1]{\label{#1}\end{equation}}
\newcommand{\bea}{\begin{eqnarray}}
\newcommand{\eea}[1]{\label{#1}\end{eqnarray}}

\def\trace{\mathop{\rm Tr}\nolimits}

\def\Tr{{\rm Tr}}
\usetikzlibrary{arrows.meta}

\newcommand{\bbox}{\lower.2ex\hbox{$\Box$}}

\csname @addtoreset\endcsname{equation}{section}
\usepackage{ifpdf}
\ifx\pdfoutput\undefined
   \pdffalse
 \else
   \pdfoutput=1
   \pdftrue
\fi

\usepackage{color}

\def\be{\begin{equation}}
\def\ee{\end{equation}}

\renewcommand{\L}{\Lambda}

\newcommand{\R}{{\rm Re}}

\ifx\pdfoutput\undefined
   \pdffalse
\else
   \pdfoutput=1
   \pdftrue
\usepackage[sorting=none]{biblatex}
\newcommand{\Ydiag}{%
  \mathord{%
    \tikz[baseline=-0.9ex,scale=1.7,line cap=round,line join=round]{
      \draw[very thick] (0,0) -- (2.0,0)
        node[pos=0.42, above=2pt] {\scriptsize $G(x-y)$};

      \draw[very thick,dashed] (2.0,0) -- (3.0,0.9)
        node[pos=0.78, above right=0pt, xshift=3pt, fill=white, inner sep=1pt] {\scriptsize $c(y)$};

      \draw[very thick,dashed] (2.0,0) -- (3.0,-0.9)
        node[pos=0.78, below right=0pt, xshift=3pt, fill=white, inner sep=1pt] {\scriptsize $c(y)$};
    }%
  }%
}
\newcommand{\Tadpole}{%
  \mathord{%
    \tikz[baseline=-0.9ex,scale=1.7,line cap=round,line join=round]{
      \draw[very thick] (0,0) -- (2.0,0)
        node[pos=0.42, above=2pt] {\scriptsize $G(x-y)$};

      \draw[very thick] (3,0) circle[radius=1];

      \node[right=7pt] at (2.5,0) {\scriptsize $G(y-y)$};
    }%
  }%
}
\usepackage{tikz}
\usepackage{amsmath,amssymb}
\usetikzlibrary{calc}

\newcommand{\dder}[1]{\lhd\,\partial^{#1}}

\tikzset{
    solid/.style={line width=0.8pt},
    dashedline/.style={dash pattern=on 2pt off 2pt, line width=0.8pt},
    tri/.style={dashedline},
    circ/.style={
        draw, circle, fill=white,
        minimum size=0.42cm,
        inner sep=0pt,
        line width=0.8pt
    }
}

\newcommand{\diag}[1]{%
\tikz[baseline=-0.55ex,scale=0.85,line cap=round,line join=round]{#1}%
}

\newcommand{\Dprop}{%
\diag{\draw[dashedline] (0,0)--(0.85,0);}%
}

\newcommand{\DCircle}{%
\diag{
    \draw[dashedline] (0,0)--(0.55,0);
    \node[circ] at (0.82,0) {};
}%
}

\newcommand{\DVertex}{%
\diag{
    \draw[dashedline] (0,0.47)--(0,0.05);
    \draw[tri] (0,0.05)--(-0.32,-0.30);
    \draw[tri] (0,0.05)--(0.32,-0.30);
}%
}

\newcommand{\DCross}{%
\diag{
    \draw[tri] (0,0.38)--(0.65,-0.38);
    \draw[tri] (0.65,0.38)--(0,-0.38);
}%
}

\newcommand{\BranchCircle}{%
\diag{
    \draw[tri] (0,0.38)--(0.55,0);
    \draw[tri] (0,-0.38)--(0.55,0);
    \node[circ] at (0.86,0) {};
}%
}

\newcommand{\DCircleD}{%
\diag{
    \draw[dashedline] (0,0)--(0.55,0);
    \node[circ] at (0.85,0) {};
    \draw[dashedline] (1.15,0)--(1.85,0);
}%
}

\newcommand{\TwoCircles}{%
\diag{
    \node[circ] (cA) at (0,0) {};
    \node[circ, anchor=west] (cB) at (cA.east) {};
}%
}

\newcommand{\HalfCircle}{%
\diag{
    \node[circ] at (0.25,0) {};
    \draw[solid] (0.04,0)--(0.46,0);
}%
}

\newcommand{\SBranch}{%
\diag{
    \draw[solid] (0,0)--(0.70,0);
    \draw[tri] (0.70,0)--(1.15,0.34);
    \draw[tri] (0.70,0)--(1.15,-0.34);
}%
}

\newcommand{\SCircle}{%
\diag{
    \draw[solid] (0,0)--(0.65,0);
    \node[circ] at (0.95,0) {};
}%
}

\newcommand{\SSBranch}{%
\diag{
    \draw[solid] (0,0)--(0.70,0);
    \draw[tri] (0.70,0)--(1.3,0);
    \draw[tri] (0.70,0)--(1.15,0.34);
    \draw[tri] (0.70,0)--(1.15,-0.34);
}%
}

\newcommand{\SVertBranch}{%
\diag{
    \draw[solid] (0,0)--(1.20,0);
    \draw[dashedline] (0.62,0)--(0.62,0.58);
    \draw[tri] (1.20,0)--(1.65,0.34);
    \draw[tri] (1.20,0)--(1.65,-0.34);
}%
}

\newcommand{\SVrtCircle}{%
\diag{
    \draw[solid] (0,0)--(0.48,0);
    \draw[dashedline] (0.48,0)--(0.48,0.58);
    \node[circ] at (0.78,0) {};
}%
}

\newcommand{\SVertCircle}{%
\diag{
    \draw[solid] (0,0)--(1.05,0);
    \draw[dashedline] (0.48,0)--(0.48,0.58);
    \node[circ] at (1.35,0) {};
}%
}

\newcommand{\SCircleD}{%
\diag{
    \draw[solid] (0,0)--(0.62,0);
    \node[circ] at (0.92,0) {};
    \draw[dashedline] (1.22,0)--(1.90,0);
}%
}

\newcommand{\DSolidD}{%
\diag{
    \draw[tri] (0,0.34)--(0.45,0);
    \draw[tri] (0,-0.34)--(0.45,0);
    \draw[solid] (0.45,0)--(1.20,0);
    \draw[tri] (1.20,0)--(1.65,0.34);
    \draw[tri] (1.20,0)--(1.65,-0.34);
}%
}

\newcommand{\DSolidCircle}{%
\diag{
    \draw[tri] (0,0.34)--(0.45,0);
    \draw[tri] (0,-0.34)--(0.45,0);
    \draw[solid] (0.45,0)--(1.10,0);
    \node[circ] at (1.40,0) {};
}%
}

\newcommand{\CircleSolidCircle}{%
\diag{
    \node[circ] at (0.25,0) {};
    \draw[solid] (0.55,0)--(1.05,0);
    \node[circ] at (1.35,0) {};
}%
}

\begin{document}

\begin{titlepage}

\hskip 1.5cm

\begin{center}

{\huge\bf{Perturbative Nicolai-Map Diagrammatics: Application to Poincar\'{e} Supergravity}}
\vskip 0.8cm  
{\bf \large Ji-Seong Chae\orcidlink{0000-0002-1890-496X}$^{\dag}$\footnote{ jiseongchae17@gmail.com}, Hun Jang\orcidlink{0000-0002-9966-0775}$^{\ddag,\star}$\footnote{hun.jang@nyu.edu (Corresponding author)} and  Junhyeok Lee\orcidlink{0009-0003-3765-637X}$^{\dag}$\footnote{ junseyu2@gmail.com}}
\vskip 0.75cm
{\it $^{\dag}$Department of Physics, Hanyang University, Seoul 04763, South Korea\\
$^{\ddag}$Center for Quantum Spacetime (CQUeST), Sogang University, Seoul 04107, South Korea, \\
$^{\star}$Yukawa Institute for Theoretical Physics (YITP), Kyoto University, Kyoto 606-8502, Japan}
\vspace{12pt}
\end{center}

\begin{abstract}
We develop a perturbative, diagrammatic framework for constructing Nicolai
maps and apply it to four-dimensional $\mathcal{N}=1$ Poincar\'e supergravity
expanded around flat Minkowski space. It provides an alternative to the
coupling-flow-operator construction, which faces several obstructions when
extended to local supersymmetry. Expanding the bosonic effective action and
the Nicolai map jointly in the gravitational coupling $\kappa$ and the
loop-counting parameter $\hbar$, we derive the Nicolai-map defining conditions, i.e. the free-action and determinant-matching conditions, order by order. The diagrammatics enumerates
all admissible local terms in the Nicolai-map ansatz from the effective-action
diagrams and reduces the construction to a finite system of nonlinear
polynomial equations. Carried through order $\kappa^{2}$, the resulting
constraints are found to be independent of the detailed bosonic input and
hierarchical, order-$\kappa^{2}$ consistency further restricting the
order-$\kappa$ data. A consistent Nicolai-map construction for the Einstein--Hilbert
graviton sector is found to require the Rarita--Schwinger gravitino already at
this order: Einstein gravity admits a Nicolai map only through its
$\mathcal{N}=1$ supersymmetric completion, Poincar\'e supergravity, supporting Nicolai's
characterization of supersymmetry.
\end{abstract}

\vskip 1 cm
\vspace{24pt}
\end{titlepage}
\tableofcontents

\section{Introduction}
The Nicolai map, introduced by Hermann Nicolai in 1980, is a nonlocal, nonlinear field redefinition that sends an interacting supersymmetric theory to a free one for the Gaussian description \cite{Nicolai:1979nr, Nicolai:1980jc}. More concretely, for any operator $Y[\phi]$ built from the bosonic fields $\phi$ (and their derivatives), the Nicolai map $T_g$ relates the interacting theory expectation value at a certain coupling $g$ to a free theory correlator in the way:
\begin{align}
    \langle Y[\phi]\rangle_g=\langle Y[T_g^{-1}\phi]\rangle_0,
\end{align}
where $\langle \cdots \rangle_g$ and $\langle \cdots \rangle_0$ are the expectation values in interacting and free theories, respectively. Practically, the Nicolai map rewrites bosonic correlators of the interacting theory in terms of free field correlators, thereby replacing the original fermionic and ghost sector by a purely bosonic description \cite{Nicolai:1980jc,Arrighi:2025eym}. Thus, investigating the Nicolai map is of central importance to supersymmetry \cite{Lechtenfeld:2023gcq}.
In 1984, Lechtenfeld derived a formalism that yields a formal power series of the Nicolai map \cite{Lechtenfeld:1984me, Dietz:1984hf}. Recently, Lechtenfeld and Rupprecht proposed the coupling-flow-operator formalism \cite{Lechtenfeld:2021uvs}, whose Nicolai-map uniqueness can be ensured at certain magical values of a $\theta$ parameter in a topological $\theta$ term added to the action \cite{Lechtenfeld:2022qpa}, which is defined as
\begin{align}
    R_g[\phi]=\int d^dx \left(\partial_g T_g^{-1} \cdot T_g\right)\phi(x) \frac{\delta}{\delta \phi(x)}
\end{align}
where ‘$x$’ represents all coordinates the fields depend on. The infinitesimal version of the inverse Nicolai map, 
\begin{align}  
    \partial_g \langle Y[\phi]\rangle_g=\partial_g \langle Y[T_g^{-1}\phi]\rangle_0=\langle \partial_g Y[\phi]\rangle_g +\left\langle \int \left(\partial_g T_g^{-1}\phi\right)\frac{\delta Y}{\delta \phi}\left[T_g^{-1}\phi\right]\right\rangle_0=\langle\left(\partial_g+R_g[\phi]\right) Y[\phi]\rangle_g
\end{align}
is generated by the coupling flow operator $R_g$. The coupling flow operator $R_g$ can be obtained by using the fact that off-shell supersymmetric action is written as a supervariation and the supersymmetry Ward identity; hence $R_g$ enables an order-by-order reconstruction of $T_g$ and $T_g^{-1}$.
This formalism has been studied in various supersymmetric theories: off-shell $\mathcal{N}=1$ super Yang-Mills theory in general $D$ dimensions  \cite{Lechtenfeld:2021yjb} and its improvement \cite{Lechtenfeld:2022lvb} through the topological theta term \cite{Lechtenfeld:2022qpa}, off-shell $D =4$ $\mathcal N=1$ super Yang-Mills theory in general gauges \cite{Malcha:2021ess}, on-shell $\mathcal N=1$ super Yang-Mills constrained to the Landau gauge in $D = 3,4,6$ and 10 \cite{Ananth:2020lup, Ananth:2020jdr}, on-shell $\mathcal N=4$ super Yang-Mills in $D=4$ obtained from the 10 dimensional $\mathcal N=1$ super Yang-Mills by means of dimensional reduction \cite{Nicolai:2020tgo},  supersymmetric nonlinear sigma model in $D=4$ with four fermion interaction \cite{Lechtenfeld:2024ecf}, and super Yang-Mills theory on the light cone \cite{Lechtenfeld:2024uhi}. Moreover, the coupling-flow-operator formalism suggests a new route to explore quantization of supermembrane theory \cite{Lechtenfeld:2021zgd, Lechtenfeld:2022fxs}. 

In the meantime, a large order perturbative formulation of the Nicolai map using the coupling-flow-operator for supersymmetric theories has been investigated in Ref.~\cite{Lechtenfeld:2022qed}. However, efforts to obtain the supergravity Nicolai map within this formalism have been obstructed by issues, which are mentioned in detail in Sec.~\ref{comparison}, stemming from local supersymmetry \cite{Arrighi:2025eym}. We therefore set aside the coupling-flow-operator formalism and pursue {\it a perturbative diagrammatic construction of the Nicolai map in $D=4$, $\mathcal N=1$ pure supergravity}. 

Our strategy is to develop a diagrammatic framework to read off Nicolai maps perturbatively (see Sec.~\ref{section3}). In the absence of a workable
coupling-flow operator, we return to the defining identity of the Nicolai
map in Eq.~\eqref{Nicolai_map_condition} directly, expand both sides jointly in the loop parameter $\hbar$ and
the gravitational coupling $\kappa$, and match both the admissible diagrams and expansion coefficients order by
order. Our diagrammatic language developed in Sec.~\ref{section3} renders the enumeration of these
candidates both finite and self-consistent. The order-by-order defining identities for the Nicolai map in Eqs.\eqref{Condition_zero_zero}--\eqref{Condition_two_two} acquire direct graphical readings, and the
``anomalous'' diagrams that are absent in the effective action side but generated by lower-order data on the ansatz side
are forced to cancel against counterterm diagrams that are in turn
independently demanded by higher-order conditions.

The paper is organized as follows. In Sec.~\ref{section2}, we review the action of $D=4$, $\mathcal N=1$ pure supergravity and its local symmetries, and we introduce the Faddeev–Popov procedure to resolve the redundant integrations implied by these symmetries. We then apply the background field method to derive the quadratic (kinetic) terms around flat space and to organize the interaction terms. Next, we examine BRST symmetry, noting that it fails off-shell and motivates the inclusion of additional terms. 

In Sec.~\ref{section3}, we develop a perturbative diagrammatic construction of the Nicolai map, keeping $\hbar$ explicit, by considering the expansion parameters $(\hbar,\kappa)$ with identifying the ``order-by-order'' defining conditions for constructing the Nicolai map. We take advantage of a compact diagrammatic language by introducing elementary diagrammatic operations that make the diagrammatic bookkeeping and loop counting transparent. 

In Sec.~\ref{section4}, we algorithmically generate all the admissible structures including the relevant undetermined coefficients of the expansion for the Nicolai map $T_\kappa c$ to a given order in $\kappa$ (here we focus on the second order in $\kappa$, for instance), and implement a numerical technique for computing a finite set of coupled nonlinear polynomial equations for the undetermined coefficients that characterize the Nicolai map expansion and satisfy the defining conditions. We consider the Einstein-Hilbert action (with our gauge fixing) as the only bosonic sector, and study the resulting constraints on the coefficients of the Nicolai map. In this setting, the $\mathcal O(\kappa^2)$ constraints are universal depending only on the bosonic sector, and thus provide a sharp diagnostic of the admissible $\mathcal O(\kappa)$ gravitino couplings, which, in particular, agree with those induced by the $\mathcal N=1$ pure supergravity action. In Sec.~\ref{comparison}, we compare our diagrammatic construction
with the coupling-flow analysis of Ref.~\cite{Arrighi:2025eym}, explaining
how the obstructions identified there are circumvented in the present
approach and discussing the limitations common to both. We conclude in
Sec.~\ref{conclusion}.

In Appendix \ref{Appendix A}, we provide an explicit listing of all interaction terms to the second order in $\kappa$, say $\mathcal O(\kappa^2)$. In Appendix~\ref{Bosonic effective action}, we explicitly present the bosonic effective action to the order of $\mathcal O(\kappa^2)$, which we use to formulate the Nicolai map condition. Although our construction is carried out through order $\kappa^2$, the explicit defining conditions are already too lengthy to be displayed in the main text even at order $\kappa$. We therefore present the defining conditions up to order $\kappa$ in Appendix~\ref{Appendix C}. In Appendix~\ref{Appendix D}, we spell out the ``monster'' solution of the Nicolai map for pure supergravity to the second order in $\kappa$.

\section{On-shell BRST action for the 4D $\mathcal N=1$ pure supergravity}
\label{section2}
In this section, we present the on-shell BRST action for the pure 4D $\mathcal{N}=1$ supergravity in Ref.~\cite{StermanTownsendVanNieuwenhuizen:1978,Freedman:1976xh} following the conventions of the standard supergravity textbook by Freedman and Van Proeyen \cite{Freedman:2012zz}.

\noindent {\bf  Faddeev–Popov gauge fixing.} We begin with the first-order formalism for a pure 4D $\mathcal N=1$ supergravity action at on-shell:
\begin{align}
\label{action}
    S=\frac{1}{2\kappa^2} \int d^4x \ e \ e^{a\mu}e^{b\nu} R_{\mu\nu ab}(\omega)-\frac{1}{2}\int d^4x \ e \bar\psi_\mu \gamma^{\mu\nu\rho}D_\nu\psi_\rho,
\end{align}
where the curvature tensor and gravitino covariant derivative are given by
\begin{align}
\begin{gathered}
     R_{\mu\nu ab}(\omega) =\partial_\mu \omega_{\nu ab}-\partial_\nu \omega_{\mu ab} +\omega_{\mu ac} {{\omega_\nu}^c}_b-\omega_{\nu ac}{{\omega_\mu}^c}_b,
    \\  D_\nu\psi_\rho=\partial_\nu \psi_\rho+\frac{1}{4}\omega_{\nu ab}\gamma^{ab}\psi_\rho.
\end{gathered}
\end{align}
Varying the action~\eqref{action} with respect to the spin connection $\omega$ yields its equation of motion, whose solution is found to be
\begin{align}
    {\omega_\mu}^{ab}&={\omega_\mu}^{ab}(e)+{K_\mu}^{ab}
    \\ \nonumber &=\frac{1}{2}\left(e^{a\nu}\partial_\mu {e^b}_\nu - e^{a\nu}\partial_\nu {e^b}_\mu-e^{b\nu}\partial_\mu {e^a}_\nu+e^{b\nu}\partial_\nu {e^a}_\mu-e^{a\nu}e^{b\sigma}e_{c\mu}\partial_\nu{e^c}_\sigma+e^{b\nu} e^{a\sigma}e_{c\mu}\partial_\nu{e^c}_\sigma\right)
    \\ \nonumber &+\frac{\kappa^2}{4}\left( \bar\psi^a \gamma_\mu \psi^b +\bar\psi_\mu \gamma^a \psi^b+\bar\psi^a\gamma^b \psi_\mu\right),
\end{align}
where ${\omega_\mu}^{ab}(e)$ is the torsion-free Levi-Civita connection built from ${e^a}_\mu$ and ${K_\mu}^{ab}$ is the contorsion bilinear in the gravitino $\psi_\rho$. Thus ${\omega_\mu}^{ab}={\omega_\mu}^{ab}({e^a}_\mu, \psi_\rho)$.  Substituting the solution ${\omega_\mu}^{ab}={\omega_\mu}^{ab}({e^a}_\mu, \psi_\rho)$ into the first-order action~\eqref{action} gives the action in the second order formalism
\begin{align}
\label{action2}
    S&=\frac{1}{2\kappa^2}\int d^4x \ e \ e^{a\mu}e^{b\nu} \left(\partial_\mu \omega_{\nu ab}(e)-\partial_\nu \omega_{\mu ab}(e)+\omega_{\mu ac} (e) {{\omega_\nu}^c}_b(e) -\omega_{\nu ac}(e) {{\omega_\mu}^c}_b (e)\right)
    \\ \nonumber &+\frac{i}{2}\int d^4x \ \varepsilon^{\mu\nu\rho\sigma} \bar\psi_\mu \gamma_*\gamma_\sigma \left(\partial_\nu\psi_\rho+\frac{1}{4}\omega_{\nu ab}(e) \gamma^{ab} \psi_\rho\right)
    \\ \nonumber &-\frac{\kappa^2}{32}\int d^4x \ e\left(\left(\bar\psi^\rho\gamma^\mu\psi^\nu\right)\left(\bar\psi_\rho\gamma_\mu\psi_\nu+2\bar\psi_\rho\gamma_\nu\psi_\mu\right)-4\left(\bar\psi_\mu\gamma\cdot \psi\right)\left(\bar\psi^\mu \gamma \cdot \psi\right)\right).
\end{align}
The action~\eqref{action2} possesses three local symmetries, whose infinitesimal transformation laws are:
\begin{itemize}
    \item (Supersymmetry) the gauge parameter $\epsilon(x)$ is a Majorana spinor
\begin{align}
    \delta {e^a}_\mu=\frac{\kappa}{2}\bar\epsilon \gamma^a \psi_\mu, \qquad \delta\psi_\mu =\frac{1}{\kappa}D_\mu\epsilon.
\end{align}
    \item (Diffeomorphism) the gauge parameter $\xi^\lambda(x)$ is a vector field 
\begin{align}
    \delta{e^a}_\mu=\left(\xi^\lambda \partial_\lambda {e^a}_\mu+\partial_\mu \xi^\lambda {e^a}_\lambda\right),\qquad \delta\psi_\mu =\left(\xi^\lambda \partial_\lambda \psi_\mu+\partial_\mu \xi^\lambda \psi_\lambda\right).
\end{align}
    \item (Local Lorentz symmetry) the gauge parameter $\lambda_{ab}(x)$ is an antisymmetric local Lorentz tensor
\begin{align}
    \delta {e^a}_\mu =\left({\lambda^a}_b e^b_\mu\right),\qquad  \delta\psi_\mu =\frac{1}{4}\left(\lambda_{ab}\gamma^{ab}\psi_\mu\right).
\end{align}
\end{itemize}
Note that these local symmetries of the action lead to redundant integration over gauge-equivalent field configurations in the path integral:
\begin{align}
    \int \mathcal D{e^a}_\mu \mathcal D\psi_\mu \ e^{iS[\kappa; {e^a}_\mu, \psi_\mu]}.
\end{align}
We therefore implement gauge fixing via the Faddeev–Popov procedure. Since the local gauge symmetries are parametrized by fourteen independent functions $\phi^B(x)$, the Faddeev–Popov procedure requires an equal number of gauge conditions $F^A({e^a}_\rho, \psi_\mu)=0$ for $A=1,\dots,14$. Concretely, we choose the standard gamma-trace (Rarita–Schwinger) gauge for the gravitino, the de Donder (harmonic) gauge in the vielbein language for diffeomorphisms, and the symmetric-vielbein gauge to fix the local Lorentz one:
\begin{equation}
    F(\psi_\mu)=\gamma^a\eta^\mu_a \psi_\mu=0
\end{equation}
\begin{equation}
    \label{harmonic gauge}
    F^\nu({e^a}_\rho)=\partial_\mu\left(\eta^\mu_a {e^a}_\rho \eta^{\rho\nu}+\eta^\nu_a{e^a}_\rho\eta^{\rho\mu}-\eta^{\mu\nu}{e^a}_\rho \eta^\rho_a\right)=0
\end{equation}
\begin{equation}
\label{symmetry guage}
    F^{ab}({e^a}_\mu)={e^a}_\mu\eta^{\mu b}-{e^b}_\mu\eta^{\mu a}=0
\end{equation}
where $F^{ab}$ is antisymmetric. Note that the gauge conditions have commuting and anticommuting character: $F$ is anticommuting (odd in the spinor), while $F^\nu$ and $F^{ab}$ are commuting (even in the spinor). 

We insert the following identity into the path integral:
\begin{align}
1=\int \mathcal D\phi \;
\prod_A \delta\!\left(F^A\!\left(\left(e^a{}_\rho\right)^\phi,\left(\psi_\mu\right)^\phi\right)-\omega^A(x)\right)
\,
\mathrm{sdet}\!\left(
\frac{\delta F^A\!\left(\left(e^a{}_\rho\right)^\phi,\left(\psi_\mu\right)^\phi\right)}
{\delta \phi^B}
\right),
\end{align}
where \(\left(e^a{}_\rho\right)^\phi\) and \(\left(\psi_\mu\right)^\phi\) denote the fields obtained by the gauge transformations with parameters \(\phi^B(x)\), \(\omega^A(x)\) is an arbitrary function, and
\(\mathrm{sdet}\!\left(\delta F^A/\delta\phi^B\right)\) is the superdeterminant of the corresponding supermatrix containing both commuting and anticommuting components.

Assuming that the action and the functional measure are invariant under the local symmetries, we may change integration variables from \(e^a{}_\rho,\psi_\mu\) to their gauge-transformed fields \(\left(e^a{}_\rho\right)^\phi,\left(\psi_\mu\right)^\phi\). Since the integration variables are dummy, we may then relabel them again by \(e^a{}_\rho,\psi_\mu\). For the linearized gauge transformations considered here, the Faddeev--Popov operator \(\delta F^A/\delta\phi^B\) is independent of \(\phi^B\), so that the factor \(\int \mathcal D\phi\) contributes only an overall constant. Hence, the path integral becomes
\begin{align}
\int \mathcal D e^a{}_\rho \,\mathcal D\psi_\mu \; e^{iS[\kappa;e^a{}_\rho,\psi_\mu]}
&=
\int \mathcal D\phi \int \mathcal D e^a{}_\rho \,\mathcal D\psi_\mu \;
e^{iS[\kappa;e^a{}_\rho,\psi_\mu]}
\prod_A \delta\!\left(F^A(e^a{}_\rho,\psi_\mu)-\omega^A(x)\right)
\nonumber\\
&\qquad\times
\mathrm{sdet}\!\left(
\frac{\delta F^A\!\left(\left(e^a{}_\rho\right)^\phi,\left(\psi_\mu\right)^\phi\right)}
{\delta \phi^B}
\right).\label{Path_int}
\end{align}
Since this relation holds for arbitrary \(\omega^A(x)\), we may multiply to the right side of Eq.~\eqref{Path_int} by a normalized functional weight \cite{Faddeev:1967fc} given by
\begin{align}
1=N(\xi)\int \mathcal D\omega^A \,
\exp\left[
i\frac{\xi}{2}\int d^4x\, \omega^A (x) Y_{AB}\omega^B (x)
\right],
\end{align}
where \(N(\xi)\) is an unimportant normalization factor, \(Y_{AB}\) denotes the gauge fixing kernel. The delta functional \(\delta(F^A-\omega^A)\) then enforces \(\omega^A=F^A\), so that the $\omega^A$ integration yields the gauge fixing Lagrangian
\begin{align}
\label{gauage_fixing_term}
\mathcal L_{\rm gauge-fixing}
&=
\frac{1}{4}\bar\psi_\nu \eta^\nu_b \gamma^b\gamma^c \eta_c^\rho \partial_\rho(\gamma^a\eta^\mu_a\psi_\mu)
-\frac{1}{4\kappa^2} \eta_{\nu\sigma}\partial_\mu\left(\eta^\mu_a e^a{}_\rho\eta^{\rho\nu}+\eta^\nu_a e^a{}_\rho\eta^{\rho\mu}-\eta^{\mu\nu}e^a{}_\rho\eta^\rho_a\right)
\nonumber\\
&\quad\times
\partial_\alpha\left(\eta^\alpha_b e^b{}_\beta \eta^{\beta\sigma}+\eta^\sigma_b e^b{}_\beta \eta^{\beta\alpha}-\eta^{\alpha\sigma} e^b{}_\beta \eta^\beta_b\right)
-\frac{\lambda}{\kappa^4}\eta_{ac}\eta_{bd}
\left(e^a{}_\mu \eta^{\mu b}-e^b{}_\mu \eta^{\mu a}\right)
\left(e^c{}_\nu \eta^{\nu d}-e^d{}_\nu \eta^{\nu c}\right),
\end{align}
which is added to the original Lagrangian. We may choose $\lambda$ to be any arbitrary constant that is later taken to infinity, making the vielbein perturbation field a symmetric tensor.

The superdeterminant can be expressed in terms of ghost fields as
\begin{align}
    \text{sdet}\left(\frac{\delta F^A}{\delta \phi^B}\right)=\prod_{A,B}\int \mathcal D \bar C_A \mathcal D C^B \ \exp\left(i\int d^4x \ \bar C_A \frac{\delta F^A}{\delta \phi^B}C^B\right)
\end{align}
where $\bar C_A$ and $C^B$ are ghost fields: they are commuting when the subscript corresponds to a fermionic index and anticommuting when it corresponds to a bosonic one. Thus, we may add the ghost term to the original Lagrangian. Explicitly, we have
\begin{align}
\label{ghost_lagrangian}
    \mathcal L_{\rm ghost}&=B\gamma^a\eta^\mu_a D_\mu C-\kappa B \gamma^a\eta^\mu_a(C^\lambda\partial_\lambda\psi_\mu+\partial_\mu C^\lambda \psi_\lambda)-\frac{1}{4}\kappa B \gamma^a \eta^\mu_a(C_{ef}\gamma^{ef}\psi_\mu)
    \\ \nonumber &-\frac{\kappa}{2}{C^*}_\nu \partial_\mu \left(\eta^\mu_a\left(\bar\psi_\rho\gamma^a C\right)\eta^{\rho\nu}+\eta^\nu_a \left(\bar\psi_\rho \gamma^a C\right)\eta^{\rho\mu}-\eta^{\mu\nu} \left(\bar\psi_\rho \gamma^a C\right) \eta^\rho_a\right)
    \\ \nonumber &+ {C^*}_\nu\partial_\mu \left[\eta^\mu_a\left( C^\lambda \partial_\lambda {e^a}_\rho+\partial_\rho C^\lambda {e^a}_\lambda\right)\eta^{\rho\nu}+\eta^\nu_a \left( C^\lambda \partial_\lambda{e^a}_\rho +\partial_\rho C^\lambda {e^a}_\lambda\right)\eta^{\rho\mu}-\eta^{\mu\nu}\left(C^\lambda \partial_\lambda {e^a}_\rho+\partial_\rho C^\lambda {e^a}_\lambda \right)\eta^\rho_a\right]
    \\ \nonumber &+C^*_\nu \partial_\mu \left[ \eta^\mu_a \left({C^a}_e {e^e}_\rho\right)\eta^{\rho\nu} +\eta^\nu_a\left({C^a}_e {e^e}_\rho\right)\eta^{\rho\mu}-\eta^{\mu\nu} \left({C^a}_e {e^e}_\rho\right)\eta^\rho_a \right]
    \\ \nonumber &-\frac{\kappa}{2} {C^*}_{ab}\left(\left(\bar\psi_\mu \gamma^a C\right)\eta^{\mu b}-\left(\bar\psi_\mu\gamma^b C\right)\eta^{\mu a}\right)
    \\ \nonumber &+ {C^*}_{ab}\left(\left(C^\lambda \partial_\lambda {e^a}_\mu+\partial_\mu C^\lambda{e^a}_\lambda \right)\eta^{\mu b}-\left(C^\lambda \partial_\lambda {e^b}_\mu +\partial_\mu C^\lambda {e^b}_\lambda\right)\eta^{\mu a}\right)
    \\ \nonumber &+{C^*}_{ab}\left(\left( {C^a}_e {e^e}_\mu\right)\eta^{\mu b}-\left( {C^b}_e {e^e}_\mu\right)\eta^{\mu a}\right), \label{gauge fixing lagrangian}
\end{align}
where $B$ and $C$ are commuting spinor fields, ${C^*}_\nu$ and $C^\lambda$ are complex anticommuting general vector fields, and ${C^*}_{ab}$ and $C_{ab}$ are complex anticommuting Lorentz antisymmetric tensor fields. 

\noindent {\bf Background field method for vielbein.} We employ the background field method to expand the Lagrangian perturbatively \cite{DeWitt:1967ub, DeWitt:1967uc}. In this approach, the vielbein field ${e^a}_\mu$ is expanded in the gravitational coupling $\kappa$ as 
\begin{eqnarray}
\begin{gathered}
    {e^a}_\mu={\eta^a}_\mu +\kappa {c^a}_\mu,
    \\ {e_a}^\mu={\eta^\mu}_a-\kappa {c^\mu}_a+\kappa^2 {c^\mu}_c {c^c}_a-\kappa^3 {c^\mu}_d {c^d}_c {c^c}_a +\cdots
    \end{gathered}
\end{eqnarray}
where ${c^a}_\mu$ is a perturbed field for the vielbein around a flat background. So far, we have kept general covariance and local Lorentz invariance distinct, but from this point on, we adopt a flat background, making it unnecessary to distinguish between Greek and Latin indices. 

Taking $\lambda \rightarrow \infty$ in the third term of the gauge-fixing Lagrangian~\eqref{gauage_fixing_term}, the functional from the third term 
\begin{equation}
    \exp\left[-i\int d^4x \ \frac{\lambda}{\kappa^2}\left(c^{ab}-c^{ba}\right)\left(c_{ab}-c_{ba}\right)\right]
\label{lambda_gauge_fixing_term}
\end{equation}
oscillates rapidly except near $c^{ab}=c^{ba}$ \cite{vanNieuwenhuizen:1981uf}. Consequently, the path integral is effectively localized on $c^{ab}=c^{ba}$, so the vielbein perturbation field ${c^a}_\mu$ can be treated as a symmetric tensor. 

The sum of the gauge fixing term $\mathcal L_{\rm gauge-fixing}$~\eqref{gauage_fixing_term} and kinetic terms $\mathcal L_{kin}$ for the vielbein perturbation field ${c^a}_\mu$ and the gravitino $\psi_\mu$ from the action~\eqref{action2} gives the following remnant as free kinetic Lagrangians for the fields ${c^\nu}_a, \psi_{\mu}$ in the limit $\lambda \rightarrow \infty$:
\begin{align}
 \mathcal L_{kin} + \mathcal L_{\rm gauge-fixing} =\eta^{\gamma\mu}\partial_\mu {c^\nu}_a\partial_\gamma {c^a}_\nu-\frac{1}{2}\eta^{\gamma\mu}\partial_\mu {c^a}_a \partial_\gamma {c^b}_b+\frac{1}{4}\bar\psi_\mu \gamma^\rho\gamma^\nu\gamma^\mu \partial_\nu \psi_\rho.
\end{align}
Next, we note that the kinetic term of the ghost field from Eq.~\eqref{ghost_lagrangian}, 
\begin{align}
    \mathcal L_{\rm free-ghost} = B\gamma^\mu \partial_\mu C+ {C^*}_\nu \partial_\mu \partial^\mu C^\nu +{C^*}_{ab}\left(\partial^b C^a-\partial^a C^b\right)+{C^*}_{ab}\left(C^{ab}-C^{ba}\right),
\end{align}
is not diagonalized. Meanwhile, by redefining the ghost tensor in the way 
\begin{align}
    C'^{ab}=C^{ab}+\frac{1}{2}\partial^bC^a-\frac{1}{2}\partial^aC^b,
\end{align}
we can diagonalize the ghost kinetic term as follows:
\begin{align}
    \mathcal L_{\rm free-ghost}=B \gamma^\mu \partial_\mu C+{C^*}_\nu \partial_\mu\partial^\mu C^\nu+2 {C^*}_{ab}C'^{ab}.
\end{align}
where $2 {C^*}_{ab}C'^{ab}$ is written in terms of the non-propagating ghost tensors, which is used in Ref.~\cite{PhysRevD.22.2995}.
Finally, propagators of the gravitino $\psi_\mu$ and the ghost field are found to be
\begin{align}
    \left\langle \psi_\rho(x) \bar\psi_\mu(y) \right\rangle=-\frac{1}{2}\int\frac{d^4k}{(2\pi)^4} \frac{\gamma_\mu k^\alpha \gamma_\alpha\gamma_\rho \ e^{-ik\cdot(x-y)}}{k^2},
    \\ \langle C(x)B(y)\rangle=-\int\frac{d^4k}{(2\pi)^4} \frac{k^\alpha \gamma_\alpha e^{-ik\cdot(x-y)}}{k^2},
    \\ \left\langle C^\nu(x) {C^*}_\rho(y)\right\rangle=-i\delta^\nu_\rho \int \frac{d^4k}{(2\pi)^4} \frac{e^{-ik\cdot(x-y)}}{k^2},
    \\ \left\langle C'^{ab}(x){C^*}_{cd}(y)\right\rangle=\frac{1}{4}i\delta(x-y)\left(\delta^a_c\delta^b_d-\delta^a_d\delta^b_c\right).
\end{align}
After taking the background field method, the interaction Lagrangian can be presented to a given order in $\kappa$ as shown in Appendix \ref{Appendix A} (where we present the results of order $\mathcal O(\kappa^2)$). 

\noindent {\bf BRST invariance of the action.} We now consider BRST invariance of the action. The BRST transformations are given by 
\begin{align}
    \delta {e^a}_\mu&=\kappa^2\left(-\frac{\kappa}{2}\left(\bar\psi_\mu\gamma^a C\right) \Lambda+\left( C^\lambda\partial_\lambda {e^a}_\mu+\partial_\mu C^\lambda {e^a}_\lambda\right)\Lambda+\left( {C^a}_e {e^e}_\mu\right)\Lambda\right),
    \\ \delta\psi_\mu&=\kappa^2\left(\frac{1}{\kappa} D_\mu C\Lambda-\left(C^\lambda\partial_\lambda \psi_\mu+\partial_\mu C^\lambda \psi_\lambda\right)\Lambda-\frac{1}{4}\left(C_{ef}\gamma^{ef}\psi_\mu\right)\Lambda\right),
    \\ \delta B&=\frac{1}{2}\kappa \partial_\rho\left(\bar\psi_\nu \eta^\nu_b \gamma^b\right)\gamma^c {\eta_c}^\rho \Lambda,
    \\ \delta{C^*}_\sigma&=-\frac{1}{2}\eta_{\nu\sigma}\partial_\mu \left(\eta^\mu_a {e^a}_\rho \eta^{\rho\nu}+\eta^\nu_a{e^a}_\rho \eta^{\rho\mu}-\eta^{\mu\nu} {e^a}_\rho \eta^\rho_a\right)\Lambda,
    \\ \delta {C^*}_{ab}&=-\frac{\lambda}{\kappa^2} \eta_{ac}\eta_{bd}\left({e^c}_\nu \eta^{\nu d}-{e^d}_\nu \eta^{\nu c}\right)\Lambda,
    \\ \delta C&=\kappa^2 \left(C^\nu\partial_\nu C+\frac{1}{4}\gamma^{ab}C_{ab} C\right)\Lambda-\kappa^3\frac{1}{4}\psi_\mu \left(\bar C\gamma^\mu C\right)\Lambda,
    \\ \delta C^\lambda&=\kappa^2\left(-C^\nu\partial_\nu C^\lambda\right)\Lambda+\kappa^2\frac{1}{4}\left(\bar C\gamma^\lambda C\right)\Lambda,
    \\ \delta C^{ab}&=\kappa^2\left(-C^{\nu}\partial_\nu C^{ab}-C^{ac}{C_c}^b\right)\Lambda +\kappa^2 \frac{1}{4}{{\omega_\mu}^{ab}}\left(\bar C\gamma^\mu C\right) \Lambda, 
\end{align}
where $\Lambda$ is the constant anticommuting BRST parameter, $\lambda$ is the gauge-fixing parameter appearing in the third term of the gauge-fixing Lagrangian, as shown in Eq.~\eqref{lambda_gauge_fixing_term}, and $\bar C=-iC^T \gamma^3\gamma^1$ is the Majorana conjugate of $C$. However, BRST invariance fails off-shell \cite{Freedman:1976py}. In fact, the variation of the Lagrangian under BRST transformation is 
\begin{align}
    \delta_{\textrm{BRST}} \mathcal L 
    &=\frac{1}{64}\kappa^3 B \bigg(-6\gamma_\lambda\gamma_\sigma\gamma^{\lambda\mu\nu}D_\mu \psi_\nu\left(\bar C\gamma^\sigma C\right)+8 g_{\sigma\lambda} \gamma^{\lambda \mu\nu} D_\mu\psi_\nu\left(\bar C \gamma^\sigma C\right) \nonumber\\
    &\qquad\qquad\qquad-\gamma_\mu\gamma_{\lambda\rho}\gamma^{\mu\nu\sigma} D_\nu\psi_\sigma \left(\bar C\gamma^{\lambda\rho} C\right)\bigg)\Lambda.\label{BRST_anomaly}
\end{align}
If the action is not invariant under the BRST transformation, the unitarity of the S-matrix is not guaranteed. To restore BRST invariance, we modify the BRST transformation of the gravitino $\psi_\mu$ by adding the following additional term to its original BRST transformation:
\begin{align}
    \delta'\psi_\mu=-\frac{1}{64e}\kappa^3\left(-6\gamma_\sigma \gamma_\mu \bar B\left(\bar C \gamma^\sigma C\right)+8 g_{\sigma\mu} \bar B\left(\bar C \gamma^\sigma C\right)- \gamma_{\lambda\rho}\gamma_\mu \bar B\left(\bar C \gamma^{\lambda\rho} C\right)\right)\Lambda,\label{add_var_gravitino}
\end{align}
where $\bar B=-i\gamma^3\gamma^1B^T$ is the Majorana conjugate of $B$. The additional variation of the gravitino $\psi_\mu$ yields the following contribution from the kinetic term:
\begin{align}
    -\frac{1}{2}e \delta'(\bar\psi_\mu\gamma^{\mu\nu\rho}D_\nu\psi_\rho)=-\, e \, D_\nu \bar{\psi}_\rho \, \gamma^{\mu\nu\rho} \, \delta'\psi_\mu,
\end{align}
which precisely cancels $\delta_{\mathrm{BRST}} \mathcal{L}$~\eqref{BRST_anomaly}. However, the modified transformation~\eqref{add_var_gravitino} unfortunately gives rise to an unpleasant remnant from the first term of the gauge-fixing Lagrangian~\eqref{gauage_fixing_term}:
\begin{align}
\label{unpleasant_remnant}
\frac{1}{4}\delta'(\bar\psi_\nu \eta^\nu_b \gamma^b\gamma^c \eta_c{}^\rho \partial_\rho(\gamma^a\eta^\mu_a\psi_\mu))
&=-\frac{1}{2}\partial_\rho(\bar\psi_\mu \eta^\mu_a \gamma^a)\gamma^c \eta_c{}^\rho \gamma^b\eta^\nu_b\delta'\psi_\nu\nonumber\\
&=\frac{5}{32}\kappa^3 \partial_\rho \left(\bar{\psi}_\mu \eta^\mu_a \gamma^a\right)\gamma^c \eta^\rho_c \gamma_d \bar{B}\left(\bar{C}\gamma^d C\right)\Lambda+\mathcal{O}(\kappa^4).
\end{align}
 To cancel this term, we introduce the following additional interaction term \cite{StermanTownsendVanNieuwenhuizen:1978, Kallosh:1978de}:
\begin{align}
    \mathcal L_{\rm additional} =\frac{5}{32}\kappa^2 B\gamma_d\bar B\left( \bar C \gamma^d C\right), \label{add2}
\end{align}
whose BRST variation is induced by the transformation of $B$,
\begin{align}
    \delta_{BRST,B}\mathcal L_{\rm additional}=\frac{5}{16}\,\kappa^2 \delta B \,\gamma_d \,\bar{B}\,\bigl(\bar{C} \gamma^d C\bigr)
\end{align}
compensates the remnant~\eqref{unpleasant_remnant} at $\mathcal O(\kappa^3)$. The variation of Eq.~\eqref{add2} induced by the BRST transformation of the field $C$ appears at the higher order of $\mathcal{O}(\kappa^4)$. Up to the second order in $\kappa$, we can ignore such a fourth contribution by the BRST transformation of the field $C$. Similarly, by iteratively introducing compensating terms into the action and the BRST transformations in order for the variation to vanish at each order in $\kappa$, the action can be rendered into the BRST invariant in the order-by-order manner. 

\noindent {\bf BRST invariance of the functional measure.} The BRST invariance of the gauge-fixed action is not by itself sufficient to guarantee the BRST invariance of the generating functional. The functional integration measure must also be invariant. Accordingly, BRST invariance must extend to the full path-integral measure for the corresponding Ward identities to follow \cite{StermanTownsendVanNieuwenhuizen:1978}.

However, the naive functional measure is not invariant under the generalized BRST transformations. Jacobian of the BRST transformations is

\begin{equation}
    \text{sdet}\left(\frac{\delta \Phi'_B(y)}{\delta \Phi_A(x)}\right)=1-2 \kappa^3 \delta(0) \int d^4x \bar{C} \gamma^\mu \psi_\mu \Lambda ,
\end{equation}

\noindent where $\Phi_A(x)$ collectively denotes all bosonic and fermionic fields appearing in the functional integral and $\Phi'_B(y)$ are BRST transformed fields. The nontrivial contribution to this Jacobian arises from the BRST transformation of the commuting spinor ghost field $C$ and from the torsion-dependent part of the gravitino $\psi_\mu$ transformation. Thus, an explicit factor is required in the functional measure in order to compensate for the non-unit Jacobian. The required factor is 

\begin{equation}
\prod_x e(x)^{-4}=\exp\left[-4\,\delta(0)\int d^4x\,\ln e(x)\right], 
\end{equation}

\noindent where the BRST variation of this factor is

\begin{equation}
\delta \prod_x[e(x)]^{-4}=2 \kappa^3 \delta(0) \int d^4x \bar{C} \gamma^\mu \psi_\mu \Lambda \prod_x[e(x)]^{-4}.   
\end{equation}

Alternatively, the explicit factor $e^{-4}$ may be exponentiated by introducing four copies of a complex commuting ghost pair $u^I,u_I^*, (I=1,\ldots,4)$. In particular,

\begin{equation}
\prod_x[e(x)]^{-4}=\mathcal N_u \int \prod_{I=1}^{4} \mathcal D u_I^* \mathcal D u^I exp\left[i\int d^4x e u_I^*u^I\right],
\end{equation}

\noindent where $\mathcal N_u$ is a field-independent normalization constant. The generating functional may therefore be written as

\begin{equation}
\int \mathcal D\Phi \mathcal D u^* \mathcal D u \exp\left[ iS_{BRST}[\Phi]+iS_u[e,u^*,u]\right],
\end{equation}

\noindent where $S_{BRST}[\Phi]$ is BRST invariant action composed of original fields $\Phi$ and the additional action $S_u[e,u^*,u]$ is

\begin{equation}
S_u[e,u^*,u] =\int d^4x e u_I^*u^I.
\end{equation}

The fields $u^I$ and $u_I^*$ are commuting and ultralocal: they possess no derivative kinetic term. Their functional integration simply reproduces the measure factor $e^{-4}$; they do not constitute an additional Faddeev–Popov ghost sector.

\section{Perturbative Nicolai-map diagrammatics: application to Poincar\'{e} supergravity}
\label{section3}
In this section, we set up the perturbative construction of the Nicolai map and work out its
order-by-order structure. Motivated by 4D $\mathcal N=1$ pure supergravity, the perturbative framework is developed here, but it can be applied more broadly to 4D gravitational theories with couplings to gravitino fields, in which $\kappa$ is the coupling constant. To keep the notation general and uncluttered, hereafter \(c\) denotes the vielbein perturbation field with its indices suppressed.

\subsection{Perturbation of Nicolai map and its order-by-order defining conditions}
\noindent {\bf Order-by-order defining conditions for the Nicolai map.} We establish the condition for the Nicolai map $T_\kappa c$ of the 4D $\mathcal{N}=1$ pure supergravity according to Ref.~\cite{Nicolai:1980jc}, which is given by
\begin{align}
    S[c;\kappa]=S[T_\kappa c, 0, 0, 0; 0]-i\hbar\trace \ln \left(\frac{\delta T_\kappa c}{\delta c}\right), \label{Nicolai_map_condition}
\end{align}
where $S[c, \psi, \bar C, C; \kappa]$ is the gauge-fixed action for the vielbein perturbation field $c$, the gravitino field $\psi$, the ghost fields $\bar C, C$ and $S[c;\kappa]$ is the bosonic effective action obtained by functionally integrating out all fields except the vielbein perturbation field $c$. Once a map $T_\kappa c$ satisfying condition~\eqref{Nicolai_map_condition} is constructed, we can compute the bosonic vacuum expectation value $\langle X[c]\rangle_\kappa$ as
\begin{align}
    \langle X[c]\rangle_\kappa
    =
    \langle X[T^{-1}_\kappa c]\rangle_0 ,
\end{align}
where the subscript $0$ denotes the vacuum expectation value in the free theory.

In order to find the condition \eqref{Nicolai_map_condition} order-by-order, we start with the $r$-loop contribution to the bosonic effective action $S[c;\kappa]$ that scales as $\hbar^r$.
Accordingly, we expand the bosonic effective action in powers of $\hbar$ ordered by loop number,
\begin{align}
\label{bosonic_effective_action_hbar_expansion}
    S[c;\kappa]=\sum^{\infty}_{r=0} \hbar^r\, S^{(r)}[c;\kappa],
\end{align}
where $S^{(r)}[c;\kappa]$ denotes the $r$-loop contribution.
Since we work perturbatively in the coupling $\kappa$, we further expand each loop contribution as
\begin{align}
    S^{(r)}[c;\kappa]=\sum^{\infty}_{n=0} \kappa^n\, S^{(r,n)}[c],
\end{align}
leading to the full expansion of the action $S$ in the two parameters $\hbar$ and $\kappa$
\begin{eqnarray}
    S[c;\kappa]=\sum^{\infty}_{r=0}\sum^{\infty}_{n=0} \hbar^r \kappa^n\, S^{(r,n)}[c].\label{full_expansion_of_S}
\end{eqnarray}
On the other hand, as introduced in Ref.~\cite{Lechtenfeld:2024ecf}, we can also take the ``quantum'' Nicolai map with the loop corrections by expanding it in the parameter $\hbar$,
\begin{align}
    T_\kappa c
    =\sum_{r=0}^\infty \hbar^r\, T^{(r)}_\kappa c,
\end{align} 
where we consider the perturbations with respect to the parameter $\kappa$ as follows
\begin{eqnarray}
  T^{(r)}_\kappa c=\sum_{n=0}^\infty \kappa^n\, T^{(r,n)} c,  
\end{eqnarray}
and thus have the full expansion of the Nicolai map in the two parameters $\hbar$ and $\kappa$
\begin{eqnarray}
    T_\kappa c=\sum_{r=0}^\infty\sum_{n=0}^\infty \hbar^r \kappa^n\, T^{(r,n)} c.\label{full_expansion_of_T}
\end{eqnarray}
Finally, upon inserting the expansions from Eqs.~\eqref{full_expansion_of_S} and \eqref{full_expansion_of_T} into condition~\eqref{Nicolai_map_condition}, we obtain the recursion relation for the coefficients $S^{(r,n)}$ and $T^{(r,n)}c$ of the expansions of the action $S$ and Nicolai map $T_\kappa c$ as follows
\begin{align}
\Big(S[c;\kappa]\Big)^{(r,n)}=\left(S\left[T_\kappa c, 0, 0, 0; 0\right]\right)^{(r,n)}-i\hbar \left( \trace \ln \left(\frac{\delta T_\kappa c}{\delta c}\right)\right)^{(r-1,n)}. \label{recursion_relation}
\end{align}
Now we are in a position to derive the order-by-order defining conditions for Nicolai map by expanding each of the terms on both sides of condition~\eqref{Nicolai_map_condition} (or equivalently the recursion relation \eqref{recursion_relation}) in powers of $\hbar$ and $\kappa$, and then matching coefficients order-by-order. To do this, we first denote the bosonic free action for simplicity
\begin{align}
    S[c,0,0,0;0]=-\frac{1}{2}\int d^4x \left(-c_{ab}\,\square\, c^{ab}
    +\frac{1}{2} {c^a}_a\,\square\, {c^b}_b \right) \equiv c\boxtimes  c,\label{free_action}
\end{align}
where $\square$ is the (relativistic) d'Alembertian operator, and we introduce another ``box'' operator, for simplicity, denoted by $\boxtimes$ for representing the integral operation in Eq.~\eqref{free_action}.
Then, according to Eq.~\eqref{full_expansion_of_T}, we can write the Nicolai map perturbatively in the powers of the two parameters $\hbar$ and $\kappa$ as
\begin{align}
T_\kappa c
=
c+\kappa\,T^{(0,1)}c+\hbar\kappa\,T^{(1,1)}c
+\kappa^2\,T^{(0,2)}c+\hbar\kappa^2\,T^{(1,2)}c+\cdots,
\label{real_component_expansion}
\end{align}
where $T^{(0,0)}c = c$ is trivially determined. Substituting this into the bosonic free action gives
\begin{align}
S[T_\kappa c,0,0,0;0]
=(T_\kappa c)\boxtimes (T_\kappa c) &= c\boxtimes c
+2\kappa\,c\boxtimes T^{(0,1)}c
+2\hbar\kappa\,c\boxtimes T^{(1,1)}c \nonumber\\
&\quad
+\kappa^2\!\left(2c\boxtimes T^{(0,2)}c+T^{(0,1)}c\boxtimes T^{(0,1)}c\right)\nonumber\\
&\quad
+\hbar\kappa^2\!\left(2c\boxtimes T^{(1,2)}c+2T^{(0,1)}c\boxtimes T^{(1,1)}c\right)\nonumber\\
&\quad
+\hbar^2\kappa^2\,T^{(1,1)}c\boxtimes T^{(1,1)}c+\cdots,\label{action_expansion}
\end{align}
which is a perturbative expansion of the Nicolai-mapped action to be used for finding the order-by-order Nicolai-map defining conditions.
Next, we expand the functional derivative of the Nicolai map with respect to the vielbein perturbation field in the way
\begin{align}
\frac{\delta T_\kappa c}{\delta c}
=
1+\kappa A+\hbar\kappa B+\kappa^2 C+\hbar\kappa^2 D+\cdots ,
\end{align}
where we define
\begin{align}
A\equiv\frac{\delta T^{(0,1)}c}{\delta c},
\qquad
B\equiv\frac{\delta T^{(1,1)}c}{\delta c},
\qquad
C\equiv\frac{\delta T^{(0,2)}c}{\delta c},
\qquad
D\equiv\frac{\delta T^{(1,2)}c}{\delta c}.
\end{align}
Then, using the relation $\Tr\ln(1+X)=\Tr\!\left(X-\frac12X^2+\cdots\right)$, we obtain the expansion
\begin{align}
\Tr\ln\!\left(\frac{\delta T_\kappa c}{\delta c}\right)&=
\kappa\,\Tr A
+\hbar\kappa\,\Tr B
+\kappa^2\,\Tr\!\left(C-\frac12A^2\right)
+\hbar\kappa^2\,\Tr (D-AB)
+\cdots ,\nonumber\\
&=
\kappa\,\Tr \frac{\delta T^{(0,1)}c}{\delta c}
+\hbar\kappa\,\Tr \frac{\delta T^{(1,1)}c}{\delta c}
+\kappa^2\,\Tr\!\left(\frac{\delta T^{(0,2)}c}{\delta c}-\frac12\Big(\frac{\delta T^{(0,1)}c}{\delta c}\Big)^2\right)\nonumber\\
&\qquad +\hbar\kappa^2\,\Tr\bigg( \frac{\delta T^{(1,2)}c}{\delta c} - \frac{\delta T^{(0,1)}c}{\delta c}\frac{\delta T^{(1,1)}c}{\delta c}\bigg)
+\cdots, \label{Trace_expansion}
\end{align}
which is a perturbative expansion of the trace part to be used for finding the conditions.
By inserting the three expansions~\eqref{full_expansion_of_S}, \eqref{action_expansion}, and \eqref{Trace_expansion} into the recursion relation in Eq.~\eqref{recursion_relation}, we can compare and match the coefficients at each order in $\hbar$ and $\kappa$. At order $(r,n)=(0,0)$, we find the free action as a trivial solution:
\begin{align}
S^{(0,0)}=c\boxtimes c .
\end{align}
At order $(r,n)=(0,1)$, we read
\begin{align}
S^{(0,1)}=2c\boxtimes T^{(0,1)}c .
\end{align}
At order $(r,n)=(1,1)$, we have $2c\boxtimes T^{(1,1)}c$ from the action, while we read
$-i\,\Tr\!\bigl(\delta T^{(0,1)}c/\delta c\bigr)$ from the trace part; so that, we obtain
\begin{align}
S^{(1,1)}
=
2c\boxtimes T^{(1,1)}c
-i\,\Tr\!\left(\frac{\delta T^{(0,1)}c}{\delta c}\right).
\end{align}
At order $(r,n)=(0,2)$, we see
\begin{align}
S^{(0,2)}
=
2c\boxtimes T^{(0,2)}c
+
T^{(0,1)}c\boxtimes T^{(0,1)}c .
\end{align}
At order $(r,n)=(1,2)$, we read
\(
2c\boxtimes T^{(1,2)}c
+
2T^{(0,1)}c\boxtimes T^{(1,1)}c
\),
and
\(
-i\,\Tr\!\bigl(\delta T^{(0,2)}c/\delta c
-\frac12(\delta T^{(0,1)}c/\delta c)^2\bigr)
\)
from the action and trace part, respectively. Thus, we identify
\begin{align}
S^{(1,2)}
&=
2c\boxtimes T^{(1,2)}c
+
2T^{(0,1)}c\boxtimes T^{(1,1)}c 
-i\,\Tr\!\left(
\frac{\delta T^{(0,2)}c}{\delta c}
-\frac12\left(\frac{\delta T^{(0,1)}c}{\delta c}\right)^2
\right).
\end{align}
Finally, at order $(r,n)=(2,2)$, we have
\(
T^{(1,1)}c\boxtimes T^{(1,1)}c
\)
from the action, and 
\(
-i\,\Tr\!\bigl(\delta T^{(1,2)}c/\delta c\bigr)
\)
from the trace part, leading to
\begin{align}
S^{(2,2)}
=
T^{(1,1)}c\boxtimes T^{(1,1)}c
-i\,\Tr\!\left(\frac{\delta T^{(1,2)}c}{\delta c}\right)+\,i\Tr\left( \frac{\delta T^{(0,1)}c}{\delta c}\frac{\delta T^{(1,1)}c}{\delta c}\right).
\end{align}
Collecting these six results, we find the {\it defining conditions for the Nicolai map} \(T_\kappa c\) up to order $\kappa^2$, i.e. the free-action and determinant-matching conditions, which is given by the following set of simultaneous equations
\begin{align}
    &\quad S^{(0,0)}= c  \boxtimes  c,\label{Condition_zero_zero}
    \\ \label{Condition_zero_one} &\quad S^{(0,1)}=2c  \boxtimes  T^{(0,1)} c,
    \\  \label{Condition_one_one} &\quad S^{(1,1)}=2c  \boxtimes  T^{(1,1)} c -i \trace \left(\frac{\delta T^{(0,1)} c}{\delta c}\right),
    \\ \label{Condition_zero_two} &\quad S^{(0,2)}=2c  \boxtimes  T^{(0,2)} c+T^{(0,1)}c  \boxtimes  T^{(0,1)} c,
    \\  \label{Condition_one_two} &\quad S^{(1,2)}=2c  \boxtimes  T^{(1,2)} c +2T^{(0,1)}c  \boxtimes  T^{(1,1)} c-i \trace \left(\frac{\delta T^{(0,2)} c}{\delta c}\right)+\frac{1}{2}i\trace\left(\left(\frac{\delta T^{(0,1)} c}{\delta c}\right)^2\right),
    \\ \label{Condition_two_two} &\quad S^{(2,2)}=T^{(1,1)}c  \boxtimes  T^{(1,1)} c-i \trace \left(\frac{\delta T^{(1,2)} c}{\delta c}\right)+\,i\Tr\left( \frac{\delta T^{(0,1)}c}{\delta c}\frac{\delta T^{(1,1)}c}{\delta c}\right). 
\end{align}
Note that while these six conditions consider the six bosonic effective actions from $S^{(0,0)}$ in Eq.~\eqref{Condition_zero_zero} to $S^{(2,2)}$ in Eq.~\eqref{Condition_two_two}, the four non-trivial components of Nicolai map, $T^{(0,1)}$, $T^{(0,2)}$, $T^{(1,1)}$, and $T^{(1,2)}$ are involved in the conditions. Again, $T^{(0,0)}c = c$ is trivially fixed.

For the case of 4D \(\mathcal{N}=1\) pure supergravity, the bosonic effective action denoted by $S^{(r,n)}$'s can be found as done in Appendix \ref{Bosonic effective action}. More generally, in other theories as well, the bosonic effective action can be constructed by the method described in Appendix \ref{Bosonic effective action}.

\noindent {\bf Nicolai map expansion ansatz.} The only unknowns at this point are the four components of Nicolai map \(T^{(0,1)}c\), \(T^{(1,1)}c\), \(T^{(0,2)}c\), and \(T^{(1,2)}c\). To solve the simultaneous equations from Eqs.~\eqref{Condition_zero_zero} to ~\eqref{Condition_two_two}, we suppose that a component of $a$-th order in $\hbar$ and $b$-th order in $\kappa$ of Nicolai map, denoted by $T^{(a,b)}c$, is given by a linear combination of {\it candidate terms} with {\it undetermined coefficients}: 
\begin{eqnarray}
    \bigl(T^{(a,b)}c\bigr)^{\mu\nu}(x) \;=\; \sum_A M^{(a,b)}_A  \cdot \mathcal{T}_A^{\mu\nu}(x) 
    \label{expansion_ansatz}
\end{eqnarray}
where $A$ is a collective index, $\sum_A$ is a collective sum over the index $A$, $M^{(a,b)}_A$ is a $A$-th {\it undetermined coefficient} of the $a$-th order in $\hbar$ and $b$-th order in $\kappa$, and $\mathcal{T}_A$ is a $A$-th {\it candidate} term. We shall refer to the Nicolai map expansion in Eq.~\eqref{real_component_expansion} that can be expressed in terms of the proposed terms in Eq.~\eqref{expansion_ansatz} as the {\it Nicolai map expansion ansatz}. Since \(T^{(a,b)}c\) is a component of the Nicolai map for the symmetric tensor field \(c^{\mu\nu}\), \(\bigl(T^{(a,b)}c\bigr)^{\mu\nu}(x)\) is also understood to be symmetric in \(\mu\) and \(\nu\). Accordingly, only symmetric candidate terms are included in the sum. We now specify the ingredients from which these candidate terms are built. Each candidate term is understood as the symmetrization in the free indices \(\mu\) and \(\nu\) of a single product of these ingredients. Guided by the defining conditions for the Nicolai map, and in particular by the ingredients appearing on the bosonic effective action side, we take the candidate terms to be constructed from derivatives \(\partial\), integration measures \(\int d^4y\), vielbein perturbation fields \(c\), and massless scalar propagators \(G(y-z)\) satisfying equation \(\square G(x-y)=\delta(x-y)\). We then classify candidate terms according to their {\it basic structures} (such as the number of derivatives, the number of fields located at each coordinate, and the coordinate structure of the propagators) in Sec.~\ref{Diagrammatics}.

\subsection{Perturbative Nicolai-map diagrammatics: Poincar\'{e} supergravity}\label{Diagrammatics}
This classification reduces the selection of admissible candidate terms to the following key question: {\it which basic structures of candidate terms are compatible with the defining conditions for the Nicolai map in Eqs.~\eqref{Condition_zero_zero}--\eqref{Condition_two_two}?} To address this question, we introduce a diagrammatic representation. Our representation is inspired by the diagrammatic notation introduced in Ref.~\cite{Lechtenfeld:2024ecf}.

\paragraph{Basic structures for diagram.} We define the {\it basic structures} for diagram. An illustrative diagram with the basic structures is shown in Fig.~\ref{example_diagram}. The definitions of the basic structure are as follows:
\begin{enumerate}
    \item (Lines: dashed \& solid lines and solid loop) Dashed lines denote the vielbein perturbation field, while solid lines denote propagators, which may produce a loop if a propagator is evaluated at a single point only.
    \item (Endpoints of solid line) The endpoints of a solid line represent coordinates connected by the corresponding propagator. The open endpoint indicates a free coordinate that is not integrated over. Its closed endpoint corresponds to a vertex.
    \item (Endpoints of dashed line) The open endpoint of a dashed line for vielbein perturbation field is not assigned to any coordinate. Its closed endpoint corresponds to a vertex.
    \item (Integration vertices) Whenever two or more lines are attached to a point, the coordinate associated with that point is understood to be integrated over. A dashed line meets another line or loop at a vertex assigned to a coordinate.
    \item (Triangle operator ``$\triangleleft$'' for derivatives $\partial$'s) 
    When the diagram representing the basic structure of the candidate terms is supplemented on its right by ``\(\triangleleft ~\partial^{N_{\partial}}\)'', the corresponding candidate terms are understood to be acted on by \(N_{\partial}\) derivatives, while their positions are left unspecified.
\end{enumerate}

\begin{figure}[t!]
\centering
\begin{tikzpicture}[
    font=\small,
    >={Stealth[length=2.2mm]},
    inputbox/.style={
        rectangle, rounded corners=2pt,
        draw=teal!70!black, thin,
        fill=teal!8,
        text width=10cm,
        align=center,
        inner sep=6pt,
        minimum height=0.7cm
    },
    stepbox/.style={
        rectangle, rounded corners=2pt,
        draw=violet!70!black, thin,
        fill=violet!8,
        text width=10cm,
        align=center,
        inner sep=6pt,
        minimum height=0.7cm
    },
    outputbox/.style={
        rectangle, rounded corners=2pt,
        draw=orange!75!black, thin,
        fill=orange!12,
        text width=10cm,
        align=center,
        inner sep=6pt,
        minimum height=0.7cm
    },
    arrowstyle/.style={
        ->, thick, draw=black!60,
        shorten >=2pt, shorten <=2pt
    }
]

\node[inputbox] (boxInput) at (0,0)
    {\textbf{Input: Nicolai-map defining condition at order $(r,n)$}};

\node[stepbox, below=8mm of boxInput] (box1)
    {\textbf{Step 1: Identify the diagrams present in the effective-action side}};

\node[stepbox, below=8mm of box1] (box2)
    {\textbf{Step 2: Read off the diagrams of $T^{(r,n)}c$ that can reproduce the effective-action-side diagrams}};

\node[stepbox, below=8mm of box2] (box3)
    {\textbf{Step 3: Detect any anomalous diagrams that are absent in the effective-action side}};

\node[stepbox, below=8mm of box3] (box4)
    {\textbf{Step 4: Add the counterterm diagrams to $T^{(r,n)}c$ that can cancel out the anomalous ones}};

\node[stepbox, below=8mm of box4] (box5)
    {\textbf{Step 5: Verify self-consistency with the $T^{(r,n)}c$}};

\node[outputbox, below=8mm of box5] (boxOut)
    {\textbf{Output: admissible basic structures of $T^{(r,n)}c$}};

\draw[arrowstyle] (boxInput.south) -- (box1.north);
\draw[arrowstyle] (box1.south)     -- (box2.north);
\draw[arrowstyle] (box2.south)     -- (box3.north);
\draw[arrowstyle] (box3.south)     -- (box4.north);
\draw[arrowstyle] (box4.south)     -- (box5.north);
\draw[arrowstyle] (box5.south)     -- (boxOut.north);

\end{tikzpicture}
\caption{%
General procedure for reading off the basic structures of the Nicolai-map
ansatz $T^{(r,n)}c$ from the bosonic effective-action diagrams via the
defining condition at order $(r,n)$.
}
\label{fig:general-procedure}
\end{figure}

\paragraph{Perturbative Nicolai-map diagrammatics.} In Fig.~\ref{fig:general-procedure}, we describe the general procedure of our perturbative Nicolai-map diagrammatics. Given the defining condition at order $(r,n)$ in Eqs.~\eqref{Condition_zero_zero}--\eqref{Condition_two_two}, whose the left hand side of the conditions is the bosonic
effective action $S^{(r,n)}$ and whose the right hand side of the conditions contains a leading term carrying
$T^{(r,n)}c$ (either $T^{(r,n)}c \boxtimes T^{(r,n)}c$, $c \boxtimes T^{(r,n)}c$ or
$\mathrm{Tr}\!\left(\prod \delta T^{(r,n)}c/\delta c\right)$) together with
$\boxtimes$ products and $\mathrm{Tr}$ terms built only from lower-order
$T^{(a,b)}c$ with $(a,b)\neq(r,n)$, the basic structures of $T^{(r,n)}c$ are
extracted through the following steps.
 
\paragraph{Step 1: Identify the diagrams present in the effective-action side.}
Enumerate the diagrams of $S^{(r,n)}$ from Appendix~\ref{Appendix A} and
Appendix~\ref{Bosonic effective action}, together with their dashed lines,
solid lines, loops, vertices, and derivative counts. These are the targets that the Nicolai-map expansion ansatz side must reproduce.

\paragraph{Step 2: Read off the diagrams of $T^{(r,n)}c$ that can reproduce the effective-action-side diagrams.}
Knowing that the leading operation adds (the factor $c$ supplies one dashed
line) or removes ($\boxtimes$ a solid propagator, or $\delta/\delta c$ a
dashed line) while preserving the derivative count, deduce the minimal basic
structure that $T^{(r,n)}c$ must carry in order for the operation to reproduce
each target diagram.
 
\paragraph{Step 3: Detect any anomalous diagrams that are absent in the effective-action side.}
Substitute the already-determined lower-order $T^{(a,b)}c$ into the $\boxtimes$ products and $\mathrm{Tr}$ terms in the Nicolai-map expansion ansatz side, and evaluate them
diagrammatically. Any resulting diagram absent from the effective-action side at order $(r,n)$
is anomalous and must be canceled within the same condition.
 
\paragraph{Step 4: Add the counterterm diagrams to $T^{(r,n)}c$ that can cancel out the anomalous ones.}
Besides the class found in Step~2, introduce additional admissible classes
$A,\,B,\,\ldots$ in $T^{(r,n)}c$ whose image under the leading operation
precisely cancels each anomalous diagram identified in Step~3. These extra
classes serve as counterterms internal to the Nicolai-map ansatz.
 
\paragraph{Step 5: Verify self-consistency with the $T^{(r,n)}c$.}
Confirm that every class introduced in Step~4 is independently required by
the $\mathrm{Tr}$ term entering another defining condition at the same or
higher order, ensuring that the full set of admissible basic structures is
mutually consistent.
 
\noindent
The output is the complete set of admissible basic structures of
$T^{(r,n)}c$ --- the classes $A,\,B,\,\ldots$ entering the explicit
Nicolai-map expansion ansatz at order $(r,n)$, ready for coefficient
determination.

\begin{figure}[t!]
\begin{center}
\centering
\resizebox{!}{6.5\baselineskip}{%
\begin{tikzpicture}[line cap=round,line join=round]

\def\Vx{8.4}   
\def\R{1.8}    

\coordinate (Xpt) at (0,0);       
\coordinate (Ypt) at (4.2,0);     
\coordinate (Zpt) at (\Vx,0);     
\coordinate (C)   at (\Vx+\R,0);  

\draw[very thick] (Xpt) -- (Zpt);

\draw[dashed,very thick] (Ypt) -- ++(0,-2.5);

\draw[very thick] (C) circle (\R);

\fill (Xpt) circle (2.3pt);
\fill (Ypt) circle (2.3pt);
\fill (Zpt) circle (2.3pt);

\node[above left=2pt]  at (Xpt) {$\X$};
\node[above right=2pt] at (Ypt) {$\Y$};
\node[above right=2pt] at (Zpt) {$\Z$};

\node[above=6pt] at (2.0,0) {$G(\X-\Y)$};
\node[above=6pt] at (6.3,0) {$G(\Y-\Z)$};
\node[right=3pt] at (4.2,-1.25) {$c(\Y)$};

\node at (C) {$G(\Z-\Z)$};

\node[anchor=west] at (\Vx+2.22*\R,0)
  {\scalebox{1.7}{$\triangleleft\,\partial^4$}};

\path (-0.5,-2.8) rectangle (\Vx+3.8*\R,2.4);

\end{tikzpicture}%
}
\end{center}
\caption{
An illustrative diagram containing one free endpoint $\X$, two integration vertices $\Y$ and $\Z$, one vielbein insertion, two propagators, and one loop. It represents $\int d^4\Y\, G(\X-\Y)\, c(\Y)\int d^4 \Z\, G(\Y-\Z)\, G(\Z-\Z)$
with the potential four derivatives $N_\partial=4$.
For pedagogical clarity, the points $\X,\Y,\Z$ are displayed explicitly only in this figure.
}
\label{example_diagram}
\end{figure}

\paragraph{Identifying the basic structures of the diagrams from the effective-action side.} Our purpose in this section is to explore possible diagrams of the candidate terms within the Nicolai-map expansion ansatz by reading off the existing diagrams within the bosonic-effective-action side of the Nicolai-map defining conditions. Following the diagrammatic rules that we have established above, we begin by identifying the diagrams present in the bosonic effective action of \(\mathcal N=1\) pure supergravity given in Appendix ~\ref{Bosonic effective action}.  First, we find a diagram corresponding to $S^{(0,0)}$ of Eq.~\eqref{Condition_zero_zero} as follows:
\begin{equation}
S^{(0,0)}  \quad \supset \hspace{0.5cm}
\label{zero_zero_conditon}
\begin{tikzpicture}[baseline=(current bounding box.center), line cap=round,line join=round]
  \useasboundingbox (-2,-0.5) rectangle (2,0.5);
  \draw[dashed, very thick] (-1.75,0) -- (1.75,0);
  \node[anchor=west] at (2.2,0) {$\triangleleft\,\partial^{2}.$};
\end{tikzpicture}
\end{equation}
We then find the non-trivial diagrams corresponding to the bosonic effective actions as follows: \\ for $S^{(0,1)}$ of Eq.~\eqref{Condition_zero_one},
\begin{equation}
\label{zero_one_conditon}
S^{(0,1)}  \quad \supset \hspace{-1.8cm}
\begin{tikzpicture}[baseline=(current bounding box.center), line cap=round,line join=round]
  \useasboundingbox (-4,-1) rectangle (4,2);
  \def\L{2.0}
  \coordinate (O) at (0,0);
  \draw[dashed, very thick] (O) -- ++(90:\L);
  \draw[dashed, very thick] (O) -- ++(210:\L);
  \draw[dashed, very thick] (O) -- ++(330:\L);
  \node[anchor=west] at (2.2,0.7) {$\triangleleft\,\partial^{2}$};
\end{tikzpicture}
\end{equation}
where, for example, this diagram represents the terms such as
\begin{gather*}
\int d^4x\,
\partial^\nu c_{\rho \nu}(x)\,
c^{\lambda\alpha}(x)\,
\partial_\lambda c_{\alpha}{}^{\rho}(x),\nonumber\\ 
\int d^4x\,
\partial_\lambda c_{\rho \nu}(x)\,
c^{\lambda\alpha}(x)\,
\partial^\nu c_{\alpha}{}^{\rho}(x),\ \nonumber\\ 
\int d^4x\,
\partial_\nu c^{\mu \nu}(x)\,
c_{ef}(x)\,
\partial_\mu c^{ef}(x);
\nonumber
\label{zero_one_example_terms}
\end{gather*}
for $S^{(1,1)}$ of Eq.~\eqref{Condition_one_one},
\begin{equation}
S^{(1,1)}  \quad \supset \hspace{-1.8cm}
\label{one_one_conditon}
\begin{tikzpicture}[baseline=(current bounding box.center), line cap=round,line join=round]
  \useasboundingbox (-4,-1) rectangle (4,1);
  \draw[dashed, very thick] (-1.75,0) -- (0,0);
  \draw[very thick] (1,0) circle (1);
  \node[anchor=west] at (2.2,0) {$\triangleleft\,\partial^{2}$};
\end{tikzpicture}
\end{equation}
where, for example, this diagram represents terms such as 
\begin{gather*}
\int d^4 x\,c^a{}_a(x)\,\partial^b \partial_b G(0), \nonumber\\ 
\int d^4 x\,c^b{}_a(x)\,\partial^b \partial_a G(0);
\end{gather*}
for $S^{(0,2)}$ of Eq.~\eqref{Condition_zero_two},
\begin{equation}
S^{(0,2)}  \quad \supset \hspace{-1.8cm}
\label{zero_two_conditon_A}
\begin{tikzpicture}[baseline=(current bounding box.center), line cap=round,line join=round]
  \useasboundingbox (-4,-1.5) rectangle (4,1.5);
  \def\L{2.0}
  \coordinate (O) at (0,0);
  \draw[dashed, very thick] (O) -- ++(45:\L);
  \draw[dashed, very thick] (O) -- ++(135:\L);
  \draw[dashed, very thick] (O) -- ++(225:\L);
  \draw[dashed, very thick] (O) -- ++(315:\L);
  \node[anchor=west] at (2.2,0) {$\triangleleft\,\partial^{2}$};
\end{tikzpicture}
\end{equation}
where, for example, this diagram represents terms such as
\begin{align}
\int d^4x\,
\partial_\mu c^{\mu \nu}(x)\,
c_\alpha{}^\rho(x)\,
c^{\lambda\alpha}(x)\,
\partial_\lambda c_{\rho \nu}(x), \nonumber
\\
\int d^4x\,
\partial_\mu c^{\mu\lambda}(x)\,
c^{\nu\alpha}(x)\,
c_{\rho \nu}(x)\,
\partial_\lambda c_\alpha{}^\rho(x), \nonumber
\\
\int d^4x\,
\partial_\mu c_{\rho\alpha}(x)\,
c^{\alpha\rho}(x)\,
c^\lambda{}_\nu(x)\,
\partial_\lambda c^{\mu \nu}(x), \nonumber
\end{align}
for $S^{(1,2)}$ of Eq.~\eqref{Condition_one_two},
\begin{equation}
S^{(1,2)}  \quad \supset \hspace{-1.8cm}
\label{one_two_conditon_A}
\begin{tikzpicture}[baseline=(current bounding box.center), line cap=round,line join=round]
  \useasboundingbox (-4,-1.5) rectangle (4,1.5);
\def\L{1.75}
  \coordinate (O) at (0,0);
  \draw[dashed, very thick] (O) -- ++(135:\L);
  \draw[dashed, very thick] (O) -- ++(225:\L);
  \draw[very thick] (1,0) circle (1);
  \node[anchor=west] at (2.2,0) {$\triangleleft\,\partial^{2},$};
\end{tikzpicture}
\end{equation}
\begin{equation}
S^{(1,2)}  \quad \supset \hspace{-0.8cm}
\label{one_two_conditon_C}
\begin{tikzpicture}[baseline=(current bounding box.center), line cap=round,line join=round]
  \useasboundingbox (-4,-1) rectangle (4,1);
  \draw[dashed, very thick] (-2.75,0) -- (-1,0);
  \draw[very thick] (0,0) circle (1);
  \draw[dashed, very thick] (1,0) -- (2.75,0);
  \node[anchor=west] at (2.95,0) {$\triangleleft\,\partial^{4}$};
\end{tikzpicture}
\end{equation}
where the diagram in Eq.~\eqref{one_two_conditon_A} represents, for example, terms such as
\begin{gather*}
\int d^4 x c_{o}{}^{\nu}(x)\,c_{\nu}{}^{o}(x)\,\delta(0),
\nonumber\\ 
\int d^4 x c^{e\nu}(x)\,\partial_{\pi}\partial_{\nu}c^{\pi}{}_{e}(x)\,G(0),
\nonumber\\
\int d^4 xc^{e\mu}(x)\,c^{f}{}_{e}(x)\,\partial_{\mu}\partial_{f}G(0),
\end{gather*}
and the diagram in Eq.~\eqref{one_two_conditon_C} represents, for example, terms such as
\begin{gather*}
\int d^4 x\,d^4 y\,
       \,\partial_{o}{c^a}_a(x)\,\partial^{\nu}G(x-y)\,G(x-y)\,\partial^{o}\partial_{\nu}{c^b}_b(y),\nonumber
\\
\int d^4 x\,d^4 y\,
     \,\partial_{\lambda}{c^a}_a(x)\,\partial^{\nu}G(x-y)\,G(x-y)\,\partial_{\pi}\partial_{\nu}c^{\pi\lambda}(y),\nonumber
\\
\int d^4 x\,d^4 y\,
       \,\partial_{o}{c^a}_b(x)\,\partial^{o}\partial^{\nu}G(x-y)\,G(x-y)\,\partial_{\nu}{c^b}_a(y),\nonumber
\end{gather*}
for $S^{(2,2)}$ of Eq.~\eqref{Condition_two_two},
\begin{equation}
S^{(2,2)}  \quad \supset \hspace{-1.8cm}
\label{two_two_conditon_A}
\begin{tikzpicture}[baseline=(current bounding box.center), line cap=round,line join=round]
  \useasboundingbox (-4,-1.2) rectangle (4,1.2);
  \draw[very thick] (-1,0) circle (1);
  \draw[very thick] (1,0) circle (1);
  \node[anchor=west] at (2.2,0) {$\triangleleft\,\partial^{2},$};
\end{tikzpicture}
\end{equation}
\begin{equation}
S^{(2,2)}  \quad \supset \hspace{-1.8cm}
\label{two_two_conditon_C}
\begin{tikzpicture}[baseline=(current bounding box.center), line cap=round,line join=round]
  \useasboundingbox (-4,-1) rectangle (4,1);
  \draw[very thick] (-1,0) -- (1,0);
  \draw[very thick] (0,0) circle (1);
  \node[anchor=west] at (1.2,0) {$\triangleleft\,\partial^{4},$};
\end{tikzpicture}
\end{equation}
where the diagram in Eq.~\eqref{two_two_conditon_A} represents, for example, terms such as
\begin{gather*}
\int d^4 x \partial^a G(0)\partial_aG(0),
\nonumber\\
\int d^4 x \partial_a \partial^a G(0)G(0);
\nonumber
\end{gather*}
and the diagram in Eq.~\eqref{two_two_conditon_C} represents, for example, terms such as
\begin{gather*}
\int d^4 x\,d^4 y\,G(x-y)\,\partial^\alpha \partial_t \partial^t G(x-y)\,\partial_\alpha G(x-y),
\nonumber\\
\int d^4 x\,d^4 y\, G(x-y)\,\partial^\alpha \partial_t G(x-y)\,\partial_\alpha \partial^t G(x-y),
\nonumber\\
\int d^4 x\,d^4 y\,\partial^\mu G(x-y)\, \partial^\alpha G(x-y)\,\partial_\mu \partial_\alpha G(x-y),
\nonumber
\end{gather*}
\paragraph{Strategy to explore the candidate terms of Nicolai map as diagrams.} Now, our next strategy consists of two steps. The first step is to assign a {\it diagrammatic factor} to each candidate term $\mathcal{T}_A$ of the Nicolai-map expansion ansatz \eqref{expansion_ansatz}, and to express the ansatz as a fully diagrammatic expansion in terms of the relevant diagrammatic factors $\mathcal{T}_A$'s. The second step is to investigate the relevant diagrammatic factors with their basic structures that can produce the diagrammatic products which match the diagrams from the effective-action side in Eqs.~\eqref{zero_zero_conditon}--\eqref{two_two_conditon_C}. As such, in this way, the diagrammatic factors can be compatible with the defining conditions for the Nicolai map. That is, those diagrams from the effective actions can be the hints about what forms the diagrammatic factors of Nicolai map should have.

\paragraph{Two operations centered on diagram.} We need to determine which basic structures of diagrammatic factors of the Nicolai map expansion ansatz are compatible with the defining conditions for the Nicolai map. To do so, we have to clarify how to perform, on diagrams, the two diagrammatic operations centered on
\begin{eqnarray}
    T c \boxtimes T c , \qquad \trace\!\left(\prod \frac{\delta T c}{\delta c}\right) \nonumber,
\end{eqnarray}
which appear in the defining conditions. Later on, these two diagrammatic operations will allow us to identify which diagrammatic factors have to appear in the Nicolai map expansion. 

\begin{figure}[h!]

\centering
\begin{tikzpicture}[
    scale=1.15,
    line cap=round,
    line join=round,
    every node/.style={font=\large}
]

\tikzset{
    solid/.style={line width=0.8pt},
    dashedline/.style={dash pattern=on 2pt off 2pt, line width=0.8pt},
    tri/.style={dashedline},
    circ/.style={draw, circle, minimum size=0.42cm, inner sep=0pt, line width=0.8pt}
}

\node at (-2,  1.6) {$T_1c =$};
\node at (-2,  0.45) {$T_2c =$};
\node at (-2.5, -1.05) {$T_1c\boxtimes T_2c =$};
\node at (-1.73, -2.15) {$=$};

\draw[solid] (-1.2,1.6) -- (-0.35,1.6);
\node[circ, minimum size=0.75cm] at (0,1.6) {};
\draw[dashedline] (0.35,1.6) -- (1.25,1.6);
\node at (2.0,1.6) {$\lhd\, \partial^4$};

\draw[dashedline] (-0.25,1.05) -- (-0.25,0.45);
\draw[solid] (-1.2,0.45) -- (0.45,0.45);
\draw[tri] (0.45,0.45) -- (1.05,0.95);
\draw[tri] (0.45,0.45) -- (1.05,-0.05);
\node at (1.55,0.45) {$\lhd\, \partial^4$};

\draw[dashedline] (-1.2,-1.05) -- (-0.31,-1.05);
\node[circ, minimum size=0.75cm]  at (0,-1.05) {};
\draw[solid] (0.32,-1.05) -- (1.2,-1.05);

\node at (2.3,-1.05) { $\qquad \lhd\, \partial^4 \quad \times \quad \qquad$};

\draw[solid] (3.0,-1.05) -- (4.20,-1.05);
\draw[tri] (4.20,-1.05) -- (4.80,-0.55);
\draw[tri] (4.20,-1.05) -- (4.80,-1.55);

\node at (5.5,-1.05) {$\lhd\, \partial^4$};

\draw[dashedline] (-1.2,-2.15) -- (-0.31,-2.15);
\node[circ, minimum size=0.75cm]  at (0,-2.15) {};

\draw[solid] (0.35,-2.15) -- (1.65,-2.15);

\draw[dashedline] (0.9,-2.1) -- (0.9,-1.5);
\draw[tri] (1.65,-2.15) -- (2.30,-1.65);
\draw[tri] (1.65,-2.15) -- (2.30,-2.65);

\node at (3.05,-2.15) {$\lhd\, \partial^8$};

\end{tikzpicture}
\caption{Example diagram of $T_1c\boxtimes T_2c$.}\label{Figure_2}
\end{figure}

In Fig.~\ref{Figure_2}, we present how to diagrammatically compute the operation \(T c \boxtimes  T c\). For example, let us assume that we have two distinct components of Nicolai map, $T_1c$ and $T_2c$ (the subscripts are not values of the coupling constant $\kappa$ but indices for representing different types). Then, we remove a solid line for propagator of the left (or right) component $T_1c$ (or $T_2c$), and attach this to the free-coordinate endpoint of the solid line of the right (or left) component $T_2c$ (or $T_1c$). That is, the x-box ``$\boxtimes$'' operation acts on diagrams by removing one propagator and reconnecting the two resulting endpoints with another propagator, thereby producing a single connected diagram. As for the power of the triangle operator, the number of derivatives for the resulting diagrammatic factor \(T c \boxtimes  T c\) is equal to the sum of the numbers of derivatives for the two sub-diagrammatic factors $T_1c$ and $T_2c$.

\begin{figure}[ht]
\centering
\begin{tikzpicture}[
    scale=0.95,
    line cap=round,
    line join=round,
    every node/.style={font=\large}
]

\tikzset{
    solid/.style={line width=0.8pt},
    dashedline/.style={dash pattern=on 2pt off 2pt, line width=0.8pt},
    tri/.style={dashedline},
    circ/.style={draw, circle, minimum size=0.70cm, inner sep=0pt, line width=0.8pt},
    dot/.style={fill, circle, minimum size=2.2pt, inner sep=0pt},
    loop/.style={
        dashedline,
        {Stealth[length=2mm,width=1.4mm]}-{Stealth[length=2mm,width=1.4mm]}
    }
}

\node[anchor=east] at (-2.65,  2.20) {$\dfrac{\delta T_1c}{\delta c} =$};
\node[anchor=east] at (-2.65,  0.95) {$\dfrac{\delta T_2c}{\delta c} =$};
\node[anchor=east] at (-2.65, -0.65) 
{$\operatorname{Tr}\!\left(\dfrac{\delta T_1c}{\delta c}
\dfrac{\delta T_2c}{\delta c}\right) =$};
\node[anchor=east] at (-2.65, -2.15) {$=$};

\draw[solid] (-2.25,2.20) -- (-1.45,2.20);
\node[circ] at (-1.10,2.20) {};
\node[dot]  at (-0.75,2.20) {};
\node at (0.05,2.20) {$\lhd\,\partial^4$};

\draw[solid] (-2.25,0.95) -- (-1.15,0.95);
\node[dot] at (-1.65,0.95) {};
\draw[tri] (-1.15,0.95) -- (-0.55,1.45);
\draw[tri] (-1.15,0.95) -- (-0.55,0.45);
\node at (0.02,0.95) {$\lhd\,\partial^4$};

\node at (0.72,0.95) {$+$};

\draw[solid] (1.20,0.95) -- (2.85,0.95);
\draw[dashedline] (1.95,0.95) -- (1.95,1.55);
\draw[dashedline] (2.85,0.95) -- (2.85,1.55);
\node[dot] at (2.85,0.95) {};
\node at (3.45,0.95) {$\lhd\,\partial^4$};


\node[anchor=east] at (-2.65, -0.65) 
{$\operatorname{Tr}\!\left(\dfrac{\delta T_1c}{\delta c}
\dfrac{\delta T_2c}{\delta c}\right) =$};

\draw[solid] (-2.20,-0.65) -- (-1.78,-0.65);
\node[circ] (trA) at (-1.43,-0.65) {};

\coordinate (trAloopend) at (-0.42,-0.65);
\draw[loop]
    (trA.east)
    .. controls (-0.95,-0.25) and (-0.58,-0.25) ..
    (trAloopend);

\draw[solid] (trAloopend) -- (0.55,-0.65);

\draw[tri] (0.55,-0.65) -- (1.05,-0.15);
\draw[tri] (0.55,-0.65) -- (1.05,-1.15);

\draw[loop]
    (-2.20,-0.65)
    .. controls (-1.55,-1.28) and (-0.35,-1.28) ..
    (0.32,-0.65);

\node at (1.45,-0.65) {$\lhd\,\partial^8$};

\node at (2.05,-0.65) {$+$};

\draw[solid] (2.55,-0.65) -- (2.98,-0.65);
\node[circ] (trB) at (3.33,-0.65) {};

\coordinate (trBloopend) at (4.18,-0.65);
\draw[loop]
    (trB.east)
    .. controls (3.80,-0.25) and (4.04,-0.25) ..
    (trBloopend);

\draw[solid] (trBloopend) -- (5.12,-0.65);

\draw[dashedline] (4.55,-0.65) -- (4.55,0.10);
\draw[dashedline] (5.12,-0.65) -- (5.12,0.10);

\draw[loop]
    (2.55,-0.65)
    .. controls (3.15,-1.28) and (4.55,-1.28) ..
    (5.12,-0.65);

\node at (5.92,-0.65) {$\lhd\,\partial^8$};

\node[circ] at (-1.40,-2.15) {};
\draw[solid] (-1.40,-2.50) -- (-1.40,-1.80);

\draw[solid] (-1.05,-2.15) -- (0.30,-2.15);
\draw[tri] (0.30,-2.15) -- (0.90,-1.65);
\draw[tri] (0.30,-2.15) -- (0.90,-2.65);

\node at (1.45,-2.15) {$\lhd\,\partial^8$};

\node at (2.10,-2.15) {$+$};

\node[circ] at (3.60,-2.15) {};
\draw[solid] (3.60,-2.50) -- (3.60,-1.80);

\draw[tri] (3.95,-2.15) -- (4.55,-1.65);
\draw[tri] (3.95,-2.15) -- (4.55,-2.65);

\node at (5.10,-2.15) {$\lhd\,\partial^8$};

\end{tikzpicture}
\caption{Example diagram of 
$\operatorname{Tr}\!\left(\dfrac{\delta T_1c}{\delta c}
\dfrac{\delta T_2c}{\delta c}\right)$.}
\label{Figure_3}
\end{figure}

In Fig.~\ref{Figure_3}, we illustrate how to diagrammatically perform the operation $\trace \left(\prod\frac{\delta T c}{\delta c}\right)$. First, the functional derivative $\delta/\delta c$ acts on the component $T_1c$ by detaching a dashed line for vielbein perturbation field and leaving a dot at the position at which the dashed line has been removed. Second, the product of the two functional derivatives of the Nicolai map components $T_1c$ and $T_2c$ acts on them by following the distributive law. Third, the trace $\textrm{Tr}(\bullet)$ acts on the $\bullet$ by connecting each dot of one diagram to each free coordinate of the propagator of another diagram (or the same diagram if $\bullet$ includes a single diagram). In Fig.~\ref{Figure_3}, the dashed ``arc'' marked by left–right arrows denotes the attachment. As for the power of the triangle operator, the number of derivatives for the resulting diagrammatic factor $\trace \left(\prod\frac{\delta T c}{\delta c}\right)$ is equal to the sum of the numbers of derivatives for the two sub-diagrammatic factors $\delta T_1c/\delta c$ and $\delta T_2c/\delta c$.

\paragraph{Inspecting the candidate terms of Nicolai-map expansion ansatz as diagrams.} We are now in a position to inspect the diagrammatic factors for candidate terms with their basic structures of Nicolai-map expansion ansatz as diagrams. To do this, we need to look at the left and right sides of the defining conditions. Here, we consider the defining conditions for the Nicolai map up to order $\kappa^2$ starting from Eq.~\eqref{Condition_zero_one}. We say that the left is the effective-action side, while the right is the (Nicolai-map) ansatz side. 

\noindent{\bf \circled{1} The 1st non-trivial condition \eqref{Condition_zero_one}.} The diagram with basic structures on the effective-action side of the first non-trivial condition in Eq.~\eqref{Condition_zero_one} is shown in the diagram for $S^{(0,1)}$ in Eq.~\eqref{zero_one_conditon}: a Y-shaped vertex with three dashed lines. This basic structure must be reproduced from the operation \(c \boxtimes T^{(0,1)}c\), which is the ansatz side of Eq.~\eqref{Condition_zero_one}. Since one dashed line is already supplied by the factor \(c\) in \(c \boxtimes T^{(0,1)}c\), and since the \(\boxtimes\) operation removes one solid (propagator) line of the diagram for \(T^{(0,1)}c\), the ``required'' basic structure of the diagram for \(T^{(0,1)}c\) must at least contain two dashed and one solid line. The one endpoint of the solid line is the free coordinate, whose other endpoint is an integrated coordinate at which two dashed lines are located. In addition, according to the basic structure of $S^{(0,1)}$ in Eq.~\eqref{zero_one_conditon}, the number of derivatives has to be two on the ansatz side. Thus, the possible basic structure of \(T^{(0,1)}c\) can be represented by the following diagram:
\begin{equation}
T^{(0,1)}c  \quad \supset \hspace{-1.8cm}
\begin{tikzpicture}[baseline=(current bounding box.center), line cap=round, line join=round]
  \useasboundingbox (-4,-1.5) rectangle (4,1.5);
  \def\L{2.0}
  \coordinate (O) at (0,0);
  \draw[dashed, very thick] (O) -- ++(45:\L);
  \draw[ very thick] (-1.75,0) -- (0,0);
  \draw[dashed, very thick] (O) -- ++(315:\L);
  \node[anchor=west] at (2.2,0) {$\triangleleft\,\partial^{2}.$};
\end{tikzpicture}
\label{zero_one_ansatz}
\end{equation}

\noindent{\bf \circled{2} The 2nd non-trivial condition \eqref{Condition_one_one}.}  The diagram with basic structures on the effective-action side of the second non-trivial condition in Eq.~\eqref{Condition_one_one} is shown in the diagram for $S^{(1,1)}$ in Eq.~\eqref{one_one_conditon}: a tadpole shape with one dashed line and solid loop (say, dashed tadpole). In fact, given the diagram of \(T^{(0,1)}c\) in Eq.~\eqref{zero_one_ansatz}, the operation \(\trace\!\left(\frac{\delta T^{(0,1)}c}{\delta c}\right)\) appearing in the condition \eqref{Condition_one_one} can generate the basic structure of the dashed-tadpole diagram in Eq.~\eqref{one_one_conditon}. That is to say, the operation indicates removing one dashed line from $T^{(0,1)}$ and attaching the free endpoint of the solid line to the dot at which the dashed line has been removed, leading to the dashed tadpole. Accordingly, we can speculate that the basic structure of \(T^{(1,1)}c\) must require the operation \(c \boxtimes T^{(1,1)}c\) to reproduce the basic structure of the dashed-tadpole diagram in Eq.~\eqref{one_one_conditon}. Since one dashed line is already supplied by the factor \(c\), and since the \(\boxtimes\) operation removes one solid line of the diagram for \(T^{(1,1)}c\), the required basic structure of the diagram for \(T^{(1,1)}c\) must contain a solid line whose one endpoint is the free coordinate and whose other endpoint is an integrated coordinate at which one solid loop is attached. Thus, the possible basic structure of \(T^{(1,1)}c\) can be represented by the following diagram:

\begin{equation}
\label{one_one_ansatz}
T^{(1,1)}c  \quad \supset \hspace{-1.8cm}
\begin{tikzpicture}[baseline=(current bounding box.center), line cap=round,line join=round]
  \useasboundingbox (-4,-1) rectangle (4,1);
  \draw[very thick] (-1.75,0) -- (0,0);
  \draw[very thick] (1,0) circle (1);
  \node[anchor=west] at (2.2,0) {$\triangleleft\,\partial^{2}.$};
\end{tikzpicture}
\end{equation}

\noindent{\bf \circled{3} The 3rd non-trivial condition \eqref{Condition_zero_two}.}  The diagram with basic structures on the effective-action side of the third non-trivial condition in Eq.~\eqref{Condition_zero_two} is shown in the diagram for $S^{(0,2)}$ in Eq.~\eqref{zero_two_conditon_A}: an X-shaped vertex with four dashed lines. First, we look at the second term given by \(T^{(0,1)}c \boxtimes T^{(0,1)}c\) in Eq.~\eqref{Condition_zero_two}. By using the diagram for \(T^{(0,1)}c\) with the basic structures in Eq.~\eqref{Condition_zero_two} and taking the $\boxtimes$-product, we find that

\begin{equation}
\label{zero_two_conditon_B}
T^{(0,1)}c \boxtimes T^{(0,1)}c \quad \supset \hspace{-1cm} 
\begin{tikzpicture}[baseline=(current bounding box.center), line cap=round,line join=round]
  \useasboundingbox (-4,-1.5) rectangle (4,1.5);
  \def\L{2.0}
  \coordinate (1) at (1,0);
  \coordinate (-1) at (-1,0);
  \draw[dashed, very thick] (1) -- ++(45:\L);
  \draw[very thick] (-1,0) -- (1,0);
  \draw[dashed, very thick] (1) -- ++(315:\L);
  \draw[dashed, very thick] (-1) -- ++(135:\L);
  \draw[dashed, very thick] (-1) -- ++(225:\L);
  \node[anchor=west] at (3.2,0) {$\triangleleft\,\partial^{4}.$};
\end{tikzpicture}
\end{equation}
At first glance, one may think that this type of diagram is pathological because it does not appear in the diagram for $S^{(0,2)}$ in Eq.~\eqref{zero_two_conditon_A}. However, in fact, the contribution \eqref{zero_two_conditon_B} can be canceled out by the compensating contribution as ``counterterm'' from the diagram for $T^{(0,2)}$. Keeping this in mind, let us pay attention to the first term, $2c\boxtimes T^{(0,2)}c$, in Eq.~\eqref{Condition_zero_two}. To obtain the X-shaped diagram from the bosonic effective action side, we have to allow $T^{(0,2)}$ to possess the following, named {\it class A}, 

\begin{equation}
\label{zero_two_ansatz_A}
T^{(0,2)}c|_{\textrm{class} A}  \quad \supset \hspace{-1.8cm}
\begin{tikzpicture}[baseline=(current bounding box.center), line cap=round,line join=round]
  \useasboundingbox (-4,-1.5) rectangle (4,1.5);
  \def\L{2.0}
  \coordinate (O) at (0,0);
  \draw[dashed, very thick] (O) -- ++(45:\L);
  \draw[ very thick] (-1.75,0) -- (0,0);
  \draw[dashed, very thick] (O) -- ++(0:\L);
  \draw[dashed, very thick] (O) -- ++(315:\L);
  \node[anchor=west] at (2.4,0) {$\triangleleft\,\partial^{2}.$};
\end{tikzpicture}
\end{equation}
Notice that the operation $2c\boxtimes T^{(0,2)}c$ with the class A of $T^{(0,2)}c$ produces the X-shaped diagram by removing one solid line of the diagram \eqref{zero_two_ansatz_A} and attaching a dashed line for $c$ to the point at which the solid line has been truncated. On the other hand, let us recall that the operation \(T^{(0,1)}c \boxtimes T^{(0,1)}c\) generates the unnecessary diagram \eqref{zero_two_conditon_B}. As we mentioned above, in order to eliminate it, we require the basic structures of $T^{(0,2)}$ to have

\begin{equation}
\label{zero_two_ansatz_B}
T^{(0,2)}c|_{\textrm{class} B}  \quad \supset \hspace{-1.8cm}
\begin{tikzpicture}[baseline=(current bounding box.center), line cap=round,line join=round]
  \useasboundingbox (-4,-1.5) rectangle (4,1.5);
  \def\L{2.0}
  \coordinate (O) at (1.2,0);
  \coordinate (-1) at (-0.25,0);
  \draw[dashed, very thick] (-1) -- ++(90:1.4);
  \draw[very thick] (-1.7,0) -- (1.2,0);
  \draw[dashed, very thick] (O) -- ++(45:\L);
  \draw[dashed, very thick] (O) -- ++(315:\L);
  \node[anchor=west] at (3,0) {$\triangleleft\,\partial^{4},$};
\end{tikzpicture}
\end{equation}
which is called {\it class B}. Notice that the operation $2c\boxtimes T^{(0,2)}c$ with the class B of $T^{(0,2)}c$ can produce a diagram as the counterterm to cancel the unnecessary diagram \eqref{zero_two_conditon_B} by removing one solid line of the diagram \eqref{zero_two_ansatz_B} and attaching a dashed line for $c$ to the point at which the solid line has been truncated. Consequently, the final basic structures of \(T^{(0,2)}c\) must be given by the class A \eqref{zero_two_ansatz_A} and class B \eqref{zero_two_ansatz_B}:
\begin{equation}
\label{zero_two_ansatz_tot}
T^{(0,2)}c \quad \supset \hspace{-1.8cm}
\begin{tikzpicture}[baseline=(current bounding box.center), line cap=round,line join=round]
  \useasboundingbox (-4,-1.5) rectangle (4,1.5);
  \def\L{2.0}
  \coordinate (O) at (0,0);
  \draw[dashed, very thick] (O) -- ++(45:\L);
  \draw[ very thick] (-1.75,0) -- (0,0);
  \draw[dashed, very thick] (O) -- ++(0:\L);
  \draw[dashed, very thick] (O) -- ++(315:\L);
  \node[anchor=west] at (2.4,0) {$\triangleleft\,\partial^{2}$};
\end{tikzpicture}
+\hspace{-1.5cm}
\begin{tikzpicture}[baseline=(current bounding box.center), line cap=round,line join=round]
  \useasboundingbox (-4,-1.5) rectangle (4,1.5);
  \def\L{2.0}
  \coordinate (O) at (1.2,0);
  \coordinate (-1) at (-0.25,0);
  \draw[dashed, very thick] (-1) -- ++(90:1.4);
  \draw[very thick] (-1.7,0) -- (1.2,0);
  \draw[dashed, very thick] (O) -- ++(45:\L);
  \draw[dashed, very thick] (O) -- ++(315:\L);
  \node[anchor=west] at (3,0) {$\triangleleft\,\partial^{4}.$};
\end{tikzpicture}
\end{equation}


\noindent{\bf \circled{4} The 4th non-trivial condition \eqref{Condition_one_two}.}  The diagrams with basic structures on the effective-action side of the fourth non-trivial condition in Eq.~\eqref{Condition_one_two} are shown in the diagrams for $S^{(1,2)}$ in Eqs.~\eqref{one_two_conditon_A} and \eqref{one_two_conditon_C}: a solid loop with a single vertex attached to two dashed lines, and another solid loop with two vertices, each attached to one dashed line. First, looking at the diagram \eqref{Condition_one_two} for $S^{(1,2)}$, let us consider the operation \(T^{(0,1)}c \boxtimes T^{(1,1)}c\) in Eq.~\eqref{Condition_one_two}. The diagrams of $T^{(0,1)}c$ and $T^{(1,1)}c$ are given by Eqs.~\eqref{zero_one_ansatz} and \eqref{one_one_ansatz}, respectively. By taking their $\boxtimes$-product, we find that


\begin{equation}
\label{one_two_conditon_B}
T^{(0,1)}c \boxtimes T^{(1,1)}c \quad \supset \hspace{-1cm} 
\begin{tikzpicture}[baseline=(current bounding box.center), line cap=round,line join=round]
  \useasboundingbox (-4,-1.5) rectangle (4,1.5);
  \def\L{1.75}
  \coordinate (1) at (1,0);
  \coordinate (-1) at (-1,0);
  \draw[very thick] (-1,0) -- (1,0);
  \draw[very thick] (2,0) circle (1);
  \draw[dashed, very thick] (-1) -- ++(135:\L);
  \draw[dashed, very thick] (-1) -- ++(225:\L);
  \node[anchor=west] at (3.2,0) {$\triangleleft\,\partial^{4}.$};
\end{tikzpicture}
\end{equation}
We see that the resulting diagram \eqref{one_two_conditon_B} is different from any of the diagrams in Eqs.~\eqref{one_two_conditon_A} and \eqref{one_two_conditon_C} for $S^{(1,2)}$, signaling that it must be canceled out by a proper counterterm as well. 


We next examine the contribution from \(\trace\!\left(\frac{\delta T^{(0,2)}c}{\delta c}\right)\) in Eq.~\eqref{Condition_one_two}. The diagrams of $T^{(0,2)}c$ are found in Eqs.~\eqref{zero_two_ansatz_A} and \eqref{zero_two_ansatz_B}. For the class A of \(T^{(0,2)}c\), all dashed lines are located at the same coordinate. Therefore, removing any one of these dashed lines and attaching the free coordinate endpoint of the solid propagator line to the coordinate at which the dashed line was removed gives the same diagram, namely the diagram in Eq.~\eqref{one_two_conditon_A}.

For the class B of \(T^{(0,2)}c\) in Eq.~\eqref{zero_two_ansatz_B}, however, the dashed lines are located at two different integrated coordinates. If the dashed line at the first integrated coordinate is removed and the free coordinate endpoint of the solid propagator is attached to that coordinate, we obtain the diagram in Eq.~\eqref{one_two_conditon_B}. On the other hand, if one of the dashed lines at the second integrated coordinate is removed and the free coordinate endpoint of the solid propagator is attached to that coordinate, we obtain the diagram in Eq.~\eqref{one_two_conditon_C}.

The operation \(\trace\!\left[\left(\frac{\delta T^{(0,1)}c}{\delta c}\right)^2\right]\) in Eq.~\eqref{Condition_one_two} is implemented as follows. We take two identical copies of the diagram for \(T^{(0,1)}c\) in Eq.~\eqref{zero_one_ansatz}, with one dashed line removed from each copy. For each copy, the free endpoint of the solid line for one copy is then attached to the point at which the dashed line was removed in the other copy. The resulting diagram is precisely the one shown in Eq.~\eqref{one_two_conditon_C}.

The last piece we need to check is the operation $2c  \boxtimes  T^{(1,2)} c$ in Eq.~\eqref{Condition_one_two}. Since we are able to obtain the diagrams in Eqs.~\eqref{Condition_one_two} and \eqref{one_two_conditon_B} through the operations but $2c  \boxtimes  T^{(1,2)} c$ in Eq.~\eqref{Condition_one_two}, we require $T^{(1,2)}c$ to have the following three basic structures

\begin{equation}
\label{one_two_ansatz_A}
T^{(1,2)}c|_{\textrm{class} A} \quad \supset \hspace{-1.8cm}
\begin{tikzpicture}[baseline=(current bounding box.center), line cap=round,line join=round]
  \useasboundingbox (-4,-1) rectangle (4,1);
  \draw[very thick] (-1.75,0) -- (0,0);
  \draw[very thick] (1,0) circle (1);
  \draw[dashed, very thick] (0,1) -- (0,0);
  \node[anchor=west] at (2.2,0) {$\triangleleft\,\partial^{2}.$};
\end{tikzpicture}
\end{equation}
\begin{equation}
\label{one_two_ansatz_B}
T^{(1,2)}c|_{\textrm{class} B}  \quad \supset \hspace{-1.8cm}
\begin{tikzpicture}[baseline=(current bounding box.center), line cap=round,line join=round]
  \useasboundingbox (-4,-1.5) rectangle (4,1.5);
  \def\L{2.0}
  \coordinate (O) at (1.2,0);
  \coordinate (-1) at (-0.25,0);
  \draw[dashed, very thick] (-1) -- ++(90:1);
  \draw[very thick] (-1.7,0) -- (1.2,0);
  \draw[very thick] (2.2,0) circle (1);
  \node[anchor=west] at (3.4,0) {$\triangleleft\,\partial^{4}.$};
\end{tikzpicture}
\end{equation}
\begin{equation}
\label{one_two_ansatz_C}
T^{(1,2)}c|_{\textrm{class} C}  \quad \supset \hspace{-1.8cm}
\begin{tikzpicture}[baseline=(current bounding box.center), line cap=round,line join=round]
  \useasboundingbox (-4,-1) rectangle (4,1);
  \draw[very thick] (-1.5,0) -- (0,0);
  \draw[very thick] (1,0) circle (1);
  \draw[dashed, very thick] (2,0) -- (3,0);
  \node[anchor=west] at (3.2,0) {$\triangleleft\,\partial^{4}.$};
\end{tikzpicture}
\end{equation}

The operations $2c  \boxtimes  T^{(1,2)} c$ with each of the classes A and C of $T^{(1,2)} c$ in Eqs.~\eqref{one_two_ansatz_A} and \eqref{one_two_ansatz_C} can produce the diagrams \eqref{one_two_conditon_A} and \eqref{one_two_conditon_C}, respectively. On the contrary, the operation $2c  \boxtimes  T^{(1,2)} c$ with the class B of $T^{(1,2)} c$ in Eq.~\eqref{one_two_ansatz_B} can produce a diagram to be used as the counterterm to the diagram \eqref{one_two_conditon_B} generated from the operation \(T^{(0,1)}c \boxtimes T^{(1,1)}c\). Hence, the final basic structures of $T^{(1,2)}c$ must be given by the classes A, B, and C in Eqs.~\eqref{one_two_ansatz_A}--\eqref{one_two_ansatz_C}:
\begin{eqnarray}
\label{one_two_ansatz_tot}
T^{(1,2)}c \quad &\supset& \hspace{-1.8cm}
\begin{tikzpicture}[baseline=(current bounding box.center), line cap=round,line join=round]
  \useasboundingbox (-4,-1) rectangle (4,1);
  \draw[very thick] (-1.75,0) -- (0,0);
  \draw[very thick] (1,0) circle (1);
  \draw[dashed, very thick] (0,1) -- (0,0);
  \node[anchor=west] at (2.2,0) {$\triangleleft\,\partial^{2}$};
\end{tikzpicture}
+ \hspace{-1.5cm}
\begin{tikzpicture}[baseline=(current bounding box.center), line cap=round,line join=round]
  \useasboundingbox (-4,-1.5) rectangle (4,1.5);
  \def\L{2.0}
  \coordinate (O) at (1.2,0);
  \coordinate (-1) at (-0.25,0);
  \draw[dashed, very thick] (-1) -- ++(90:1);
  \draw[very thick] (-1.7,0) -- (1.2,0);
  \draw[very thick] (2.2,0) circle (1);
  \node[anchor=west] at (3.4,0) {$\triangleleft\,\partial^{4}$};
\end{tikzpicture}
\nonumber\\
&&+ \hspace{-1.5cm}
\begin{tikzpicture}[baseline=(current bounding box.center), line cap=round,line join=round]
  \useasboundingbox (-4,-1) rectangle (4,1);
  \draw[very thick] (-1.5,0) -- (0,0);
  \draw[very thick] (1,0) circle (1);
  \draw[dashed, very thick] (2,0) -- (3,0);
  \node[anchor=west] at (3.2,0) {$\triangleleft\,\partial^{4}.$};
\end{tikzpicture}
\end{eqnarray}

Throughout \circled{1}--\circled{4}, we have identified all the possible diagrams of the four Nicolai map components $T^{(0,1)}c$, $T^{(1,1)}c$, $T^{(0,2)}c$, and $T^{(1,2)}c$, which are exactly what we need when solving the equations from the non-trivial Nicolai map defining conditions in Eqs.~\eqref{Condition_zero_one}--\eqref{Condition_two_two}.

\noindent{\bf \circled{5} The 5th non-trivial condition \eqref{Condition_two_two}.} The diagrams with basic structures on the effective-action side of the fifth non-trivial condition in Eq.~\eqref{Condition_two_two} are shown in the diagrams for $S^{(2,2)}$ in Eqs.~\eqref{two_two_conditon_A} and \eqref{two_two_conditon_C}: 8-shaped and minus-screw-shaped loops. First, let us start with the operation $T^{(1,1)}c  \boxtimes  T^{(1,1)} c$ in Eq.~\eqref{Condition_two_two}. By using the diagram of $T^{(1,1)}c$ found in Eq.~\eqref{one_one_ansatz}, we find a dumbbell-shaped diagram given by
\begin{equation}
\label{two_two_conditon_B}
T^{(1,1)}c \boxtimes T^{(1,1)}c \quad \supset \hspace{-1cm} 
\begin{tikzpicture}[baseline=(current bounding box.center), line cap=round,line join=round]
  \useasboundingbox (-4,-1.5) rectangle (4,1.5);
  \draw[very thick] (2.1,0) circle (1);
  \draw[very thick] (-0.9,0) -- (1.1,0);
  \draw[very thick] (-1.9,0) circle (1);
  \node[anchor=west] at (3.3,0) {$\triangleleft\,\partial^{4},$};
\end{tikzpicture}
\end{equation}
which is the result after we get rid of one solid line of one copy of $T^{(1,1)}c$ and attach it to another copy of $T^{(1,1)}c$ in accordance with the $\boxtimes$ multiplication. However, this type of diagram does not belong to any of the diagrams for $S^{(2,2)}$ in Eqs.~\eqref{two_two_conditon_A} and \eqref{two_two_conditon_C}. This means that we need a counterterm to cancel out the anomalous diagram \eqref{two_two_conditon_B}. In fact, such counterterm can be found through the other operation \(\trace\!\left(\frac{\delta T^{(1,2)}c}{\delta c}\right)\) in Eq.~\eqref{Condition_two_two}. 

We find that the classes A, B, and C of $T^{(1,2)}c$ in Eqs.~\eqref{one_two_ansatz_A}--\eqref{one_two_ansatz_C} can produce the 8-shaped loop \eqref{two_two_conditon_A}, the dumbbell-shaped one \eqref{two_two_conditon_B}, and the minus-screw-shaped loop \eqref{two_two_conditon_C}, respectively. Remember that the operation \(\trace\!\left(\frac{\delta T^{(1,2)}c}{\delta c}\right)\) indicates removing one dashed line of each class of $T^{(1,2)}c$ and reconnecting the endpoint of the solid line to the point at which the dashed line has been detached. Here, we observe that the operation \(\trace\!\left(\frac{\delta T^{(1,2)}c}{\delta c}\right)\) with the class B of $T^{(1,2)}c$ is able to reproduce a diagram as the counterterm to eliminate the anomalous dumbbell-shaped one \eqref{two_two_conditon_B} coming from $T^{(1,1)}c \boxtimes T^{(1,1)}c$. Therefore, we check that the basic structures of $T^{(1,2)}c$ found in Eqs.~\eqref{one_two_ansatz_A}--\eqref{one_two_ansatz_C} are self-consistent.

Finally, let us consider the remaining operation given by $\trace\!\left(
\frac{\delta T^{(0,1)}c}{\delta c}
\frac{\delta T^{(1,1)}c}{\delta c}
\right)$ in Eq.~\eqref{Condition_two_two}. However, the diagram of \(T^{(1,1)}c\) found in Eq.~\eqref{one_one_ansatz} contains no vielbein perturbation field. Therefore, the functional derivative $\frac{\delta T^{(1,1)}c}{\delta c}$ vanishes, and consequently the operation $\trace\!\left(
\frac{\delta T^{(0,1)}c}{\delta c}
\frac{\delta T^{(1,1)}c}{\delta c}
\right)$ does not produce any contribution.

\begin{figure}[h!]
\centering
\small
\[
\begin{aligned}
S
&= S^{(0,0)}
 + \kappa S^{(0,1)}
 + \hbar\kappa S^{(1,1)}
 + \kappa^2 S^{(0,2)}
 + \hbar\kappa^2 S^{(1,2)}
 + \hbar^2\kappa^2 S^{(2,2)}
\\[2pt]
&= \bigl(\Dprop\,\dder{2}\bigr)
 + \hbar\kappa\bigl(\DCircle\,\dder{2}\bigr)
 + \kappa\bigl(\DVertex\,\dder{2}\bigr)
 + \kappa^2\bigl(\DCross\,\dder{2}\bigr)
\\[2pt]
&\quad
 + \hbar\kappa^2
 \bigl(
    \BranchCircle\,\dder{2}
    + \DCircleD\,\dder{4}
 \bigr)
 + \hbar^2\kappa^2
 \bigl(
    \TwoCircles\,\dder{2}
    + \HalfCircle\,\dder{4}
 \bigr),
\\[5pt]
T_\kappa c
&= c
 + \kappa T^{(0,1)}c
 + \hbar\kappa T^{(1,1)}c
 + \kappa^2 T^{(0,2)}c
 + \hbar\kappa^2 T^{(1,2)}c
\\[2pt]
&= \bigl(\Dprop\bigr)
 + \kappa\bigl(\SBranch\,\dder{2}\bigr)
 + \hbar\kappa\bigl(\SCircle\,\dder{2}\bigr)
 + \kappa^2
 \bigl(
    \SSBranch\,\dder{2}
    + \SVertBranch\,\dder{4}
 \bigr)
\\[2pt]
&\quad
 + \hbar\kappa^2
 \bigl(
    \SVrtCircle\,\dder{2}
    + \SVertCircle\,\dder{4}
    + \SCircleD\,\dder{4}
 \bigr).
\end{aligned}
\]
\caption{Diagrams of the bosonic effective action $S_B[c;\kappa]$ and Nicolai-map expansion ansatz $T_\kappa c$.}
\label{Figure_4}
\end{figure}

\paragraph{Diagrammatic representations of the effective actions, Nicolai map, and its defining conditions.} We have determined the admissible basic structures of the four Nicolai map components $T^{(0,1)}c$, $T^{(1,1)}c$, $T^{(0,2)}c$, and $T^{(1,2)}c$ of the Nicolai map expansion ansatz. These structures, as well as those of the effective actions, are summarized in Fig.~\ref{Figure_4}. Upon inserting the diagrammatic basic structures of the Nicolai map expansion ansatz and the bosonic effective actions summarized in Fig.~\ref{Figure_4} into the defining conditions in Eqs.~\eqref{Condition_zero_zero}--\eqref{Condition_two_two}, we obtain their diagrammatic representation as shown in Fig.~\ref{Figure_5}.

\begin{figure}[h!]
\small
\[
\begin{aligned}
&(3.22):\quad \Dprop\,\dder{2}
=
\Dprop\,\dder{2},
\\[4pt]
&(3.23):\quad \DVertex\,\dder{2}
=
\DVertex\,\dder{2},
\\[4pt]
&(3.24):\quad \DCircle\,\dder{2}
=
\DCircle\,\dder{2}
+
\DCircle\,\dder{2},
\\[4pt]
&(3.25):\quad\DCross\,\dder{2}
=
\bigl(
    \DCross\,\dder{2}
    +
   {\color{red} \DSolidD\,\dder{4} }
\bigr)
+
{\color{red} \DSolidD\,\dder{4}},
\\[6pt]
&(3.26):\quad \BranchCircle\,\dder{2}
+
\DCircleD\,\dder{4}\\[2pt]
&\qquad\qquad=
\bigl(
    \BranchCircle\,\dder{2}
    +
    {\color{red} \DSolidCircle\,\dder{4}}
    +
    \DCircleD\,\dder{4}
\bigr)
+
{\color{red} \DSolidCircle\,\dder{4}}
\\[2pt]
&\qquad\qquad
+
\bigl(
    \BranchCircle\,\dder{2}
    +
   {\color{red} \DSolidCircle\,\dder{4}}
    +
    \DCircleD\,\dder{4}
\bigr)
+
\DCircleD\,\dder{4},
\\[6pt]
&(3.27):\quad \TwoCircles\,\dder{2}
+
\HalfCircle\,\dder{4}
=
{\color{red} \CircleSolidCircle\,\dder{4}}
+
\bigl(
    \HalfCircle\,\dder{4}
    +
    \TwoCircles\,\dder{2}
    +
   {\color{red} \CircleSolidCircle\,\dder{4}}
\bigr).
\end{aligned}
\]
\caption{Diagrammatic representation of Nicolai map defining conditions in Eqs.~\eqref{Condition_zero_zero}--\eqref{Condition_two_two}. The left side of the equality comes from the effective action, while the right comes from the Nicolai-map expansion ansatz. The diagrams in red are the anomalous ones to be canceled out by one another.}
\label{Figure_5}
\end{figure}

\subsection{Proposal for explicit Nicolai map expansion ansatz}
\noindent {\bf Explicit Nicolai map expansion ansatz.} Here we propose an explicit version of the Nicolai map expansion ansatz in Eq.~\eqref{expansion_ansatz}, which reflects the effect of the triangle operator $\triangleleft$ with the derivatives $\partial$'s. Since the basic structures that can appear in the map ansatz are given diagrammatically in Fig.~\ref{Figure_4}, the next step is to generate the explicit ansatz by assigning all possible index assignments and derivative arrangements to the given basic structures. Accordingly, for a given basic structure of ansatz of loop order $a$, \(\kappa\)-order $b$ and class $X$, we introduce two operations: \(I_j\), which specifies an index assignment, and \(D_i\), which specifies a derivative arrangement. In addition, we denote a certain basic structure of a specific class $X$ at the $a$-th order in $\hbar$ and $b$-th order in $\kappa$ by $\mathcal B^{(a,b),X}$. Next, we define a combined operation $D_i \circ I_j$ that is to perform the index assignment $I_j$ and then derivative arrangement $D_i$. By acting this on the basic structures $\mathcal B^{(a,b),X}$, we denote the explicit diagrammatic factor obtained in this way by $(D_i \circ I_j)\mathcal B^{(a,b),X}$. Then, we assign an undetermined coefficient denoted by $M^{(a,b),X}_{i,j}$ to that term $(D_i \circ I_j)\mathcal B^{(a,b),X}$. With these conventions, at fixed loop order $a$ and $\kappa$-order $b$, we expand $T^{(a,b)}c$ as a linear combination of the terms $(D_i \circ I_j)\mathcal B^{(a,b),X}$ with coefficients $M_{i,j}^{(a,b),X}$:
\begin{eqnarray}
    T^{(a,b)}c \;=\; \sum_{X}\sum_{i}\sum_{j} M_{i,j}^{(a,b),X}\, (D_i \circ I_j)\mathcal B^{(a,b),X} = \sum_X \mathcal{B}^{(a,b),X} \triangleleft~\partial^{N_{\partial}},
    \label{eq:general-ansatz-expansion}
\end{eqnarray}
which we call an {\it explicit Nicolai map expansion ansatz} constructed from these components \(T^{(a,b)}\). The index $X$ runs from A to the number of possible classes, $i$ runs from 1 to the number of derivative arrangements, and $j$ runs from 1 to the number of index assignments. In particular, at the second equality of Eq.~\eqref{eq:general-ansatz-expansion}, we introduce the triangle operator with $N_{\partial}$ derivatives denoted by ``$\triangleleft~\partial^{N_{\partial}}$'' as follows:
\begin{eqnarray}
 \textrm{A diagrammatic factor at the orders $(a,b)$ and class $X$}~\Leftrightarrow~\mathcal B^{(a,b),X},  
 \end{eqnarray}
\begin{eqnarray} 
\textrm{Triangle operator corresponding to a basic structure,}~ \triangleleft~\partial^{N_{\partial}} \Leftrightarrow  \sum_{i}\sum_{j} M_{i,j}^{(a,b),X}\, (D_i \circ I_j). \label{tri_with_deriv_operator}
\end{eqnarray}
Notice that the basic structures $\mathcal{B}^{(0,1),A}$, $\mathcal{B}^{(1,1), A}$, $\mathcal{B}^{(0,2),A}$, $\mathcal{B}^{(0,2),B}$, $\mathcal{B}^{(1,2),A}$, $\mathcal{B}^{(1,2),B}$, and $\mathcal{B}^{(1,2),C}$ correspond to the diagrams in Eqs.\eqref{zero_one_ansatz}, \eqref{one_one_ansatz}, \eqref{zero_two_ansatz_A}, \eqref{zero_two_ansatz_B}, \eqref{one_two_ansatz_A}, \eqref{one_two_ansatz_B}, and \eqref{one_two_ansatz_C}, respectively. 

\noindent {\bf Example of the explicit Nicolai map expansion ansatz.} As an illustrative example, we construct an explicit expansion ansatz for $T^{(0,1)}c$ and $T^{(1,1)}c$. The ansatz for $T^{(0,2)}c$ and
$T^{(1,2)}c$ are obtained in the same way, but are omitted
here because of its length. The basic structure of $T^{(0,1)}c$ is
\begin{equation}
\label{eq:T10c-ansatz}
T^{(0,1)}c = \quad \Ydiag \scalebox{1.0}{$\triangleleft ~\partial^2$}.
\end{equation}
The $D_i$ derivative arrangement and the $I_j$ index assignment lists for the ansatz for $T^{(0,1)}c$ are shown in Table~\ref{tab:deriv-and-index}.  The last column, $I_j[\eta]$, is used only when the corresponding index assignment contains an explicit flat metric tensor $\eta_{\mu\nu}$. Here $\mu,\nu$ denote the map indices. Hence, it should be noticed that this ansatz \eqref{eq:T10c-ansatz} contains $108$ terms (from $9\times 12$). Since the ordering of the partial derivatives $\partial$ and the vielbein perturbation fields is irrelevant, $40$ of these are duplicates. Removing them leaves $68$ distinct terms. Following the ansatz in Eq.~\eqref{eq:general-ansatz-expansion}, we find the following explicit expression. In what follows, every term is understood to be symmetrized with respect to the free map indices \(\mu\) and \(\nu\). We suppress the parentheses \((\mu\nu)\) in order not to overload the notation:
\begin{align}
T^{(0,1)} c_{\mu\nu} = \int d^4 y\bigg(~& 
 M^{(0,1)}_{1,1}\,
   \partial_\mu\partial_\nu G(x-y)\,{c^a}_b(y){c^b}_a(y)
 + M^{(0,1)}_{2,1}\,
   \partial_\mu G(x-y)\,\partial_\nu {c^a}_b(y){c^b}_a(y)  \nonumber\\
&\quad
 + M^{(0,1)}_{5,1}\,
   G(x-y)\,\partial_\mu\partial_\nu {c^a}_b(y){c^b}_a(y)
 + M^{(0,1)}_{6,1}
   G(x-y)\,\partial_\mu {c^a}_b(y)\,\partial_\nu {c^b}_a(y) \nonumber\\[4pt]
&\quad
 +  M^{(0,1)}_{1,2}\,
   \partial_\mu\partial_\nu G(x-y)\,{c^a}_a(y){c^b}_b(y)
 + M^{(0,1)}_{2,2}\,
   \partial_\mu G(x-y)\,\partial_\nu {c^a}_a(y)\,{c^b}_b(y) \nonumber\\
&\quad
 + M^{(0,1)}_{5,2}\,
   G(x-y)\,\partial_\mu\partial_\nu {c^a}_a(y)\,{c^b}_b(y)
 + M^{(0,1)}_{6,2}\,
   G(x-y)\,\partial_\mu {c^a}_a(y)\,\partial_\nu {c^b}_b(y) \nonumber\\[4pt]
&\quad
 + M^{(0,1)}_{1,3}\,
   \partial_\mu\partial_a G(x-y)\,{c^a}_\nu(y)\,{c^b}_b(y)
 + M^{(0,1)}_{2,3}\,
   \partial_\mu G(x-y)\,\partial_a {c^a}_\nu(y)\,{c^b}_b(y) \nonumber\\
&\quad
 + M^{(0,1)}_{3,3}\,
   \partial_\mu G(x-y)\,{c^a}_\nu(y)\,\partial_a {c^b}_b(y)
 + M^{(0,1)}_{4,3}\,
   \partial_a G(x-y)\,\partial_\mu {c^a}_\nu(y)\,{c^b}_b(y) \nonumber\\
&\quad
 + M^{(0,1)}_{5,3}\,
   G(x-y)\,\partial_\mu\partial_a {c^a}_\nu(y)\,{c^b}_b(y)
 + M^{(0,1)}_{6,3}\,
   G(x-y)\,\partial_\mu {c^a}_\nu(y)\,\partial_a {c^b}_b(y) \nonumber\\
&\quad
 + M^{(0,1)}_{7,3}\,
   \partial_a G(x-y)\,{c^a}_\nu(y)\,\partial_\mu {c^b}_b(y)
 + M^{(0,1)}_{8,3}\,
   G(x-y)\,\partial_a {c^a}_\nu(y)\,\partial_\mu {c^b}_b(y) \nonumber\\
&\quad
 + M^{(0,1)}_{9,3}\,
   G(x-y)\,{c^a}_\nu(y)\,\partial_\mu\partial_a {c^b}_b(y) \nonumber\\[4pt]
&\quad
 + M^{(0,1)}_{1,4}\,
   \partial_\mu\partial_a G(x-y)\,{c^b}_\nu(y)\,{c^a}_b(y)
 + M^{(0,1)}_{2,4}\,
   \partial_\mu G(x-y)\,\partial_a {c^b}_\nu(y)\,{c^a}_b(y) \nonumber\\
&\quad
 + M^{(0,1)}_{3,4}\,
   \partial_\mu G(x-y)\,{c^b}_\nu(y)\,\partial_a {c^a}_b(y)
 + M^{(0,1)}_{4,4}\,
   \partial_a G(x-y)\,\partial_\mu {c^b}_\nu(y)\,{c^a}_b(y) \nonumber\\
&\quad
 + M^{(0,1)}_{5,4}\,
   G(x-y)\,\partial_\mu\partial_a {c^b}_\nu(y)\,{c^a}_b(y)
 + M^{(0,1)}_{6,4}\,
   G(x-y)\,\partial_\mu {c^b}_\nu(y)\,\partial_a {c^a}_b(y) \nonumber\\
&\quad
 + M^{(0,1)}_{7,4}\,
   \partial_a G(x-y)\,{c^b}_\nu(y)\,\partial_\mu {c^a}_b(y)
 + M^{(0,1)}_{8,4}\,
   G(x-y)\,\partial_a {c^b}_\nu(y)\,\partial_\mu {c^a}_b(y) \nonumber\\
&\quad
 + M^{(0,1)}_{9,4}\,
   G(x-y)\,{c^b}_\nu(y)\,\partial_\mu\partial_a {c^a}_b(y) \nonumber\\[4pt]
&\quad
 + M^{(0,1)}_{1,5}\,
   \partial_a\partial^{b} G(x-y)\,c_{\mu\nu}(y)\,{c^a}_b(y)
 + M^{(0,1)}_{2,5}\,
   \partial_a G(x-y)\,\partial^{b}c_{\mu\nu}(y)\,{c^a}_b(y) \nonumber\\
&\quad
 + M^{(0,1)}_{3,5}\,
   \partial_a G(x-y)\,c_{\mu\nu}(y)\,\partial^{b}{c^a}_b(y)
 + M^{(0,1)}_{5,5}\,
   G(x-y)\,\partial_a\partial^{b}c_{\mu\nu}(y)\,{c^a}_b(y) \nonumber\\
&\quad
 + M^{(0,1)}_{6,5}\,
   G(x-y)\,\partial_a c_{\mu\nu}(y)\,\partial^{b}{c^a}_b(y)
 + M^{(0,1)}_{9,5}\,
   G(x-y)\,c_{\mu\nu}(y)\,\partial_a\partial^{b}{c^a}_b(y) \nonumber\\[4pt]
&\quad
 + M^{(0,1)}_{1,6}\,
   \partial_a\partial^{a} G(x-y)\,c_{\mu\nu}(y)\,{c^b}_b(y)
 + M^{(0,1)}_{2,6}\,
   \partial_a G(x-y)\,\partial^{a}c_{\mu\nu}(y)\,{c^b}_b(y) \nonumber\\
&\quad
 + M^{(0,1)}_{3,6}\,
   \partial_a G(x-y)\,c_{\mu\nu}(y)\,\partial^{a}{c^b}_b(y)
 + M^{(0,1)}_{5,6}\,
   G(x-y)\,\partial_a\partial^{a}c_{\mu\nu}(y)\,{c^b}_b(y) \nonumber\\
&\quad
 + M^{(0,1)}_{6,6}\,
   G(x-y)\,\partial_a c_{\mu\nu}(y)\,\partial^{a}{c^b}_b(y)
 + M^{(0,1)}_{9,6}\,
   G(x-y)\,c_{\mu\nu}(y)\,\partial_a\partial^{a}{c^b}_b(y) \nonumber\\[4pt]
&\quad
 + M^{(0,1)}_{1,7}\,
   \partial_a\partial_b G(x-y)\,{c^a}_\mu(y)\,{c^b}_\nu(y)
 + M^{(0,1)}_{2,7}\,
   \partial_a G(x-y)\,\partial_b {c^a}_\mu(y)\,{c^b}_\nu(y) \nonumber\\
&\quad
 + M^{(0,1)}_{3,7}\,
   \partial_a G(x-y)\,{c^a}_\mu(y)\,\partial_b {c^b}_\nu(y)
 + M^{(0,1)}_{5,7}\,
   G(x-y)\,\partial_a\partial_b {c^a}_\mu(y)\,{c^b}_\nu(y) \nonumber\\
&\quad
 + M^{(0,1)}_{6,7}\,
   G(x-y)\,\partial_a {c^a}_\mu(y)\,\partial_b {c^b}_\nu(y)
 + M^{(0,1)}_{8,7}\,
   G(x-y)\,\partial_b {c^a}_\mu(y)\,\partial_a {c^b}_\nu(y) \nonumber\\[4pt]
&\quad
 + M^{(0,1)}_{1,8}\,
   \partial_a\partial^{a} G(x-y)\,{c^b}_\mu(y)\,c_{b\nu}(y)
 + M^{(0,1)}_{2,8}\,
   \partial_a G(x-y)\,\partial^{a}{c^b}_\mu(y)\,c_{b\nu}(y) \nonumber\\
&\quad
 + M^{(0,1)}_{5,8}\,
   G(x-y)\,\partial_a\partial^{a}{c^b}_\mu(y)\,c_{b\nu}(y)
 + M^{(0,1)}_{6,8}\,
   G(x-y)\,\partial_a {c^b}_\mu(y)\,\partial^{a}c_{b\nu}(y) \nonumber\\[4pt]
&\quad
 + M^{(0,1)}_{1,9}\,
   \eta_{\mu\nu}\,\partial_a\partial_b G(x-y)\,c^{ab}(y){c^d}_d(y)
 + M^{(0,1)}_{2,9}\,
   \eta_{\mu\nu}\,\partial_a G(x-y)\,\partial_b c^{ab}(y){c^d}_d(y) \nonumber\\
&\quad
 + M^{(0,1)}_{3,9}\,
   \eta_{\mu\nu}\,\partial_a G(x-y)\,c^{ab}(y)\,\partial_b {c^d}_d(y)
 +M^{(0,1)}_{5,9}\,
   \eta_{\mu\nu}\,G(x-y)\,\partial_a\partial_b c^{ab}(y){c^d}_d(y) \nonumber\\
&\quad
 + M^{(0,1)}_{6,9}\,
   \eta_{\mu\nu}\,G(x-y)\,\partial_a c^{ab}(y)\,\partial_b {c^d}_d(y)
 + M^{(0,1)}_{9,9}\,
   \eta_{\mu\nu}\,G(x-y)\,c^{ab}(y)\,\partial_a\partial_b {c^d}_d(y) \nonumber\\
&\quad
 + M^{(0,1)}_{1,10}\,
   \eta_{\mu\nu}\,\partial_a\partial_b G(x-y)\,{c^a}_d(y)\,c^{bd}(y)
 + M^{(0,1)}_{2,10}\,
   \eta_{\mu\nu}\,\partial_a G(x-y)\,\partial_b {c^a}_d(y)\,c^{bd}(y) \nonumber\\
&\quad
 + M^{(0,1)}_{3,10}\,
   \eta_{\mu\nu}\,\partial_a G(x-y)\,{c^a}_d(y)\,\partial_b c^{bd}(y)
 + M^{(0,1)}_{5,10}\,
   \eta_{\mu\nu}\,G(x-y)\,\partial_a \partial_b {c^a}_d(y)\, c^{bd}(y) \nonumber\\
&\quad + M^{(0,1)}_{6,10}\,
   \eta_{\mu\nu}\,G(x-y)\,\partial_a {c^a}_d(y)\,\partial_b c^{bd}(y)
 + M^{(0,1)}_{8,10}\,
   \eta_{\mu\nu}\,G(x-y)\,\partial_b {c^a}_d(y)\,\partial_a c^{bd}(y) \nonumber\\[4pt]
&\quad + M^{(0,1)}_{1,11}\,
   \eta_{\mu\nu}\,\partial_a\partial^{a} G(x-y)\,{c^b}_d(y){c^d}_b(y)
 +  M^{(0,1)}_{2,11}\,
   \eta_{\mu\nu}\,\partial_a G(x-y)\,\partial^{a}{c^b}_d(y){c^d}_b(y) \nonumber\\
&\quad
 +  M^{(0,1)}_{5,11}\,
   \eta_{\mu\nu}\,G(x-y)\,\partial_a\partial^{a}{c^b}_d(y){c^d}_b(y)
 +  M^{(0,1)}_{6,11}\,
   \eta_{\mu\nu}\,G(x-y)\,\partial_a {c^b}_d(y)\,\partial^{a}{c^d}_b(y) \nonumber\\[4pt]
&\quad
 +  M^{(0,1)}_{1,12}\,
   \eta_{\mu\nu}\,\partial_a\partial^{a} G(x-y)\,{c^b}_b(y){c^d}_d(y)
 +  M^{(0,1)}_{2,12}\,
   \eta_{\mu\nu}\,\partial_a G(x-y)\,\partial^{a}{c^b}_b(y){c^d}_d(y) \nonumber\\
&\quad
 +  M^{(0,1)}_{5,12}\,
   \eta_{\mu\nu}\,G(x-y)\,\partial_a\partial^{a}{c^b}_b(y){c^d}_d(y)
 +  M^{(0,1)}_{6,12}\,
   \eta_{\mu\nu}\,G(x-y)\,\partial_a {c^b}_b(y)\,\partial^{a}{c^d}_d(y)\,\bigg).\label{eq:T10cmunu-long}
\end{align}

{
\setlength{\belowcaptionskip}{4pt}
\begin{table}[!htbp]
\centering
\begin{tabular}{c l | r l l l l}
\hline\hline
$i$ & $D_i$ & $j$ & $I_j[\partial]$ & $I_j[c(y)|_{\textrm{1st}}]$ & $I_j[c(y)|_{\textrm{2nd}}]$ & $I_j[\eta]$ \\
\hline
1 & $\partial^{A}\partial^{B} G(x-y)\, c(y)\,c(y)$
  & 1  & $[\partial:\mu,\nu]$ & $[y:a,b]$     & $[y:a,b]$     &  \\
2 & $\partial^{A} G(x-y)\,\partial^{B} c(y)\,c(y)$
  & 2  & $[\partial:\mu,\nu]$ & $[y:a,a]$     & $[y:b,b]$     &  \\
3 & $\partial^{A} G(x-y)\,c(y)\,\partial^{B} c(y)$
  & 3  & $[\partial:\mu,a]$   & $[y:\nu,a]$   & $[y:b,b]$     &  \\
4 & $\partial^{B} G(x-y)\,\partial^{A} c(y)\,c(y)$
  & 4  & $[\partial:\mu,a]$   & $[y:\nu,b]$   & $[y:a,b]$     &  \\
5 & $G(x-y)\,\partial^{A}\partial^{B} c(y)\,c(y)$
  & 5  & $[\partial:a,b]$     & $[y:\mu,\nu]$ & $[y:a,b]$     &  \\
6 & $G(x-y)\,\partial^{A} c(y)\,\partial^{B} c(y)$
  & 6  & $[\partial:a,a]$     & $[y:\mu,\nu]$ & $[y:b,b]$     &  \\
7 & $\partial^{B} G(x-y)\,c(y)\,\partial^{A} c(y)$
  & 7  & $[\partial:a,b]$     & $[y:\mu,a]$   & $[y:\nu,b]$   &  \\
8 & $G(x-y)\,\partial^{B} c(y)\,\partial^{A} c(y)$
  & 8  & $[\partial:a,a]$     & $[y:\mu,b]$   & $[y:\nu,b]$   &  \\
9 & $G(x-y)\,c(y)\,\partial^{A}\partial^{B} c(y)$
  & 9  & $[\partial:a,b]$     & $[y:a,b]$     & $[y:d,d]$     & $[\eta:\mu,\nu]$ \\
\multicolumn{2}{c|}{} 
  & 10 & $[\partial:a,b]$     & $[y:a,d]$     & $[y:b,d]$     & $[\eta:\mu,\nu]$ \\
\multicolumn{2}{c|}{} 
  & 11 & $[\partial:a,a]$     & $[y:b,d]$     & $[y:b,d]$     & $[\eta:\mu,\nu]$ \\
\multicolumn{2}{c|}{} 
  & 12 & $[\partial:a,a]$     & $[y:b,b]$     & $[y:d,d]$     & $[\eta:\mu,\nu]$ \\
\hline\hline
\end{tabular}
\caption{Derivative arrangements \(D_i\) and index assignments \(I_j\) for \(T^{(0,1)}\)}
\label{tab:deriv-and-index}
\end{table}
}

The basic structure of $T^{(1,1)}c$ is
\begin{equation}
\label{eq:T11c-ansatz}
T^{(1,1)}c = \quad \Tadpole \;\scalebox{1.0}{$\triangleleft ~\partial^2$}.
\end{equation}
The \(D_i\) derivative arrangement and the \(I_j\) index assignment lists for the ansatz for \(T^{(1,1)}c\) are shown in Table~\ref{tab:deriv-and-index_1}. The meaning of the \(I_j[\eta]\) column is the same as above. Since there are two partial derivatives and two possible objects on which they can act, namely one propagator and one single-propagator loop, one would naively obtain \(2^2\) derivative arrangements. However, the loop-parity condition eliminates the arrangements in which an odd number of partial derivatives acts on the single-propagator loop. Therefore, only two derivative arrangements are admissible. Thus, this ansatz \eqref{eq:T11cmunu-long} contains \(4\) terms (from \(2\times 2\)). Applying Eq.~\eqref{eq:general-ansatz-expansion}, and symmetrizing the free map indices \(\mu\) and \(\nu\) as before, we obtain
\begin{align}
T^{(1,1)} c_{\mu\nu} =  -2 \int d^4 y\bigg(~& 
 M^{(1,1)}_{1,1}\,
   \partial_\mu\partial_\nu G(x-y)\,G(y-y)
 + M^{(1,1)}_{2,1}\,
   G(x-y)\,\partial_\mu\partial_\nu G(y-y)  \nonumber\\
&\quad
 + M^{(1,1)}_{1,2}\,
   \eta_{\mu\nu}\partial_a\partial^a G(x-y)\,G(y-y)
 + M^{(1,1)}_{2,2}
   \eta_{\mu\nu}G(x-y)\,\partial_a\partial^a G(y-y) \bigg)
\label{eq:T11cmunu-long}
\end{align}

{
\setlength{\belowcaptionskip}{4pt}
\begin{table}[!htbp]
\centering
\begin{tabular}{c l | r l l l l}
\hline\hline
$i$ & $D_i$ & $j$ & $I_j[\partial]$ & $I_j[\eta]$ \\
\hline
1 & $\partial^{A}\partial^{B} G(x-y)\, G(y-y)$
  & 1  & $[\partial:\mu,\nu]$ &  \\
2 & $G(x-y)\,\partial^{A}\partial^{B} G(y-y)$
  & 2  & $[\partial:a,a]$ & $[\eta:\mu,\nu]$ \\
\hline\hline
\end{tabular}
\caption{Derivative arrangements \(D_i\) and index assignments \(I_j\) for \(T^{(1,1)}\)}
\label{tab:deriv-and-index_1}
\end{table}
}

The discussion of this section has been kept general at the level of the perturbative construction scheme. In particular, once the free bosonic action, the effective bosonic action, and the relevant field content are specified, the Nicolai map conditions can be expanded order by order and converted into algebraic constraints on a suitable ansatz. At the same time, the explicit realization of this scheme remains theory-dependent, since the precise tensor structure and allowed terms are determined by the underlying field content and symmetries. We therefore now specialize this framework to the case of pure four-dimensional  $\mathcal N=1$ supergravity, for which the general strategy developed above can be implemented concretely.

\section{Perturbative Nicolai map for Poincar\'{e} supergravity}
\label{section4}

In this section, we apply the perturbative construction scheme developed in Sec.~\ref{section3} to four-dimensional $\mathcal N=1$ pure supergravity. Our goal is to translate the general conditions for the Nicolai map into an explicit, model-dependent construction adapted to the pure supergravity system studied in this work. Section~\ref{section3} gave the method in general terms; here we feed in the
pure supergravity input and run the order-by-order analysis.

Using the diagrammatic representation, we construct an explicit expansion ansatz for the Nicolai map. After substituting this ansatz into the defining conditions for the Nicolai map, both sides of the condition can be expressed as linear combinations of diagrammatic factors as we have seen in Fig.~\ref{Figure_5}. Since the full set of such diagrammatic factors is not linearly independent, owing to integration-by-parts identities, we re-express these linear combinations in terms of a basis of diagrammatic factors. Comparing the coefficients of the basis factors then reduces the defining conditions for the Nicolai map to a system of simultaneous equations for the undetermined coefficients \(M^{(a,b),X}_{i,j}\). Because the calculation becomes combinatorially involved, we automate the entire pipeline in $\texttt{Python}$. The main text presents only the core algorithms, while the complete and reproducible implementation is provided in the supplementary materials\footnote{Code: \url{https://github.com/junseyu2/SUGRA_Nicolai_map/releases/tag/v1.1.2} (commit \texttt{63eacaf}).}.

\subsection{Part I: Computational construction of the explicit Nicolai-map expansion ansatz}\label{Part_I}

\begin{figure}[t]
\centering
\begin{adjustbox}{
    max width=0.90\textwidth,
    max totalheight=0.40\textheight,
    keepaspectratio
}
\begin{tikzpicture}[
    node distance=8mm and 14mm,
    box/.style={
        rectangle,
        rounded corners=2mm,
        draw=black!70,
        thick,
        align=center,
        text width=3.65cm,
        minimum height=9mm,
        inner sep=4pt,
        font=\small
    },
    widebox/.style={
        rectangle,
        rounded corners=2mm,
        draw=black!70,
        thick,
        align=center,
        text width=5.25cm,
        minimum height=10mm,
        inner sep=4pt,
        font=\small
    },
    arrow/.style={
        -{Latex[length=2.1mm]},
        thick,
        draw=black!75,
        shorten >=2pt,
        shorten <=2pt
    }
]

\node[box, fill=InputBlue] (b1)
    { {\bf Input} \\ Basic structures \\ of ansatz};

\node[box, fill=BranchGreen, below left=10mm and 20mm of b1] (b2)
    { {\bf Step 1} \\ Index assignments \\ Case 1 : $\eta^{\mu\nu}$ $\times$\\ Case 2 : $\eta^{\mu\nu}$
     $\ocircle$};

\node[box, fill=BranchGreen, below right=10mm and 20mm of b1] (b3)
    { {\bf Step 2} \\ Derivative placements};

\node[widebox, fill=AnsatzOrange, below=18mm of b1] (b4)
    { {\bf Step 3} \\ Remove duplicates};

\node[widebox, fill=FinalYellow, below=9mm of b4] (b5)
    { {\bf Output} \\ Explicit expansion ansatz};

\coordinate (split1) at ($(b1.south)+(0,-4mm)$);
\coordinate (merge1L) at ($(b4.west)+(-6mm,0)$);
\coordinate (merge1R) at ($(b4.east)+(6mm,0)$);

\draw[arrow] (b1.south) -- (split1);
\draw[arrow] (split1) -| (b2.north);
\draw[arrow] (split1) -| (b3.north);

\draw[arrow] (b2.east) -- (merge1L) |- (b4.west);
\draw[arrow] (b3.west) -- (merge1R) |- (b4.east);

\draw[arrow] (b4.south) -- (b5.north);

\end{tikzpicture}
\end{adjustbox}

\caption{Computational steps for constructing the explicit Nicolai-map expansion ansatz.}
\label{fig:nicolai_ansatz_construction}
\end{figure}

We first describe the algorithm used to construct the explicit expansion ansatz for the Nicolai map, as shown in Fig.~\ref{fig:nicolai_ansatz_construction}. Given the basic structures of the diagrammatic factors appearing in the ansatz, the explicit ansatz is generated exhaustively through the following steps.

\paragraph{Input: Basic structures of ansatz.}
We first input the admissible basic structures of the diagrammatic factors in the ansatz up to the desired order in \(\kappa\). In the present work, we construct the Nicolai-map ansatz through order \(\kappa^2\), and therefore use the admissible basic structures through order \(\kappa^2\) obtained in Sec.~\ref{Diagrammatics}. These structures fix the number of vielbein perturbation fields $N_c$, propagators $N_G$, partial derivatives $N_\partial$, and provide the available slots for the index assignments and derivative placements performed in the next steps.

\paragraph{Step 1: Index assignments.} 
We assign indices to the vielbein perturbation fields and to the partial derivatives whose placements have not yet been fixed. To do this systematically, we first introduce index slots associated with the given basic structure. The unfixed partial derivatives provide one unordered slot of size \(N_\partial\), corresponding to the \(N_\partial\) derivative indices. Each vielbein perturbation field provides one unordered slot of size two, reflecting the symmetry \(c^{ab}=c^{ba}\). Thus, for \(N_c\) vielbein perturbation fields, we introduce \(N_c\) such size-two slots. These size-two slots are further grouped according to their coordinates, so that permutations of vielbein perturbation fields associated with the same coordinate can be treated as equivalent.

We then fill these slots with free and dummy indices. The enumeration is divided according to whether a flat metric \(\eta^{\mu\nu}\) is present. If no \(\eta^{\mu\nu}\) is present, the two free indices \(\mu\) and \(\nu\) are assigned to the available index entries, with the exchange \(\mu\leftrightarrow\nu\) treated as equivalent because the Nicolai-map indices are symmetrized. The remaining \(N_\partial+2N_c-2\) index entries are filled with
\[
    \frac{N_\partial+2N_c-2}{2}
\]
distinct dummy-index labels, each appearing exactly twice. If a flat metric \(\eta^{\mu\nu}\) is present, the free indices \(\mu\) and \(\nu\) are already carried by the metric, and hence all \(N_\partial+2N_c\) index entries are filled with
\[
    \frac{N_\partial+2N_c}{2}
\]
distinct dummy-index labels, again with each dummy index appearing exactly twice. In this way, we generate all possible index assignments compatible with the given basic structure.

\paragraph{Step 2: Derivative placements.} 
We next fix the placement of each partial derivative \(\partial\) after the index assignments have been specified. Each derivative may act either on a propagator or on a vielbein perturbation field. Treating the \(N_\partial\) derivatives as independently placed, the number of possible derivative arrangements is
\[
    (N_G+N_c)^{N_\partial},
\]
where \(N_G\) is the number of propagators and \(N_c\) is the number of vielbein perturbation fields. In the presence of a single-propagator loop \(G(0)\), the loop integral is invariant under loop-momentum reversal, \(p\to -p\). Hence, any contribution with an odd number of derivatives acting on that loop vanishes, and only configurations with an even number of derivatives acting on the loop need to be retained.

\paragraph{Step 3: Remove duplicates.} 
We interleave the list of derivative arrangements with the list of index assignments to generate the full list of diagrammatic factors. If \(N_D\) derivative arrangements and \(N_I\) index assignments are obtained for a given basic structure, this interleaving produces \(N_D \times N_I\) candidate factors. However, this list generally contains duplicate factors, because different index assignments or derivative arrangements may represent the same diagrammatic factor. We therefore identify all duplicate factors by applying the following equivalences and keep only one representative from each equivalence class:
\begin{itemize}
    \item symmetry under the exchange of \(\mu\) and \(\nu\);
    \item relabeling equivalence of dummy indices;
    \item symmetry of the vielbein perturbation field, \(c^{ab}=c^{ba}\);
    \item commutativity of partial derivatives at the same position, \(\partial^a\partial^b=\partial^b\partial^a\);
    \item permutation equivalence of vielbein perturbation fields or propagators associated with the same coordinate;
    \item geometric symmetries of the diagram.
\end{itemize}

\paragraph{Output: Explicit expansion ansatz.} 
For the given input basic structures, we multiply each inequivalent diagrammatic factor in the resulting list by its associated undetermined coefficient \(M_{i,j}^{(a,b),X}\). Summing over all such factors, labeled by the class \(X\), the derivative arrangement \(i\), and the index assignment \(j\), yields the explicit expansion ansatz for the Nicolai map.

\subsection{Part II: Reduction of the defining conditions to a system of equations}\label{Part_II}

\begin{figure}[t]
\centering
\begin{adjustbox}{
    max width=0.98\textwidth,
    max totalheight=0.62\textheight,
    keepaspectratio
}
\begin{tikzpicture}[
    node distance=8mm and 14mm,
    box/.style={
        rectangle,
        rounded corners=2mm,
        draw=black!70,
        thick,
        align=center,
        text width=5.1cm,
        minimum height=10mm,
        inner sep=4pt,
        font=\small
    },
    smallbox/.style={
        rectangle,
        rounded corners=2mm,
        draw=black!70,
        thick,
        align=center,
        text width=3.6cm,
        minimum height=10mm,
        inner sep=4pt,
        font=\small
    },
    inputbox/.style={
        rectangle,
        rounded corners=2mm,
        draw=black!70,
        thick,
        align=center,
        text width=3.7cm,
        minimum height=10mm,
        inner sep=4pt,
        font=\small
    },
    arrow/.style={
        -{Latex[length=2.1mm]},
        thick,
        draw=black!75,
        shorten >=2pt,
        shorten <=2pt
    }
]

\path[use as bounding box] (-3.2,1.0) rectangle (13.8,-14.6);


\node[inputbox, fill=InputBlue, text width=4.0cm] (in1)
    at (0,0.2)
    {{\bf Input}\\
    Explicit expansion ansatz};

\node[inputbox, fill=InputBlue, text width=5.0cm] (in2)
    at (0,-2.9)
    {
    Basic structures\\
    in the defining conditions};

\node[inputbox, fill=BranchGreen, text width=4.0cm] (gen)
    at (0,-4.5)
    {The same procedure\\
      as in Part I};

\node[inputbox, fill=InputBlue, text width=5.0cm] (sub_out)
    at (0,-6.0)
    {List of explicit diagrammatic factors};

\node[inputbox, fill=BranchGreen, text width=4.0cm] (ibp)
    at (0,-7.8)
    {Integration-by-parts\\
    \& \\ Gauss--Jordan elimination};

\node[inputbox, fill=InputBlue, text width=5.0cm] (rules)
    at (0,-9.6)
    {Diagrammatic factor basis\\
    and elimination rules};


\node[box, fill=AnsatzOrange] (s1)
    at (8.8,0.2)
    {{\bf Step 1}\\
Substitute the ansatz\\
into the defining conditions};

\node[smallbox, fill=EvalRed] (s2a)
    at (6.5,-3.0)
    {{\bf Step 2a}\\
    Evaluate the term:\\[1mm]
    $Tc\boxtimes Tc$};

\node[smallbox, fill=EvalRed] (s2b)
    at (11.1,-3.0)
    {{\bf Step 2b}\\
    Evaluate the term:\\[1mm]
    $\displaystyle \mathrm{Tr}\!\bigg(\prod\frac{\delta Tc}{\delta c}\bigg)$};

\node[box, fill=CombinePurple] (s3)
    at (8.8,-6.0)
    {{\bf Step 3}\\
    Collect like terms using the list};

\node[box, fill=ReduceTeal] (s4)
    at (8.8,-9.6)
    {{\bf Step 4}\\
    Rewrite the condition \\ in the basis
    using rules \\ \& \\ Extract
    the basis coefficients};

\node[box, fill=FinalYellow] (out)
    at (8.8,-13.2)
    {{\bf Output}\\
    System of equations for the\\
$M^{(a,b),X}_{i,j}$};


\draw[arrow] (in1.east) -- (s1.west);

\coordinate (barL) at ($(s2a.north)+(0,6mm)$);
\coordinate (barR) at ($(s2b.north)+(0,6mm)$);
\coordinate (split1) at ($(barL)!0.5!(barR)$);

\draw[thick, draw=black!75] (s1.south) -- (split1);
\draw[arrow] (split1) -| (s2a.north);
\draw[arrow] (split1) -| (s2b.north);

\coordinate (merge1) at ($(s3.north)+(0,6mm)$);
\draw[arrow] (s2a.south) |- (merge1);
\draw[arrow] (s2b.south) |- (merge1);
\draw[arrow] (merge1) -- (s3.north);

\draw[arrow] (s3.south) -- (s4.north);
\draw[arrow] (s4.south) -- (out.north);

\draw[arrow] (in2.south) -- (gen.north);
\draw[arrow] (gen.south) -- (sub_out.north);
\draw[arrow] (sub_out.east) -- (s3.west);

\draw[arrow] (sub_out.south) -- (ibp.north);
\draw[arrow] (ibp.south) -- (rules.north);
\draw[arrow] (rules.east) -- (s4.west);

\end{tikzpicture}
\end{adjustbox}

\caption{Computational steps for reducing the defining conditions to a system of equations.}
\label{fig:nicolai_condition_reduction}
\end{figure}

In Fig.~\ref{fig:nicolai_condition_reduction}, we now describe how the defining conditions are reduced to a system of equations for the undetermined coefficients \(M^{(a,b),X}_{i,j}\).

\paragraph{Step 1: Substitution the ansatz into the defining conditions.} We substitute the explicit expansion ansatz obtained in Part I into the ansatz side of the defining conditions for the Nicolai map. This substitution requires the evaluation of the two operations appearing on the ansatz side,
\[
    Tc \boxtimes Tc
    \qquad\text{and}\qquad
    \trace\!\left(\prod \frac{\delta Tc}{\delta c}\right).
\]
These two operations are evaluated in Step 2 below. Since we work through order \(\kappa^2\), the relevant nontrivial conditions are those from Eq.~\eqref{Condition_zero_one} to Eq.~\eqref{Condition_two_two}. Thus, we use the Part I procedure for the basic structures of \(T^{(0,1)}c\), \(T^{(1,1)}c\), \(T^{(0,2)}c\), and \(T^{(1,2)}c\), and substitute the resulting explicit ansätze into these defining conditions.

\paragraph{Step 2a: Evaluation of \(Tc\boxtimes Tc\).}
To evaluate \(Tc \boxtimes Tc\) with \(\boxtimes\) defined in Eq.~\eqref{free_action}, we first relabel the integrated coordinates and dummy indices in the left and right diagrammatic factors so that no labels overlap. We then integrate the product of the two factors over the common coordinate \(x\). Integration by parts is performed so that all partial derivatives acting on \(G(x-y)\) in the right diagrammatic factor are transferred to the left diagrammatic factor. The \(\square\) appearing in \(\boxtimes\) then acts on the remaining \(G(x-y)\) in the right diagrammatic factor, producing \(\delta(x-y)\). After carrying out the \(\int d^4x\) integration contained in \(\boxtimes\), the delta function identifies the coordinate \(x\) in both diagrammatic factors with \(y\).

For the first term \(c_{\mu\nu}\boxtimes c^{\mu\nu}\) in \(Tc\boxtimes Tc\), the symmetrization in \(\mu\) and \(\nu\) must be preserved. For the second term \({c^\mu}_\mu\boxtimes {c^\nu}_\nu\), the two diagrammatic factors are traced independently. Thus, \(\mu\) and \(\nu\) are identified within each diagrammatic factor, while the trace indices remain distinct between the two factors.

\paragraph{Step 2b: Evaluation of \(\trace(\prod \delta Tc/\delta c)\).} 
We next evaluate
\[
    \trace\!\left(\prod \frac{\delta Tc}{\delta c}\right).
\]
We first relabel all variables and indices in the diagrammatic factors so that labels belonging to different factors do not overlap. We then apply the Leibniz rule within each diagrammatic factor: the functional derivative acts in turn on each occurrence of the vielbein perturbation field, and the resulting diagrammatic factors are summed. The sums obtained from the different diagrammatic factors are then multiplied using the distributive law.

Concretely, functional differentiation with respect to the vielbein perturbation field is implemented by replacing that field with Kronecker deltas that identify its indices with the free indices of the diagrammatic factor to its right, together with a Dirac delta function that identifies its coordinate with the free coordinate of the factor to its right. For the rightmost diagrammatic factor, the factor to its right is understood to be the leftmost factor, thereby implementing the trace. Thus, if one differentiates with respect to \(c^{ab}(y)\), and the diagrammatic factor to its right has free indices \(\mu,\nu\) and free coordinate \(x\), then \(c^{ab}(y)\) is replaced by
\[
    {\delta^a}_{\mu}\,{\delta^b}_{\nu}
    \int d^4x\,\delta^{(4)}(y-x).
\]
In this operation as well, the symmetrization in \(\mu\) and \(\nu\) must be respected. After this replacement, all resulting contractions are performed.

\paragraph{Step 3: Collection of terms with the same diagrammatic factor.} 
After the operations of Step 2 have been performed, the ansatz sides of the defining conditions are expressed as a sum of diagrammatic factors whose basic structures have no free coordinate, as shown in Fig.~\ref{Figure_5}. We must then collect identical diagrammatic factors and determine the coefficient of each factor as a polynomial in the undetermined coefficients of the explicit expansion ansatz.

Before carrying out this collection, we substitute the bosonic effective action obtained in Appendix~\ref{Bosonic effective action} into the bosonic effective-action sides of the defining conditions. For later convenience, we then move these terms to the ansatz side, so that the defining conditions are written entirely as expressions to be collected on the ansatz side.

For this purpose, we construct the full list of diagrammatic factors that can arise after the Step 2 operations. The required input for this construction is the set of basic structures of the diagrammatic factors appearing on the ansatz side of the defining conditions after the Step 2 operations. These basic structures, including both anomalous and non-anomalous ones, are represented by the diagrams shown in Fig.~\ref{Figure_5}. By feeding these basic structures into the procedure of Part I, we obtain the required list as the output, taking into account that there are no free indices $\mu$ and $\nu$. Using the resulting list as a reference, we identify each generated term with its corresponding diagrammatic factor, group identical factors together, and sum their coefficients.

\paragraph{Step 4: Rewriting in the basis and extracting coefficients.} 
The diagrammatic factors appearing in the defining conditions are not all linearly independent, because they are related by integration-by-parts identities. For example, the diagrammatic factor
\[
\int d^d x\,
\partial_\mu c_\alpha{}^\rho\,
c^{\mu\nu}\,
c^{\lambda\alpha}\,
\partial_\lambda c_{\rho\nu}
\]
can be rewritten, by integration by parts, as a linear combination of other diagrammatic factors:
\[
\begin{aligned}
\int d^d x\,
\partial_\mu c_\alpha{}^\rho\,
c^{\mu\nu}\,
c^{\lambda\alpha}\,
\partial_\lambda c_{\rho\nu}
={}&
-\int d^d x\,
c_\alpha{}^\rho\,
\partial_\mu c^{\mu\nu}\,
c^{\lambda\alpha}\,
\partial_\lambda c_{\rho\nu}
\\
&-\int d^d x\,
c_\alpha{}^\rho\,
c^{\mu\nu}\,
\partial_\mu c^{\lambda\alpha}\,
\partial_\lambda c_{\rho\nu}
\\
&-\int d^d x\,
c_\alpha{}^\rho\,
c^{\mu\nu}\,
c^{\lambda\alpha}\,
\partial_\mu\partial_\lambda c_{\rho\nu},
\end{aligned}
\]
up to boundary terms. We therefore generate all linear relations, arising from integration by parts, among the diagrammatic factors in the full list constructed in Step 3. We then apply Gauss--Jordan elimination to these relations. This produces elimination rules that express the non-basis diagrammatic factors as linear combinations of the chosen basis factors. Using the elimination rules, we rewrite the defining conditions in terms of the linearly independent diagrammatic-factor basis. Finally, we extract the coefficient of each basis factor and set it to zero. This yields a system of equations for the undetermined coefficients \(M^{(a,b),X}_{i,j}\).

\paragraph{Output: System of equations for the $M^{(a,b),X}_{i,j}$.} The resulting system gives a necessary and sufficient set of conditions for the ansatz to define a Nicolai map within the chosen explicit expansion ansatz. We implemented this procedure in code and carried it out up to order \(\kappa^2\), thereby obtaining the conditions that the Nicolai-map ansatz must satisfy at that order.

\subsection{Part III: Finding solutions for the Nicolai map through the computation} \label{Part_III}
In this section, we present how to solve a system of simultaneous equations for the undetermined coefficients $M^{(a,b),X}_{i,j}$. 

\paragraph{Types of the undetermined coefficients.} The variables appearing in the simultaneous equations defining the Nicolai map up to order \(\kappa^2\) can be divided into two types. The first type consists of 68 variables of type \(M^{(0,1)}\) and 4 variables of type \(M^{(1,1)}\), which we collectively denote by \(M_i^{(1)}\) \((i=1,\dots,72)\). The second type consists of \( 17042\) variables of type \(M^{(0,2)}\) and \(627\) variables of type \(M^{(1,2)}\), which we collectively denote by \(M_a^{(2)}\) \((a=1,\dots,17669)\). 

\paragraph{Classification of the simultaneous equations.} The simultaneous equations defining the Nicolai map up to order \(\kappa^2\) can also be divided into two classes. 

The first class is the equations that are generated from the two order-$\kappa$ defining conditions in Eqs.~\eqref{Condition_zero_one} and ~\eqref{Condition_one_one}, and consist of 16 equations. Only the first-type variables, $M_i^{(1)}$'s, appear in these equations, and they are linear equations of the form
\begin{equation}
g_\alpha=\sum^{72}_{i=1} F_{\alpha i} M_i^{(1)},
\qquad (\alpha=1,2,\dots,16), \label{1st_class_eqs}
\end{equation}
where \(g_\alpha\) depends on the constant coefficients multiplying the diagrammatic factors in \(S^{(0,1)}\) and \(S^{(1,1)}\) that appear on the bosonic effective action sides of Eqs.~\eqref{Condition_zero_one} and ~\eqref{Condition_one_one} and \(F_{\alpha i}\) depends on the constant appearing in the coefficient factors of the terms obtained by substituting the explicit Nicolai map expansion ansatz at order \(\kappa\), namely \(T^{(0,1)}\) and \(T^{(1,1)}\) into the ansatz sides of Eqs.~\eqref{Condition_zero_one} and ~\eqref{Condition_one_one}.

The second class consists of the equations generated from the three order-$\kappa^2$ defining conditions in Eqs.~\eqref{Condition_zero_two}, \eqref{Condition_one_two}, and \eqref{Condition_two_two}. It contains 736 equations. Both types of variables, \(M_i^{(1)}\)'s and \(M_a^{(2)}\)'s, appear in these equations.

We isolate the constant contribution and denote it by \(h_\beta\). Then each equation contains two types of non-constant contributions: one is linear in the second-type variables, and the other is quadratic in the first-type variables. Thus the second-class equations can be written as
\begin{equation}
h_\beta
=
\sum^{17669}_{a=1} S_{\beta a} M_a^{(2)}
+
\sum^{72}_{i=1}\sum^{72}_{j=1}
s_{\beta ij} M_i^{(1)} M_{j}^{(1)},
\qquad
(\beta=1,2,\dots,736),
\label{2nd_class_eqs}
\end{equation}
with \(s_{\beta ij}=s_{\beta ji}\).

Here \(h_\beta\) denotes the constant contribution coming from the bosonic effective-action sides of Eqs.~\eqref{Condition_zero_two}, \eqref{Condition_one_two}, and \eqref{Condition_two_two}. More precisely, it is determined by the constants multiplying the diagrammatic factors appearing in \(S^{(0,2)}\), \(S^{(1,2)}\), and \(S^{(2,2)}\).

The coefficients \(S_{\beta a}\) are determined by the terms obtained by substituting the order-$\kappa^2$ ansatz, namely \(T^{(0,2)}\) and \(T^{(1,2)}\), into the relevant ansatz-side terms. These are the first term in Eq.~\eqref{Condition_zero_two}, the first and third terms in Eq.~\eqref{Condition_one_two}, and the second term in Eq.~\eqref{Condition_two_two}.

Similarly, the coefficients \(s_{\beta ij}\) are determined by the terms obtained by substituting the order-$\kappa$ ansatz, namely \(T^{(0,1)}\) and \(T^{(1,1)}\), into the relevant quadratic ansatz-side terms. These are the second term in Eq.~\eqref{Condition_zero_two}, the second and fourth terms in Eq.~\eqref{Condition_one_two}, and the first term in Eq.~\eqref{Condition_two_two}.

\paragraph{Procedure for obtaining a representative solution.}
To obtain a representative solution of the equations, we use the following strategy.
\begin{itemize}
    \item {\bf Step 1:}
We treat the first-type variables as constants and write the second class equations as a linear system for the second-type variables.
    \item {\bf Step 2:}
We apply Gauss-Jordan elimination to this linear system with respect to the second-type variables and extract the residual constraints on the first-type variables from the second class equations.
    \item {\bf Step 3:}
We obtain a representative solution for the first-type variables from the extracted constraints together with the first class equations.
    \item {\bf Step 4:}
We substitute the representative solution for the first-type variables into the second-class equations and obtain a representative solution for the second-type variables.
\end{itemize}

\paragraph{Step 1.}
If the first-type variables are treated as constants, the second class equations become a linear system for the second-type variables:
\begin{equation}
V_\beta=\sum^{17669}_{a=1} S_{\beta a} M_a^{(2)},
\qquad (\beta=1,2,\dots,736), \label{Linear_eq_Mj}
\end{equation}
where we introduce a constant vector assuming the $M_i^{(1)}$'s to be constants:
\begin{eqnarray}
  V_\beta \equiv    h_\beta-\sum^{72}_{i=1}\sum^{72}_{j=1} s_{\beta ij} M_i^{(1)} M_{j}^{(1)}\qquad (\beta =1,2,\dots,736). 
\end{eqnarray}

\paragraph{Step 2.}
We then perform Gauss-Jordan elimination on the matrix $S_{\beta a}$ in the second class equations in Eq.~\eqref{Linear_eq_Mj} with respect to the second-type variables $M_a^{(2)}$'s by multiplying the Gauss-Jordan elimination matrix $R_{\gamma \beta}$ to both sides of Eq.~\eqref{Linear_eq_Mj} and taking the sum over the index $\beta$ ($\beta=1,2,3\cdots,736$). We subsequently obtain
\begin{eqnarray}
V'_\gamma = \sum^{17669}_{a=1} S'_{\gamma a} M_a^{(2)},
\qquad (\gamma=1,2,\dots,736). \label{Linear_eq_Mj_GJ_eli}   
\end{eqnarray}
where we define
\begin{eqnarray}
&& V'_\gamma \equiv h'_\gamma-\sum^{72}_{i=1}\sum^{72}_{j=1} s'_{\gamma ij} M_i^{(1)} M_{j}^{(1)},\\
&& S'_{\gamma a} \equiv \sum^{736}_{\beta=1} R_{\gamma \beta}S_{\beta a},\\
&& h'_{\gamma}  \equiv \sum^{736}_{\beta=1} R_{\gamma \beta}h_\beta,\\
&& s'_{\gamma ij}=\sum^{736}_{\beta=1} R_{\gamma \beta} s_{\beta ij}.
\end{eqnarray}

If a certain row of the matrix $S'_{\gamma a}$ still contains a pivot variable when the corresponding row of $V'_\gamma$ is also non-vanishing, then that pivot variable can always be used to absorb the non-vanishing component of the row of $V'_\gamma$, and hence a solution can be found with such row. On the other hand, there remain two vanishing rows of the matrix $S'_{\gamma a}$, which contain no pivot variables. We find
\begin{eqnarray}
\textrm{For a specific~}\gamma=1,2~: ~ S'_{\gamma a} = 0, \qquad h'_{\gamma} = 0,    
\end{eqnarray}
and thus we obtain two constraints on the first-type variable $M_i^{(1)}$'s:
\begin{equation}
\label{additional equations}
\textrm{For a specific~}\gamma=1,2~:~V'_\gamma = 0 \implies \sum^{72}_{i=1}\sum^{72}_{j=1} s'_{\gamma ij} M_i^{(1)} M_{j}^{(1)}=0.
\end{equation}
Therefore, we obtain 18 equations for the first-type variables: the 16 first-class equations in Eq.~\eqref{1st_class_eqs}, together with the two additional equations in Eq.~\eqref{additional equations}. We spell out these simultaneous equations explicitly in Appendix~\ref{Appendix C}; they are given in Eqs.~\eqref{begin_eq}--\eqref{last_eq}. Thus, the first-order part of the Nicolai map cannot be determined solely from the defining conditions up to order \(\kappa\).

\paragraph{Step 3.}
Solving the nonlinear system ~\eqref{begin_eq}--~\eqref{last_eq} with the \texttt{nonlinsolve} function in \textsc{SymPy}, we obtain the relations ~\eqref{2begin_eq}--~\eqref{2last_eq}. By setting all variables except those appearing on the right sides of Eqs.~\eqref{2begin_eq}--\eqref{2last_eq} to zero, we obtain a representative solution:
\[
M^{(0,1)}_{1,3}=-1,\quad
M^{(0,1)}_{1,4}=1,\quad
M^{(0,1)}_{2,4}=-2,\quad
M^{(0,1)}_{1,5}=-1,\quad
M^{(0,1)}_{1,6}=\tfrac12,\quad
M^{(0,1)}_{1,7}=2,\quad
M^{(0,1)}_{1,8}=-\tfrac12,
\]
with all other \(M^{(0,1)}_{i,j}\) and \(M^{(1,1)}_{i,j}\) set to zero.

\paragraph{Step 4.}
We substitute the above representative solution for the first-type variables into the remaining 734 equations in  Eq.~\eqref{Linear_eq_Mj_GJ_eli}, excluding the two equations corresponding to  Eq.~\eqref{additional equations}. Setting all variables except the pivot variables to zero, we obtain a representative solution to the combined system of equations defined by Eqs.~\eqref{1st_class_eqs} and \eqref{2nd_class_eqs}. Substituting these values into the \(\mathcal O(\kappa^2)\) ansatz, we obtain a representative Nicolai map up to order \(\kappa^2\) for four-dimensional  \(\mathcal N=1\) pure supergravity as follows:
\begin{align}
T_{\kappa} c_{\mu\nu}(x)
&= c_{\mu\nu}(x)
\nonumber\\
&\qquad
 + \kappa \int d^4 y\, \Bigl[
   -\frac{1}{2}\,\partial_a\partial^{a} G(x-y)\,{c^b}_\mu(y)\,c_{b\nu}(y)
   +\frac{1}{2}\,\partial_a\partial^{a} G(x-y)\,c_{\mu\nu}(y)\,{c^b}_b(y)
\nonumber\\
&\qquad
   +\partial_\mu\partial_a G(x-y)\,{c^b}_\nu(y)\,{c^a}_b(y)
   +2\,\partial_a\partial_b G(x-y)\,{c^a}_\mu(y)\,{c^b}_\nu(y)
\nonumber\\
&\qquad
   -2\,\partial_\mu G(x-y)\,\partial_a {c^b}_\nu(y)\,{c^a}_b(y)
   -\partial_\mu\partial_a G(x-y)\,{c^a}_\nu(y)\,{c^b}_b(y)
\nonumber\\
&\qquad
   -\partial_a\partial^{b} G(x-y)\,c_{\mu\nu}(y)\,{c^a}_b(y)
 \Bigr] + \kappa^2 T^{(0,2)}c_{\mu\nu}(x) +\hbar \kappa^2 T^{(1,2)}c_{\mu\nu}(x),\label{rep_sol}
\end{align}
where $T^{(0,2)}c_{\mu\nu}(x)$ and $T^{(1,2)}c_{\mu\nu}(x)$ are given explicitly in Appendix \ref{Appendix D}, while we find that $T^{(1,1)}c_{\mu\nu}(x)$ vanishes.

The solution presented here should be regarded only as a representative example. In fact, there exist infinitely many other solutions, including solutions with
\(
T^{(1,1)}c_{\mu\nu}(x)=0
\)
as well as solutions with
\(
T^{(1,1)}c_{\mu\nu}(x)\neq 0
\).
Reference~\cite{Arrighi:2025eym} showed that the order-\(\kappa\) Nicolai map is not fixed by the order-\(\kappa\) conditions alone. It is further anticipated that the ambiguity would be removed once the order-\(\kappa^2\) conditions are also imposed. Our analysis shows, however, that although the order-\(\kappa^2\) conditions further reduce the freedom in the order-\(\kappa\) Nicolai map, they are still not sufficient to fix it uniquely. The remaining ambiguity is likely resolved only at still higher orders.

In addition, our condition Eq.~\eqref{Condition_one_one} is formulated in terms of $S^{(1,1)}$, including both the gravitino and the Faddeev--Popov ghost contributions.
By contrast, the $\mathcal O(\kappa)$ determinant matching analysis in Ref.~\cite{Arrighi:2025eym} is presented primarily by expanding and evaluating the $\mathcal O(\kappa)$ term in the gravitino determinant.

\subsection{Key observation: Einstein gravity admits Nicolai maps only through its $\mathcal N=1$ supersymmetric completion}\label{keySection}

We now generalize the discussion to a class of gravitational theories coupled to massless fermionic fields, whose only coupling constant is \(\kappa\) and whose purely bosonic sector is fixed to the Einstein--Hilbert action for the graviton. Compared with four-dimensional \(\mathcal N=1\) pure supergravity, the pure gravity contribution is unchanged, and possible differences can arise only from the fermionic and ghost sectors. More precisely, the possible diagrammatic factors appearing in \(S^{(1,1)}\), \(S^{(1,2)}\), and \(S^{(2,2)}\) have the same basic structures as before, while their numerical coefficients may be different.

\paragraph{What if a Einstein-Dirac system is given?} For example, consider the standard second-order action for a massless Dirac fermion coupled to gravity,
\begin{equation}
\label{diraction}
S=\frac{1}{2\kappa^2}\int d^4x\, e\,e^{a\mu}e^{b\nu}R_{\mu\nu ab}(\omega(e))-\int d^4x\, e\,\bar{\psi}\gamma^\mu D_\mu\psi .
\end{equation}
As in the pure \(\mathcal{N}=1\) supergravity case discussed in Sec.~\ref{section2}, we expand this action around the flat Minkowski background. The pure gravity part is identical to that of pure \(\mathcal{N}=1\) supergravity, so its perturbative expansion is unchanged. Thus the only difference comes from the fermionic part.

Even this difference, however, disappears at the level of schematic classification. If we classify the fermionic interaction terms only by the number of perturbation vielbein field and the derivative position, while suppressing the gamma-matrix structures and index contractions, their schematic classes coincide with those appearing in pure \(\mathcal{N}=1\) supergravity. At order \(\kappa\), the schematic classes that are present as the admissible diagrams in the corresponding effective action are
\begin{equation}
c\bar\psi\partial\psi,
\qquad
\partial c\,\bar\psi\psi ,
\end{equation}
while at order \(\kappa^2\), they are
\begin{equation}
c^2\bar\psi\partial\psi,
\qquad
c\,\partial c\,\bar\psi\psi .
\end{equation}

Although the action in Eq.~\eqref{diraction} does not contain a four-fermion term, the corresponding structure can still be regarded as an allowed schematic class with a vanishing coefficient. Thus Eq.~\eqref{diraction} is simply the special case in which the coefficient of this four-fermion class is set to zero. In this sense, including such a four-fermion term in Eq.~\eqref{diraction} would not affect the following discussion.

When deriving the bosonic effective action, we functionally integrate out the fermionic fields. In this process, the fermionic fields are contracted by fermion propagators carrying gamma-matrix structures. Since these contractions form closed fermion loops, the gamma matrices from the propagators and those from the interaction vertices are multiplied and finally traced over, reducing the gamma-matrix structures to numerical coefficients. After the fermionic integration and the gamma-matrix trace, neither the fermionic fields nor the gamma-matrix structures appear explicitly in the bosonic effective action. Since our present discussion is restricted to the basic structures, we further suppress the positions of the derivatives and the index structures. Therefore, which basic structures appear in the bosonic effective action is determined by the schematic classes of the fermionic interaction terms. Since these schematic classes coincide with those in pure \(\mathcal{N}=1\) supergravity, the basic structures of the diagrammatic factors appearing in
the bosonic effective action obtained from Eq.~\eqref{diraction} are already contained in the pure \(\mathcal{N}=1\) supergravity case.

It remains to check whether the gauge-fixing and ghost sectors can introduce new basic structures. Apart from local supersymmetry, the local symmetries of the present theory are the same as those of pure \(\mathcal{N}=1\) supergravity. Hence, after imposing the same gauge conditions \eqref{harmonic gauge} and \eqref{symmetry guage}, the gauge-fixing Lagrangian is obtained from Eq.~\eqref{gauage_fixing_term} by omitting the first term, while the ghost Lagrangian is obtained from Eq.~\eqref{ghost_lagrangian} by setting \(B\) and \(C\) to zero. Consequently, the basic structures of the diagrammatic factors generated by the gauge-fixing and ghost sectors are also already contained in the pure \(\mathcal{N}=1\) supergravity case.

The above discussion of the Einstein--Dirac example is not restricted to this particular model. The same reasoning applies to the whole class of gravitational theories coupled to massless fermionic fields whose only coupling constant is \(\kappa\), as described above. Since the schematic classes of the fermionic interaction terms remain the same throughout this class, changing the fermionic sector may change the numerical coefficients of the diagrammatic factors in the bosonic effective action, but it does not introduce new basic structures.

We therefore leave the coefficients of the diagrammatic factors from the effective actions that include loop diagrams in \(S^{(1,1)}\), \(S^{(1,2)}\), and \(S^{(2,2)}\) arbitrary, and repeat the procedure described above. Following the procedure in Sec.~\ref{Part_III}, in Step 1, in addition to treating the first-type variables \(M_i^{(1)}\)'s as constants, we also treat these arbitrary coefficients as constants. Then the second-class equations are again regarded as a linear system for the second-type variables \(M_a^{(2)}\)'s.

The result of Step 2 in Sec.~\ref{Part_III} is particularly important. The same Gauss--Jordan elimination again produces the two residual constraints in Eq.~\eqref{additional equations}. In other words, the two additional constraints on the first-type variables are independent of the arbitrary coefficients in \(S^{(1,2)}\) and \(S^{(2,2)}\). In this sense, they are universal constraints following from the second-class equations.

We then perform Step 3 in Sec.~\ref{Part_III} with these same residual constraints. The only difference from the previous pure-supergravity computation is that the two constant terms determined by \(S^{(1,1)}\) in the 18 equations \eqref{begin_eq}--\eqref{last_eq} are now treated as arbitrary unknowns. These two terms appear on the left-hand sides of Eq.~\eqref{S^(1,1)_condtion_1} and Eq.~\eqref{S^(1,1)_condtion_2}.

\paragraph{Exact determination of the two coefficients in the effective action \(S^{(1,1)}\).}
We now explain why the two coefficients in Eq.~\eqref{Action with a map}
are uniquely fixed by the consistency equations.  We
replace the two numerical coefficients in \(S^{(1,1)}\) by two arbitrary
parameters \(p\) and \(q\),
\begin{align}
S^{(1,1)}_{\rm trial}
=
i\int d^4x\,
\left[
p\,c^b{}_a(x)\,\partial^b\partial_aG(0)
+
q\,c^a{}_a(x)\,\delta(0)
\right].
\label{trial_one_one_pq}
\end{align}
With this convention, the pure supergravity value in
Eq.~\eqref{Action with a map} corresponds to
\begin{align}
p=4,\qquad q=-1.
\end{align}

The first fourteen equations in Appendix~\ref{Appendix C}, beginning with
Eq.~\eqref{begin_eq}, are the equations obtained from
Eq.~\eqref{Condition_zero_one}.  The next two equations,
Eqs.~\eqref{S^(1,1)_condtion_1} and
\eqref{S^(1,1)_condtion_2}, are obtained from
Eq.~\eqref{Condition_one_one}.  In the present test we replace the
left-hand sides of these two equations by \(p\) and \(q\), respectively.
Thus Eqs.~\eqref{S^(1,1)_condtion_1} and
\eqref{S^(1,1)_condtion_2} are read as
\begin{align}
p&={\cal F}_p(M^{(0,1)},M^{(1,1)}),
\label{Fp_def}
\\
q&={\cal F}_q(M^{(0,1)},M^{(1,1)}),
\label{Fq_def}
\end{align}
where \({\cal F}_p\) and \({\cal F}_q\) denote the right-hand sides of
Eqs.~\eqref{S^(1,1)_condtion_1} and
\eqref{S^(1,1)_condtion_2}.  No Rarita--Schwinger value is assumed at this
stage.

The two remaining equations are the two
quadratic constraints corresponding to Eq.~\eqref{additional equations} which are Eq.~\eqref{last_eq1} and Eq.~\eqref{last_eq}.  Let us denote these
two quadratic residuals by
\begin{align}
Q_1(M^{(0,1)},M^{(1,1)})=0,
\qquad
Q_2(M^{(0,1)},M^{(1,1)})=0 .
\end{align}
Although they look lengthy, their structure is
simple.  Define
\begin{align}
Y_1&:=M^{(1,1)}_{2,1},
&
Y_2&:=M^{(1,1)}_{2,2}.
\end{align}
The first quadratic residual can be written exactly as a square,
\begin{align}
Q_1=-{\cal C}_1^2,
\label{Q1_square_form}
\end{align}
where
\begin{align}
{\cal C}_1
:=
Y_1-\Xi_1,
\label{C1_def}
\end{align}
with
\begin{align}
\Xi_1
={}&
4M^{(0,1)}_{5,1}
-2M^{(0,1)}_{5,2}
-\frac12M^{(0,1)}_{5,3}
+M^{(0,1)}_{5,4}
+\frac12M^{(0,1)}_{5,5}
\nonumber\\
&-4M^{(0,1)}_{6,1}
+2M^{(0,1)}_{6,2}
+\frac12M^{(0,1)}_{6,3}
-M^{(0,1)}_{6,4}
-\frac12M^{(0,1)}_{6,5}
\nonumber\\
&+\frac12M^{(0,1)}_{8,3}
-M^{(0,1)}_{8,4}
-\frac12M^{(0,1)}_{9,3}
+M^{(0,1)}_{9,4}
+\frac12M^{(0,1)}_{9,5}.
\label{Xi1_def}
\end{align}
Indeed, expanding \(-{\cal C}_1^2\) reproduces term by term the first of
the two quadratic equations displayed after Eq.~\eqref{additional equations}
in Appendix~\ref{Appendix C}.  For example, the terms linear in
\(Y_1=M^{(1,1)}_{2,1}\) are
\begin{align}
2Y_1\Xi_1
={}&
8Y_1M^{(0,1)}_{5,1}
-4Y_1M^{(0,1)}_{5,2}
-Y_1M^{(0,1)}_{5,3}
+2Y_1M^{(0,1)}_{5,4}
+Y_1M^{(0,1)}_{5,5}
\nonumber\\
&-8Y_1M^{(0,1)}_{6,1}
+4Y_1M^{(0,1)}_{6,2}
+Y_1M^{(0,1)}_{6,3}
-2Y_1M^{(0,1)}_{6,4}
-Y_1M^{(0,1)}_{6,5}
\nonumber\\
&+Y_1M^{(0,1)}_{8,3}
-2Y_1M^{(0,1)}_{8,4}
-Y_1M^{(0,1)}_{9,3}
+2Y_1M^{(0,1)}_{9,4}
+Y_1M^{(0,1)}_{9,5},
\end{align}
which are precisely the \(M^{(1,1)}_{2,1}\)-linear terms in that equation.

The second quadratic residual has the corresponding completed-square form
\begin{align}
Q_2
=
\frac12{\cal C}_1^2
+
2{\cal C}_1{\cal C}_2
+
4{\cal C}_2^2,
\label{Q2_square_form}
\end{align}
where
\begin{align}
{\cal C}_2
:=
Y_2-\Xi_2,
\label{C2_def}
\end{align}
and
\begin{align}
\Xi_2
={}&
M^{(0,1)}_{5,10}
+4M^{(0,1)}_{5,11}
-2M^{(0,1)}_{5,12}
-\frac14M^{(0,1)}_{5,5}
-\frac12M^{(0,1)}_{5,6}
+\frac14M^{(0,1)}_{5,7}
+M^{(0,1)}_{5,8}
-\frac12M^{(0,1)}_{5,9}
\nonumber\\
&-M^{(0,1)}_{6,10}
-4M^{(0,1)}_{6,11}
+2M^{(0,1)}_{6,12}
+\frac14M^{(0,1)}_{6,5}
+\frac12M^{(0,1)}_{6,6}
-\frac14M^{(0,1)}_{6,7}
-M^{(0,1)}_{6,8}
+\frac12M^{(0,1)}_{6,9}
\nonumber\\
&-M^{(0,1)}_{8,10}
-\frac14M^{(0,1)}_{8,7}
-\frac14M^{(0,1)}_{9,5}
-\frac12M^{(0,1)}_{9,6}
-\frac12M^{(0,1)}_{9,9}.
\label{Xi2_def}
\end{align}
Again, Eq.~\eqref{Q2_square_form} is an identity obtained by direct
expansion of the second quadratic equation in Appendix~\ref{Appendix C},
namely Eq.~\eqref{last_eq}. What remains is to express ${\cal C}_1$ and ${\cal C}_2$ in terms of $p$ and
$q$.

Let \(E^{(0)}_r=0\) for \(r=1,\ldots,14\) denote the first fourteen linear
equations after we move all the right-hand-side terms to the left in Appendix~\ref{Appendix C}, beginning with
Eq.~\eqref{begin_eq}. Likewise
let
\begin{align}
E_p:={\cal F}_p(M^{(0,1)},M^{(1,1)})-p,
\qquad
E_q:={\cal F}_q(M^{(0,1)},M^{(1,1)})-q ,
\end{align}
where \({\cal F}_p\) and \({\cal F}_q\) are the right-hand sides of
Eqs.~\eqref{S^(1,1)_condtion_1} and
\eqref{S^(1,1)_condtion_2}.  Thus the sixteen first-order matching
conditions are
\begin{align}
E^{(0)}_1=\cdots=E^{(0)}_{14}=E_p=E_q=0 .
\end{align}
With this convention, the following two identities hold by direct expansion:
\begin{align}
{\cal C}_1+\frac12(p-4)
={}&
\frac94 E^{(0)}_1
-2E^{(0)}_2
-E^{(0)}_3
+E^{(0)}_4
+\frac14E^{(0)}_6
-\frac12E^{(0)}_7
-\frac12E^{(0)}_8
+E^{(0)}_9
-\frac12E_p ,
\label{C1_linear_certificate}
\end{align}
and
\begin{align}
{\cal C}_2-\frac14(p+2q-2)
={}&
-E^{(0)}_1
+E^{(0)}_2
+\frac34E^{(0)}_3
-\frac12E^{(0)}_4
+\frac12E^{(0)}_5
-\frac18E^{(0)}_6
+\frac14E^{(0)}_7
+\frac14E^{(0)}_8
\nonumber\\
&-\frac12E^{(0)}_9
+\frac18E^{(0)}_{10}
+2E^{(0)}_{11}
+\frac12E^{(0)}_{12}
+\frac12E^{(0)}_{13}
-E^{(0)}_{14}
+\frac14E_p
+\frac12E_q .
\label{C2_linear_certificate}
\end{align}
On any solution of the sixteen
first-order matching equations, all quantities
\(E^{(0)}_1,\ldots,E^{(0)}_{14},E_p,E_q\) vanish. Therefore the right-hand sides of Eqs.~\eqref{C1_linear_certificate} and
\eqref{C2_linear_certificate} vanish as well. Therefore, on the solution set of the sixteen linear equations,
Eqs.~\eqref{C1_linear_certificate} and
\eqref{C2_linear_certificate} reduce to
\begin{align}
{\cal C}_1
=
-\frac12(p-4),
\label{C1_p_relation}
\\
{\cal C}_2
=
\frac14(p+2q-2).
\label{C2_pq_relation}
\end{align}
The identities \eqref{C1_linear_certificate} and \eqref{C2_linear_certificate} are exact
linear identities with rational coefficients.
These relations are obtained by ordinary Gauss--Jordan elimination of the
sixteen linear equations.  Since all coefficients in Appendix~\ref{Appendix C}
are rational numbers, this elimination is exact and involves no numerical
approximation.

Substituting Eqs.~\eqref{C1_p_relation} and
\eqref{C2_pq_relation} into Eqs.~\eqref{Q1_square_form} and
\eqref{Q2_square_form}, the two quadratic residuals become
\begin{align}
Q_1
&=
-\frac14(p-4)^2,
\label{Q1_pq_reduced}
\\
Q_2
&=
\frac18
\left(
p^2+4pq-4p+8q^2+8
\right).
\label{Q2_pq_reduced}
\end{align}
The two quantities \(Q_1\) and \(Q_2\) are the two compatibility residuals of the \(O(\kappa^2)\) matching equations
after the second-order map coefficients have been eliminated. Hence, a
second-order Nicolai-map solution can exist only if
\(Q_1=0,\ Q_2=0.\) 

From Eq.~\eqref{Q1_pq_reduced} one obtains
\begin{align}
p=4.
\end{align}
With this value inserted into Eq.~\eqref{Q2_pq_reduced}, one finds
\begin{align}
Q_2
=
\frac18
\left(
16+16q-16+8q^2+8
\right)
=
(q+1)^2.
\end{align}
Thus
\begin{align}
q=-1.
\end{align}
Consequently,
\begin{align}
p=4,\qquad q=-1.
\label{pq_unique_result}
\end{align}

This proves that, within the reduced system of Appendix~\ref{Appendix C},
the existence of a second-order Nicolai-map solution uniquely fixes the two
coefficients in \(S^{(1,1)}\).  This is not a uniqueness statement for the
Nicolai map itself: the first-order map coefficients \(M^{(0,1)}_{i,j}\) and
\(M^{(1,1)}_{2,j}\) still have residual freedom, as illustrated by
Eqs.~\eqref{2begin_eq}--\eqref{2last_eq}.  The unique object fixed by the
consistency equations is the one-loop tensor structure in
Eq.~\eqref{trial_one_one_pq}, namely the value shown in
Eq.~\eqref{Action with a map}.

\paragraph{Key observation: Poincar\'{e} supergravity is necessary for the Nicolai-map construction.} Solving the resulting generalized system, we find that the two coefficients are fixed to definite constants, which can be obtained from \(S^{(1,1)}\) that takes the following form:\footnote{We solved the resulting generalized system of 18 equations using \texttt{nonlinsolve} in Python's \texttt{sympy} package.}
\begin{align}
\label{Action with a map}
  S^{(1,1)}
  &= \int d^4 x\,
     \Bigl(
       +4 i\, c^b{}_a(x)\,\partial^b \partial_a G(0)
       - i\, c^a{}_a(x)\,\delta(0)
     \Bigr).
\end{align}
This indicates that only theories whose \(S^{(1,1)}\) is given by Eq.~\eqref{Action with a map} can admit a Nicolai map of Einstein gravity. Remarkably, Eq.~\eqref{Action with a map} coincides with \(S^{(1,1)}\) consisting of the Rarita–Schwinger action for gravitino from the four dimensional \(\mathcal N=1\) pure supergravity. Consequently, once order-$\kappa^2$ terms of Einstein gravity are taken into account, one cannot expect an arbitrary gravitational theory. We observe that a consistent Nicolai-map construction for the Einstein--Hilbert graviton sector already requires the Rarita--Schwinger action for a gravitino. Thus, in our perturbative framework, Einstein gravity admits Nicolai maps only through its $\mathcal N=1$ supersymmetric completion, namely Poincar\'{e} supergravity. 

\paragraph{Relation to Nicolai's characterization of supersymmetry.}
This observation is, we think, a perturbative instance of the characterization
of supersymmetry that Nicolai used to motivate the map in the first place
\cite{Nicolai:1979nr,Nicolai:1980jc}. There a theory counts as supersymmetric
exactly when one can find a transformation of the bosonic fields alone, in
general nonlinear and nonlocal, that does two things at once: (i)~it carries
the interacting bosonic action into the free action, and (ii)~its functional
Jacobian reproduces the determinant left over from integrating out the
fermions. We refer to these as the free-action and determinant-matching
conditions. Of the two, only the second knows anything about supersymmetry.
Condition (i) constrains the bosonic sector and is indifferent to the fermion
content; it is condition (ii) that does the real work, since supersymmetry is
what allows the fermionic integral to be traded for the Jacobian of a single
bosonic field redefinition.

The loop expansion sorts our conditions into precisely these two types. The
tree-level conditions \eqref{Condition_zero_one} and \eqref{Condition_zero_two}
contain no $\hbar$ and are built only from $\boxtimes$-products; they are the
free-action conditions. They fix the graviton (vielbein-perturbation) data and
leave the gravitino couplings open---the same bosonic-sector ambiguity already
noted in Ref.~\cite{Arrighi:2025eym}. The conditions
\eqref{Condition_one_one}, \eqref{Condition_one_two} and
\eqref{Condition_two_two} are the ones carrying the log-Jacobian
``$-i\hbar\,\mathrm{Tr}\ln(\delta T_\kappa c/\delta c)$,'' matched against the loop
effective action $S^{(1,1)}$, $S^{(1,2)}$, $S^{(2,2)}$, that is, against the
gravitino--ghost determinant. These play the role of the determinant-matching
condition and its higher-order descendants.

With this in mind, the result of Sec.~\ref{keySection} is the free-action and
determinant-matching conditions spelled out through $\mathcal O(\kappa^2)$. We fix
the bosonic sector to the gauge-fixed Einstein--Hilbert action. No amount of
free-action data then certifies a map on its own; the loop conditions become
mutually consistent through $\mathcal O(\kappa^2)$ only at $(p,q)=(4,-1)$ in
Eq.~\eqref{Action with a map}, and that value is just the Rarita--Schwinger
gravitino determinant. So among gravitational theories coupled to massless
fermions with $\kappa$ as their sole coupling, it is the determinant-matching
half of the conditions that picks out $\mathcal N=1$ Poincar\'e supergravity.
To the order we have reached, asking the map to exist is already enough to force
the supersymmetric completion of Einstein gravity, much as
Refs.~\cite{Nicolai:1979nr,Nicolai:1980jc} would lead one to expect.

\section{Comparison with the coupling-flow approach for supergravity}
\label{comparison}

It is instructive to compare the present diagrammatic framework with the
coupling-flow approach for supergravity in Ref.~\cite{Arrighi:2025eym}, which addresses the same
physical problem---the construction of a Nicolai map for four-dimensional
$\mathcal{N}=1$ pure supergravity around flat Minkowski space---through a
different formalism. The two works address the same problem by different means and end up saying
different things about it. Reference~\cite{Arrighi:2025eym} explains why the
coupling-flow construction is obstructed by local supersymmetry. Here the
obstructions do not arise, since the diagrammatic route never invokes the
structures that fail. The price is a much larger combinatorial problem.

\paragraph{Common ground.} Both works treat $D=4$, $\mathcal{N}=1$ pure Poincar\'e supergravity around the flat Minkowski background, expanded perturbatively in the gravitational coupling $\kappa$ via the background-field method \cite{DeWitt:1967ub,DeWitt:1967uc}, with the vielbein fluctuation defined by $e^{a}{}_{\mu}=\delta^{a}{}_{\mu}+\kappa\,(\cdot)^{a}{}_{\mu}$. The gauge sector is fixed through the Faddeev--Popov procedure \cite{Faddeev:1967fc} with the same three gauge conditions: the gamma-trace (Rarita--Schwinger) gauge for the gravitino, the de Donder (harmonic) gauge in the vielbein language for diffeomorphisms, and the symmetric-vielbein gauge for the local Lorentz redundancy. Finally, both works point to the same qualitative hierarchical feature: the order-$\kappa$ data of the Nicolai map are not pinned down by the order-$\kappa$ defining conditions alone. This feature is anticipated in Ref.~\cite{Arrighi:2025eym} and made explicit here, where the order-$\kappa^{2}$ conditions are shown to impose additional constraints on the order-$\kappa$ coefficients (Sec.~\ref{section4}).

\paragraph{Methodological divergence.} The two works diverge at the structural level, in two related respects.

First, in the choice of formulation. Reference~\cite{Arrighi:2025eym} retains
the auxiliary fields $(S,P,A_{\mu})$ of the off-shell minimal multiplet and
gauge-fixes with Nakanishi--Lautrup auxiliary fields, so that the BRST
transformations are nilpotent off-shell. The present work instead
eliminates the auxiliary fields and works on-shell, where off-shell BRST
invariance fails and is restored by a Kallosh-type four-ghost counterterm
$\tfrac{5}{32}\kappa^{2}\,B\gamma_{d}\bar{B}\,(\bar{C}\gamma^{d}C)$
\cite{Kallosh:1978de,StermanTownsendVanNieuwenhuizen:1978}. This counterterm is
therefore specific to the present on-shell treatment and has no analogue in the
off-shell construction of Ref.~\cite{Arrighi:2025eym}.

Second, in the construction strategy. Reference~\cite{Arrighi:2025eym} attempts
to lift the coupling-flow operator $R_{g}$ of Ref.~\cite{Lechtenfeld:2021uvs}
from rigid to local supersymmetry. In the global case this operator generates
the Nicolai map recursively from the supersymmetry Ward identity, provided two
ingredients are available: (i) the off-shell action can be written as a
supervariation, and (ii) the resulting flow operator reproduces the Euler
operator at leading order. The present work abandons $R_{g}$ and returns to the
defining identity
\begin{equation}
S[c;\kappa] \;=\; S[\,T_{\kappa}c,\,0,\,0,\,0;\,0\,]
\;-\;i\hbar\,\mathrm{Tr}\ln\!\left(\frac{\delta T_{\kappa}c}{\delta c}\right),
\label{eq:defining-id-comparison}
\end{equation}
expanding both sides simultaneously in $\hbar$ and $\kappa$ as in
Eqs.~\eqref{full_expansion_of_S} and~\eqref{full_expansion_of_T}. The matching
is translated into a diagrammatic algorithm, in which the admissible candidate
terms are enumerated by their basic structures (lines, vertices, loops, and
derivative placements) and reduced to a finite system of nonlinear polynomial
equations for the coefficients $M^{(a,b),X}_{i,j}$. Neither the Ward identity
nor a supervariation enters as input: both sides of
Eq.~\eqref{eq:defining-id-comparison} are evaluated as ordinary perturbative
objects, and consistency is enforced solely by diagram matching.

\subsection{Circumvention of the three obstructions of the coupling-flow approach}
\label{sec:obstructions}

The three obstructions identified in Ref.~\cite{Arrighi:2025eym} stem from the
algebraic requirements of the coupling-flow construction in the off-shell
formalism. We treat them in turn, and then comment separately on the on-shell
route, which Ref.~\cite{Arrighi:2025eym} explores as an independent attempt
rather than as one of the three obstructions.

\paragraph{No supervariation required: no density obstruction.}
Reference~\cite{Arrighi:2025eym} shows that the off-shell supergravity
Lagrangian $\mathcal{L}=e\,\mathcal{L}_{0}$, being a density rather than a
scalar, is not expressible as a complete supervariation under off-shell SUSY
transformations. This breaks the identity from which $R_{g}$ is derived and
propagates into the flow equation, where $R_{g}$ is accompanied by a
multiplicative term; the candidate Nicolai map is then only partial,
i.e.\ supplemented by an additional measure factor in the path integral. In the
present construction, this distinction does not enter: the bosonic effective
action $S[c;\kappa]$ on the left-hand side of
Eq.~\eqref{eq:defining-id-comparison} is built by ordinary Wick contractions in
the free theory defined by $S^{[0]}_{r}$ (Appendix~\ref{Bosonic effective action}),
and the density character of the Lagrangian enters only through ordinary
momentum-space integrals. Equation~\eqref{eq:defining-id-comparison} is the
integrated form of the Nicolai-map condition, with no flow equation and hence no
multiplicative term to absorb. We do not, however, analyze here the
$\kappa$-dependence of the vacuum energy to which the measure factor of
Ref.~\cite{Arrighi:2025eym} corresponds; that would require the
determinant-matching analysis discussed below.

\paragraph{No BRST Ward identity required: no $\{\delta_\alpha,s\}$ obstruction.}
Because local supersymmetry $\delta_{\alpha}$ is part of the gauge invariance, it no longer
graded-commutes with the Slavnov (BRST) variation $s$ used in gauge fixing,
$\{\delta_{\alpha},s\}\neq0$. Reference~\cite{Arrighi:2025eym} finds that the
BRST Ward identity can then be used in the construction of the rescaled flow
operator $\widetilde{R}_g$ only at the expense of a second multiplicative contribution. This obstruction does not arise in the present construction,
which does not invoke a BRST Ward identity. The gauge-fixed action of the gravitino and full
Faddeev--Popov ghost sector enters only through the bosonic effective
action $S^{(r,n)}$ obtained by integrating out all non-vielbein fields
(Appendix~\ref{Bosonic effective action}). The commutator $\{\delta_{\alpha},s\}$ never appears.

\paragraph{No coupling-flow and Euler operators required: no conformal-mode obstruction.} The conformal mode\footnote{The trace $c\equiv c^{a}{}_{a}$ from the metric fluctuation is what one calls the conformal mode: a local
Weyl rescaling $g_{\mu\nu}\to e^{2\sigma}g_{\mu\nu}$, linearized around the flat
background, shifts the metric fluctuation purely along the trace direction,
$\delta c_{\mu\nu}=2\sigma\,\eta_{\mu\nu}$, leaving the transverse--traceless
graviton untouched \cite{Freedman:2012zz}. Equivalently, the local volume
element is governed by the same mode,
$\sqrt{-g}=e=\det(e^{a}{}_{\mu})\simeq 1+\kappa\,c^{a}{}_{a}+O(\kappa^{2})$,
so that the Weyl factor $\Omega=e^{\sigma}$ is literally the conformal factor of
the metric.} controls the structural obstruction found in
Ref.~\cite{Arrighi:2025eym}. There, at leading order in $\kappa$, the
degree-zero part of the rescaled flow operator $\widetilde{R}$ is required to
reduce to the field-counting Euler operator
\begin{equation}
E\;\equiv\;\int\! d^{4}x\;\phi(x)\,\frac{\delta}{\delta\phi(x)},
\label{eq:euler-comparison}
\end{equation}
where $\phi$ denotes the (unrescaled) graviton fluctuation\footnote{Unlike the work in Ref.~\cite{Arrighi:2025eym}, we denote this by $c$ in this work.} of that reference,
$e^{a}{}_{\mu}=\delta^{a}{}_{\mu}+\kappa\,\phi^{a}{}_{\mu}$, related to the
rescaled field $\tilde{\phi}$ by $\tilde{\phi}=\kappa\phi$. This test fails by a
term proportional to the trace of the metric fluctuation, i.e.
$R_{0}=E+(\text{terms}\propto\phi^{a}{}_{a})$, the residual pieces descending
from the expansion of the vielbein determinant
$e\simeq 1+\kappa\,\phi^{a}{}_{a}$. The conformal factor is thereby singled out
as the structural source of the failure, and a unimodular formulation (with
$\sqrt{-g}$ fixed and the conformal mode non-dynamical) is expected to evade it. On the contrary, in our diagrammatic construction, no such external reduction test
arises. The defining conditions
\eqref{Condition_zero_zero}--\eqref{Condition_two_two} never require any operator
to collapse onto $E$, so there is no step at which a trace remnant could spoil
the matching; instead, the trace $c\equiv c^{a}{}_{a}$ enters the bosonic
effective action on the same footing as the traceless components, with no
distinguished role. Already at tree level,
$S^{(0,1)}=\int d^{4}x\,\mathcal{L}_{\text{vielbein},1}$ (Appendix~\ref{Appendix A})
contains genuine traceless tensor contractions, such as
$\int d^{4}x\,\partial^{\nu}c_{\rho\nu}\,c^{\lambda\alpha}\,\partial_{\lambda}c_{\alpha}{}^{\rho}$,
side by side with pure-trace structures, such as
$\int d^{4}x\,\eta^{\gamma\mu}\,\partial_{\mu}c\,c\,\partial_{\gamma}c$; both are
reproduced on the ansatz side through the order-$(0,1)$ condition
\eqref{Condition_zero_one}, the trace pieces being carried by the trace and
$\eta_{\mu\nu}$-type candidate terms of $T^{(0,1)}c$
(Eq.~\eqref{eq:T10cmunu-long}). The same pattern persists at loop level: the
trace structures appearing in $S^{(1,1)}$, $S^{(1,2)}$ and $S^{(2,2)}$ are
matched by the trace and $\eta_{\mu\nu}$ candidate terms of $T^{(0,1)}c$ and
$T^{(1,1)}c$ (Eqs.~\eqref{eq:T10cmunu-long} and~\eqref{eq:T11cmunu-long}) through
conditions \eqref{Condition_one_one}--\eqref{Condition_two_two}. (The one-loop
coincident-point objects $G(0)$, $\partial^{a}\partial^{b}G(0)$ and $\delta(0)$
entering, e.g., $S^{(1,1)}$ in Eq.~\eqref{Action with a map} are treated here as
independent, unregularized symbols; their eventual regularization fixes their
numerical values but does not alter this structural point.) The conformal mode is
therefore an ordinary degree of freedom of the ansatz rather than an obstruction.

\paragraph{On the on-shell route of Ref.~\cite{Arrighi:2025eym}.} Reference~\cite{Arrighi:2025eym} also reports that the on-shell supersymmetry
strategy, successful for super-Yang--Mills theory in its critical dimensions
\cite{Ananth:2020lup}, fails for supergravity because the graviton
self-interaction cannot be written as a supervariation. We record this
separately, as it is an independent attempt in Ref.~\cite{Arrighi:2025eym}, not
one of the three off-shell obstructions. In the present construction no
supervariation of any term is required: the graviton self-interaction enters as
ordinary $\mathcal{O}(\kappa^{n})$ vertices in $\mathcal{L}_{\text{vielbein},n}$
(Appendix~\ref{Appendix A}) and contributes through standard Feynman
contractions.



\paragraph{Common limitations and open issues.} Several limitations are shared by the two approaches. The order-$\kappa$ data of
the map is not unique: Ref.~\cite{Arrighi:2025eym} obtains a four-parameter
family of leading-order maps satisfying the free-action condition, while here an
infinite family of order-$\kappa^{2}$ solutions remains, parameterized by the
residual freedom in the relations \eqref{2begin_eq}--\eqref{2last_eq}; removing
it presumably requires extending the analysis to $\mathcal{O}(\kappa^{3})$.
Neither work treats the regularization of the coincident-point objects $G(0)$,
$\delta(0)$, and their derivatives, which appear throughout the loop-corrected
map (Appendix~\ref{Appendix D}); a regularization scheme, with its contact terms
and scheme dependence, must be fixed before the map can be used to compute
bosonic correlators or to test Ward identities. Finally, the
determinant-matching condition, the most stringent quantum-level test, is
reduced to an algebraic statement in Eq.~\eqref{eq:defining-id-comparison} but
has not been verified beyond the order at which we work;
Ref.~\cite{Arrighi:2025eym} stresses that this is the decisive test any
candidate map must pass.

 \paragraph{Landau gauge limit and consistency.} The limit $\lambda\to\infty$ in Eq.~\eqref{lambda_gauge_fixing_term} restricts the functional integral to $c_{\mu\nu}=c_{\nu\mu}$. In this work, we do not take Landau-gauge limit for the harmonic gauge, not imposing preservation of the harmonic-gauge slice for the Nicolai map solution we found. On the contrary, Ref.~\cite{Arrighi:2025eym} works in the Landau gauge. In fact, according to a gauge fixing condition, the corresponding inverse-kinetic operator of the vielbein can be defined. This implies that the specific form of the $\boxtimes$-operator in Eq.~\eqref{free_action} can be established in accordance with the gauge fixing condition. In principle, with a proper $\boxtimes$-operator, our diagrammatic method for the Nicolai map is applicable for any gauge choice. We have checked a consistency that our Nicolai map solutions~\cite{SUGRA_Nicolai_Code} (found through $\mathcal{O}(\kappa^2)$) in the harmonic gauge with the Landau-gauge limit are equal to that of Ref.~\cite{Arrighi:2025eym} at the leading order $\mathcal{O}(\kappa)$.

\section{Conclusion}\label{conclusion}

We have presented a perturbative diagrammatic scheme for the Nicolai map of $D=4$, $\mathcal N=1$ pure (Poincar\'{e}) supergravity around flat Minkowski space, with $\kappa$ and $\hbar$ as the two expansion parameters. The coupling-flow operator $R_g$, which works well in rigidly supersymmetric theories, is obstructed in the local case by the gauge structure of supersymmetry \cite{Arrighi:2025eym}. We therefore returned to the defining identity
\[
S[c;\kappa] = S[T_\kappa c,0,0,0;0] - i\hbar\,\trace\ln\!\left(\frac{\delta T_\kappa c}{\delta c}\right),
\]
expanded both sides in $(\hbar,\kappa)$, and matched the resulting expressions order by order. This matching procedure admits a diagrammatic translation. In this translation, the operations $Tc \boxtimes Tc$ and $\trace(\prod \delta Tc/\delta c)$ acquire direct graphical readings, and the candidate terms of the ansatz are enumerated by their basic structures, namely lines, vertices, loops, and derivative placements. The matching is thereby turned into a finite system of nonlinear polynomial equations for the coefficients $M^{(a,b),X}_{i,j}$, which we solved through $\mathcal O(\kappa^2)$ by an automated Python pipeline.

Three features of the construction deserve comment.

The order-$\kappa$ map is not fixed by the order-$\kappa$ conditions. After
Gauss--Jordan reduction of the order-$\kappa^2$ system, two non-pivot rows
survive as quadratic equations on the order-$\kappa$ coefficients, so that
extendability to higher order feeds back on the lower-order data. This is what
Ref.~\cite{Arrighi:2025eym} anticipated. Even with both the order-$\kappa$ and
order-$\kappa^2$ conditions imposed, however, the order-$\kappa$ map remains
non-unique; an infinite family of solutions survives, some with
$T^{(1,1)}c=0$ and some without. We expect $\mathcal O(\kappa^3)$ to be needed
to remove what is left.

Our main physical conclusion concerns the order-$\kappa^2$ conditions. With the
bosonic sector set to the gauge-fixed Einstein--Hilbert action, the 18
universal equations on the first-class coefficients do not depend on how the
gravitino--ghost sectors enter $S^{(1,2)}$ and $S^{(2,2)}$. They act as a
filter on the admissible $\mathcal O(\kappa)$ gravitino couplings. Equivalently,
fixing $S^{(0,1)}$ to the Einstein--Hilbert value and leaving $S^{(1,1)}$ free,
the same conditions force $S^{(1,1)}$ into the form of
Eq.~\eqref{Action with a map}---precisely the one-loop bosonic effective action
of the $\mathcal N=1$ gravitino--ghost sector. So the existence of a Nicolai
map through $\mathcal O(\kappa^2)$ already picks out the supersymmetric
completion of pure gravity from the wider class of gravity-plus-gravitino
theories with $\kappa$ as their only coupling. We take this as a perturbative
trace of the rigidity expected of locally supersymmetric extensions of gravity.

Finally, Appendix~\ref{Appendix D} gives an explicit representative solution.
We are not aware of an earlier Nicolai map for a four-dimensional locally
supersymmetric theory carried to $\mathcal O(\kappa^2)$. It contains coincident
objects $G(0)$, $\delta(0)$ and their derivatives, a residue of their loop
origin, and a regularization scheme would have to be chosen before the map
could be used to compute anything. We have not done so, our concern here being
the algebraic structure of the map and its defining conditions.

\subsection*{Acknowledgments} 
H.J. is supported by the National Research Foundation of Korea (NRF) through the Grants: RS-2020-NR049598 (CQUeST, Sogang University), RS-2023-NR077094, RS-2024-00441954, and RS-2026-25484163 (Sejong Science Fellowship). H.J. would like to thank Olaf Lechtenfeld for his support and encouragement when H.J. began his research on the Nicolai map as a Riemann Fellow at the Riemann Center for Geometry and Physics, Leibniz University Hannover, Germany, in 2022. The authors also thank Hermann Nicolai for his valuable and encouraging comments on the manuscript.

\printbibliography

\begin{appendices}
\section{The gauge-fixed action of Poincar\'{e} supergravity to $O(\kappa^2)$}
\label{Appendix A}

Throughout Appendices A and B, \(c \equiv {c^{a}}{}_{a}\) denotes the trace of the vielbein perturbation \({c^{a}}{}_{\mu}\). In this appendix, we explicitly present the total action by taking advantage of the
background–field method for $D=4$, $\mathcal{N}=1$ pure supergravity,
expanded up to $\mathcal{O}(\kappa^2)$. The full action is
\begin{equation}
  S
  = S_{\text{vielbein}}\bigl[c^a{}_\mu\bigr]
  + S_{\text{gravitino}}\bigl[c^a{}_\mu,\psi_\mu\bigr]
  + S_{\text{ghost}}\bigl[c^a{}_\mu,\psi_\mu,B,C,C^*_\nu,C^\lambda,
        C^*_{ab},C'_{ab},u_I^*,u^I\bigr] \, .
\end{equation}

The vielbein action $S_{\text{vielbein}}[c^a{}_\mu]$ is
\begin{equation}
\label{L_vielbein_0}
  S_{\text{vielbein}}\bigl[c^a{}_\mu\bigr]
  = -\frac{1}{2} \int d^4 x \Bigl(
      \eta^{\gamma\mu} \partial_\mu c^v{}_a \, \partial_\gamma c^a{}_v
      - \frac12 \eta^{\gamma\mu} \partial_\mu c \, \partial_\gamma c
      + \kappa \mathcal{L}_{\text{vielbein},1}
      + \kappa^2 \mathcal{L}_{\text{vielbein},2}
      + \mathcal{O}(\kappa^3)
    \Bigr) .
\end{equation}
The interaction densities $\mathcal{L}_{\text{vielbein},1}$ and $\mathcal{L}_{\text{vielbein},2}$
are
\begin{align}
  \mathcal{L}_{\text{vielbein},1}
  =&\;
   - \eta^{\gamma\mu} \partial_\mu c \, c \, \partial_\gamma c
   + 2 \partial_\mu c \, c \, \partial_\gamma c^{\mu\gamma}
   + 2 \partial_\mu c \, c^{\gamma\mu} \, \partial_\gamma c
   + 2 \eta^{\gamma\lambda} \partial_\gamma c \, c^{\delta v} \, \partial_\lambda c_{v\delta}
   \nonumber\\[3pt]
  &\;
   + \eta^{\gamma\mu} c \, \partial_\mu c^v{}_b \, \partial_\gamma c^b{}_v
   - 4 \partial_\gamma c^{x\gamma} c^\lambda{}_x \, \partial_\lambda c
   - 2 \partial_\gamma c \, c^{\lambda v} \, \partial_v c^\gamma{}_\lambda
   \nonumber\\[3pt]
  &\;
   - c \, \partial_\mu c^\mu{}_b \, \partial_\gamma c^{b\gamma}
   - c \, \partial_\mu c^{v\gamma} \, \partial_\gamma c^\mu{}_v
   - 2 \eta^{\gamma\mu} \partial_\mu c^v{}_c \, c^c{}_a \, \partial_\gamma c^a{}_v
   \nonumber\\[3pt]
  &\;
   + 2 \partial^v c_{\rho v} \, c_\alpha{}^\rho \, \partial_\lambda c^{\lambda\alpha}
   - 2 \partial_\mu c_{\lambda\rho} \, c^{\mu v} \, \partial_v c^{\lambda\rho}
   + 0 \, \partial^v c^{\lambda\alpha} c_\alpha{}^\rho \partial_\lambda c_{\rho v}
   \nonumber\\[3pt]
  &\;
   + 2 \partial^v c_{\rho v} \, c^{\lambda\alpha} \, \partial_\lambda c_\alpha{}^\rho
   + 4 \partial_\lambda c_{\rho v} \, c^{\lambda\alpha} \, \partial^v c_\alpha{}^\rho
   - 2 \partial_v c^{\mu v} c_{ef} \, \partial_\mu c^{ef}
   \nonumber\\[3pt]
  &\;
   + a \, \partial_\mu c \, \partial_v c^{\alpha v} c_\alpha{}^\mu
   - a \, \partial_\mu c \, c^{\alpha v} \partial_v c_\alpha{}^\mu
   + a \, c \, \partial_v c^{\alpha v} \partial_\mu c_\alpha{}^\mu
   - a \, c \, \partial_\mu c^{\alpha v} \partial_v c_\alpha{}^\mu
   \nonumber\\[3pt]
  &\;
   + b \, \partial_\lambda c^{\lambda\mu} c^x{}_\mu \partial_\gamma c_x{}^\gamma
   + b \, \partial_\lambda c^{\lambda\mu} \partial_\gamma c^x{}_\mu \, c_x{}^\gamma
   - b \, \partial_\gamma c^{\lambda\mu} c^x{}_\mu \partial_\lambda c_x{}^\gamma
   - b \, c^{\lambda\mu} \partial_\gamma c^x{}_\mu \partial_\lambda c_x{}^\gamma \, ,\label{L_vielbein_1}
\end{align}
and
\begin{align}
  \mathcal{L}_{\text{vielbein},2}
  =&\;
  - \frac12 \partial^v c \, \partial_v c \, c^2
  + 2 \partial_\mu c \, \partial_v c \, c^{\mu v} c
  + \partial_\mu c \, \partial_v c^{\mu v} c^2
  + \frac12 \partial_v c \, \partial^v c \, c^{ef} c_{ef}
  \nonumber\\[3pt]
  &\;
  - 3 \partial_\mu c \, \partial_v c \, c^{\mu\rho} c_\rho{}^v
  + 2 \partial^v c \, \partial_v c_{\alpha\beta} c^{\alpha\beta} c
  - 4 \partial_v c_\rho{}^v \, \partial_\mu c \, c^{\mu\rho} c
  \nonumber\\[3pt]
  &\;
  - 2 \partial_v c^{\mu\rho} \partial_\mu c \, c_\rho{}^v c
  + \frac12 \partial^v c_{\lambda\rho} \partial_v c^{\lambda\rho} c^2
  - \frac12 \partial_\mu c^{\mu\rho} \partial_v c_\rho{}^v c^2
  \nonumber\\[3pt]
  &\;
  - \frac12 \partial_\mu c_\rho{}^v \partial_v c^{\mu\rho} c^2
  - 2 \partial^v c \, \partial_v c_{\lambda\rho} c^{\lambda\alpha} c_\alpha{}^\rho
  - \partial_\mu c \, \partial_v c^{\mu v} c^{ef} c_{ef}
  \nonumber\\[3pt]
  &\;
  - 4 \partial_v c \, \partial_\mu c^{ef} c_{ef} c^{\mu v}
  + 6 \partial_\mu c^{\mu\alpha} c_\alpha{}^\rho c_\rho{}^v \partial_v c
  + 4 c^{\mu\alpha} \partial_\mu c_\alpha{}^\rho c_\rho{}^v \partial_v c
  \nonumber\\[3pt]
  &\;
  + 2 c^{\mu\alpha} c_\alpha{}^\rho \partial_\mu c_\rho{}^v \partial_v c
  - 2 \partial^v c_{\lambda\rho} \partial_v c^{\lambda\alpha} c_\alpha{}^\rho c
  + 2 \partial^v c_{\rho v} c_\alpha{}^\rho \partial_\lambda c^{\lambda\alpha} c
  \nonumber\\[3pt]
  &\;
  - 2 \partial_\mu c_{\lambda\rho} c^{\mu v} \partial_v c^{\lambda\rho} c
  + 0 \partial^v c^{\lambda\alpha} c_\alpha{}^\rho \partial_\lambda c_{\rho v} c
  + 2 \partial^v c_{\rho v} c^{\lambda\alpha} \partial_\lambda c_\alpha{}^\rho c
  \nonumber\\[3pt]
  &\;
  + 4 \partial_\lambda c_{\rho v} c^{\lambda\alpha} \partial^v c_\alpha{}^\rho c
  - 2 \partial_v c^{\mu v} c_{ef} \partial_\mu c^{ef} c
  - \frac12 \partial^v c^{\lambda\rho} \partial_v c_{\lambda\rho} c^{ef} c_{ef}
  - c^{\lambda\rho} \partial_v c_{\lambda\rho} \partial^v c^{ef} c_{ef}
  \nonumber\\[3pt]
  &\;
  + \frac52 \partial^v c^{\lambda\alpha} c_\alpha{}^\beta c_\beta{}^\rho \partial_v c_{\lambda\rho}
  + \frac12 c^{\lambda\alpha} \partial^v c_\alpha{}^\beta c_\beta{}^\rho \partial_v c_{\lambda\rho}
  + \frac12 \partial_\mu c^{\mu v} c^{\alpha\rho} c_{\rho\alpha} \partial_\lambda c^\lambda{}_v
  - 3 \partial_\mu c^{\mu v} c_\alpha{}^\rho c_{\rho v} \partial_\lambda c^{\lambda\alpha}
  \nonumber\\[3pt]
  &\;
  + 4 \partial_\mu c^{v\alpha} c_{\rho v} c^{\mu\lambda} \partial_\lambda c_\alpha{}^\rho
  + 3 \partial_\mu c^{\alpha\rho} c^{\mu v} c^\lambda{}_v \partial_\lambda c_{\rho\alpha}
  - \partial_\mu c_{\rho v} c^{\mu v} c^{\lambda\alpha} \partial_\lambda c_\alpha{}^\rho
  - \frac72 \partial_\mu c_\alpha{}^\rho c^{\mu v} c^{\lambda\alpha} \partial_\lambda c_{\rho v}
  \nonumber\\[3pt]
  &\;
  + \frac12 \partial_\mu c^\lambda{}_v c^{\alpha\rho} c_{\rho\alpha} \partial_\lambda c^{\mu v}
  + \frac12 \partial_\mu c^{\lambda\alpha} c_\alpha{}^\rho c_{\rho v} \partial_\lambda c^{\mu v}
  - 4 \partial_\mu c^{\mu v} c_{\rho v} c^{\lambda\alpha} \partial_\lambda c_\alpha{}^\rho
  + 4 \partial_\mu c^{\mu v} c^{\alpha\rho} c^\lambda{}_v \partial_\lambda c_{\rho\alpha}
  \nonumber\\[3pt]
  &\;
  - 2 \partial_\mu c^{\mu v} c_\alpha{}^\rho c^{\lambda\alpha} \partial_\lambda c_{\rho v}
  + 2 \partial_\mu c^{\mu\lambda} c^{v\alpha} c_{\rho v} \partial_\lambda c_\alpha{}^\rho
  + 2 \partial_\mu c_{\rho\alpha} c^{\alpha\rho} c^\lambda{}_v \partial_\lambda c^{\mu v}
  - \partial_\mu c_\alpha{}^\rho c_{\rho v} c^{\lambda\alpha} \partial_\lambda c^{\mu v}
  \nonumber\\[3pt]
  &\;
  - 6 \partial_\mu c_{\rho v} c_\alpha{}^\rho c^{\lambda\alpha} \partial_\lambda c^{\mu v}
  + 2 d \, \partial_v c_\rho{}^v \, \partial_\mu c \, c^{\mu\rho} c
  - 2 d \, \partial_v c^{\mu\rho} \partial_\mu c \, c_\rho{}^v c
  + d \, \partial_\mu c^{\mu\rho} \partial_v c_\rho{}^v c^2
  \nonumber\\[3pt]
  &\;
  - d \, \partial_\mu c_\rho{}^v \partial_v c^{\mu\rho} c^2
  + e \, \partial_\mu c^{\mu\alpha} c_\alpha{}^\rho c_\rho{}^v \partial_v c
  - e \, c^{\mu\alpha} c_\alpha{}^\rho \partial_\mu c_\rho{}^v \partial_v c
  + e \, \partial^v c_{\rho v} c_\alpha{}^\rho \partial_\lambda c^{\lambda\alpha} c
  \nonumber\\[3pt]
  &\;
  - e \, \partial^v c^{\lambda\alpha} c_\alpha{}^\rho \partial_\lambda c_{\rho v} c
  + e \, \partial^v c_{\rho v} c^{\lambda\alpha} \partial_\lambda c_\alpha{}^\rho c
  - e \, \partial_\lambda c_{\rho v} c^{\lambda\alpha} \partial^v c_\alpha{}^\rho c
  + z \, \partial_\mu c^{\mu v} c^{\alpha\rho} c_{\rho\alpha} \partial_\lambda c^\lambda{}_v
  \nonumber\\[3pt]
  &\;
  - z \, \partial_\mu c^\lambda{}_v c^{\alpha\rho} c_{\rho\alpha} \partial_\lambda c^{\mu v}
  + 2 z \, \partial_\mu c^{\mu v} c^{\alpha\rho} c^\lambda{}_v \partial_\lambda c_{\rho\alpha}
  - 2 z \, \partial_\mu c_{\rho\alpha} c^{\alpha\rho} c^\lambda{}_v \partial_\lambda c^{\mu v}
  - x \, \partial_\mu c^{\mu v} c_\alpha{}^\rho c_{\rho v} \partial_\lambda c^{\lambda\alpha}
  \nonumber\\[3pt]
  &\;
  - y \, \partial_\mu c_{\rho v} c^{\mu v} c^{\lambda\alpha} \partial_\lambda c_\alpha{}^\rho
  + y \, \partial_\mu c_\alpha{}^\rho c^{\mu v} c^{\lambda\alpha} \partial_\lambda c_{\rho v}
  + x \, \partial_\mu c^{\lambda\alpha} c_\alpha{}^\rho c_{\rho v} \partial_\lambda c^{\mu v}
  \nonumber\\[3pt]
  &\;
  + (-x-y) \partial_\mu c^{\mu v} c_{\rho v} c^{\lambda\alpha} \partial_\lambda c_\alpha{}^\rho
  + (-x+y) \partial_\mu c^{\mu v} c_\alpha{}^\rho c^{\lambda\alpha} \partial_\lambda c_{\rho v}
  \nonumber\\[3pt]
  &\;
  + (x+y) \partial_\mu c_\alpha{}^\rho c_{\rho v} c^{\lambda\alpha} \partial_\lambda c^{\mu v}
  + (x-y) \partial_\mu c_{\rho v} c_\alpha{}^\rho c^{\lambda\alpha} \partial_\lambda c^{\mu v} \, .\label{L_vielbein_2}
\end{align}

Here $a,b,d,e,z,x,y$ are arbitrary constants.  In the vielbein action
$S_{\text{vielbein}}$ there is a redundancy generated by total–derivative
terms.  Accordingly we fix the ambiguity, as far as possible, by
distributing derivatives so that each vielbein perturbation field
$c^a{}_\mu$ carries at most a single derivative $\partial$.  Nevertheless,
some ambiguity remains; for example, in $L_{\text{vielbein},1}$ one has
\begin{align}
  \partial_\mu \bigl( c \, \partial_v c^{\alpha v} c_\alpha{}^\mu \bigr)
  - \partial_\mu \bigl( c \, c^{\alpha v} \partial_v c_\alpha{}^\mu \bigr)
  = \partial_\mu c \, \partial_v c^{\alpha v} c_\alpha{}^\mu
   - \partial_\mu c \, c^{\alpha v} \partial_v c_\alpha{}^\mu
   + c \, \partial_v c^{\alpha v} \partial_\mu c_\alpha{}^\mu
   - c \, \partial_\mu c^{\alpha v} \partial_v c_\alpha{}^\mu= 0 \,.
\end{align}
To make this freedom explicit, we introduce the arbitrary coefficients
$a,b,d,e,z,x,y$.

As a cross–check, Ref.~\cite{Goroff:1985th} presents the metric–based expansion to
$\mathcal{O}(\kappa^2)$ about a general background $g^{\mu\nu}$ in terms
of the metric perturbation field $h^{\mu\nu}$.  Specializing to flat
background $g^{\mu\nu} = \eta^{\mu\nu}$ and identifying
\begin{equation}
  h^{\mu v} = 2 c^{\mu v} + \kappa c^\mu{}_b c^{b v} \, ,
\end{equation}
our $S_{\text{vielbein}}$ \eqref{L_vielbein_0} is recovered for a particular choice of these
parameters, e.g.
\begin{equation}
  a=1, \qquad
  b=-2, \qquad
  d=\frac12, \qquad
  e=2, \qquad
  z=\frac12, \qquad
  x=-3, \qquad
  y=1 \, .
\end{equation}

The gravitino action $S_{\text{gravitino}}[c^a{}_\mu,\psi_\mu]$ and the
ghost action $S_{\text{ghost}}[c^a{}_\mu,\psi_\mu,B,C,C^*_\nu,
C^\lambda,C^*_{ab},C'_{ab},u_I^*,u^I]$ are
\begin{align}
\label{gravitino_per}
  S_{\text{gravitino}}\bigl[c^a{}_\mu,\psi_\mu\bigr]
  =&\;
  \int d^4 x \Bigl(
     + \frac14 \bar\psi_\mu \gamma^p \gamma^v \gamma^\mu \partial_v \psi_p
          + \frac14 \kappa \partial^a c^b{}_v \,
           \bar\psi_\mu \gamma^{\mu v p} \gamma_{ab} \psi_p
     \nonumber\\[3pt]
  &\qquad
     + \frac18 \kappa^2 c^{\gamma a} \partial_v c^b{}_\gamma \,
           \bar\psi_\mu \gamma^{\mu v p} \gamma_{ab} \psi_p
     + \frac{i}{2} \kappa c^s{}_\sigma \epsilon^{\mu v p \sigma}
           \bar\psi_\mu \gamma_* \gamma_s \partial_v \psi_p
     \nonumber\\[3pt]
  &\qquad
     - \frac14 \kappa^2 c^{\gamma a} \partial_\gamma c^b{}_v \,
           \bar\psi_\mu \gamma^{\mu v p} \gamma_{ab} \psi_p
     - \frac18 \kappa^2 c^{\delta b} \partial^a c_{v\delta} \,
           \bar\psi_\mu \gamma^{\mu v p} \gamma_{ab} \psi_p
     \nonumber\\[3pt]
  &\qquad
     + \frac18 \kappa^2 c_{z v} \partial^a c^{z b} \,
           \bar\psi_\mu \gamma^{\mu v p} \gamma_{ab} \psi_p
     - \frac{i}{4} \kappa^2 \partial_\gamma c^b{}_v c^s{}_\sigma
           \epsilon^{\mu v p \sigma}
           \bar\psi_\mu \gamma_* \gamma_s \gamma^\gamma{}_b \psi_p
     \nonumber\\[3pt]
  &\qquad
     - \frac{\kappa^2}{32} \Bigl[
       \bigl(\bar\psi^\rho \gamma^\mu \psi^v\bigr)
       \bigl(\bar\psi_\rho \gamma_\mu \psi_v
             + 2 \bar\psi_\rho \gamma_v \psi_\mu\bigr)
       - 4 \bigl(\bar\psi_\mu \gamma\cdot\psi\bigr)
             \bigl(\bar\psi^\mu \gamma\cdot\psi\bigr)
       \Bigr]
     + \mathcal{O}(\kappa^3)
  \Bigr) ,
\end{align}
and
\begin{align}
\label{ghost_per}
  S_{\text{ghost}}&\bigl[c^a{}_\mu,\psi_\mu,B,C,
   C^*_\nu,C^\lambda,C^*_{ab},C'_{ab},u_I^*,u^I\bigr]
  =\;
  \int d^4 x \Bigl(
      B \gamma^v \partial_v C
    - \frac12 \kappa \partial^a c^b{}_v B \gamma^v \gamma_{ab} C
    \nonumber\\[3pt]
  &\qquad
    - \frac14 \kappa^2 c^{\gamma a} \partial_v c^b{}_\gamma
         B \gamma^v \gamma_{ab} C
    + \frac12 \kappa^2 c^{\gamma a} \partial_\gamma c^b{}_v
         B \gamma^v \gamma_{ab} C
    + \frac14 \kappa^2 c^{\gamma b} \partial^a c_{v\gamma}
         B \gamma^v \gamma_{ab} C
    \nonumber\\[3pt]
  &\qquad
    + \frac14 \kappa^2 c^\gamma{}_v \partial^a c^b{}_\gamma
         B \gamma^v \gamma_{ab} C
    + \frac{\kappa^2}{16}
        \bigl(\bar\psi_a \gamma_v \psi_b\bigr)
        B \gamma^v \gamma^{ab} C
    + \frac{\kappa^2}{8}
        \bigl(\bar\psi_v \gamma_a \psi_b\bigr)
        B \gamma^v \gamma^{ab} C
  \Bigr)
  \nonumber\\[3pt]
  &\;
  + \int d^4 x \Bigl(
    - \kappa B \gamma^\mu C^\lambda \partial_\lambda \psi_\mu
    - \kappa B \gamma^\mu \partial_\mu C^\lambda \psi_\lambda
    - \frac{\kappa}{4} B \gamma^\mu C'_{ef} \gamma^{ef} \psi_\mu
    + \frac{\kappa}{4} B \gamma^\mu \partial_f C_e \gamma^{ef} \psi_\mu
    \nonumber\\[3pt]
  &\qquad
    + C^*_v \partial_\mu \partial^\mu C^v
    - \frac{\kappa}{2} C^*_v
         \bigl(\partial_\mu \bar\psi^v \gamma^\mu C\bigr)
    - \frac{\kappa}{2} C^*_v
         \bigl(\bar\psi^v \gamma^\mu \partial_\mu C\bigr)
    - \frac{\kappa}{2} C^*_v
         \bigl(\partial_\mu \bar\psi^\mu \gamma^v C\bigr)
    - \frac{\kappa}{2} C^*_v
         \bigl(\bar\psi^\mu \gamma^v \partial_\mu C\bigr)
    \nonumber\\[3pt]
  &\qquad
    + \frac{\kappa}{2} C^{*\mu}
         \bigl(\partial_\mu \bar\psi_\rho \gamma^\rho C\bigr)
    + \frac{\kappa}{2} C^{*\mu}
         \bigl(\bar\psi_\rho \gamma^\rho \partial_\mu C\bigr)
    + 2 \kappa C^*_v (\partial_\mu C^\lambda \partial_\lambda c^{\mu v})
    + 2 \kappa C^*_v (C^\lambda \partial_\mu \partial_\lambda c^{\mu v})
    \nonumber\\[3pt]
  &\qquad
    - \kappa C^*_v (\partial^v C^\lambda \partial_\lambda c)
    - \kappa C^*_v (C^\lambda \partial^v \partial_\lambda c)
    - \kappa C^*_v (\partial^v \partial_\rho C^\lambda c^\rho{}_\lambda)
    - \kappa C^*_v (\partial_\rho C^\lambda \partial^v c^\rho{}_\lambda)
    \nonumber\\[3pt]
  &\qquad
    + \kappa C^*_v (\partial_\mu C'^\mu{}_e c^{ev})
    + \kappa C^*_v (C'^\mu{}_e \partial_\mu c^{ev})
    - \frac{\kappa}{2} C^*_v (\partial_\mu \partial_e C^\mu c^{ev})
    - \frac{\kappa}{2} C^*_v (\partial_e C^\mu \partial_\mu c^{ev})
    \nonumber\\[3pt]
  &\qquad
    + \frac{3}{2} \kappa C^*_v (\partial_\mu \partial^\mu C_e c^{ev})
    + \frac{3}{2} \kappa C^*_v (\partial^\mu C_e \partial_\mu c^{ev})
    + \kappa C^*_v (\partial_\mu C'^v{}_e c^{e\mu})
    + \kappa C^*_v (C'^v{}_e \partial_\mu c^{e\mu})
    \nonumber\\[3pt]
  &\qquad
    - \frac{\kappa}{2} C^*_v (\partial_\mu \partial_e C^v c^{e\mu})
    - \frac{\kappa}{2} C^*_v (\partial_e C^v \partial_\mu c^{e\mu})
    + \frac{3}{2} \kappa C^*_v (\partial_\mu \partial^v C_e c^{e\mu})
    + \frac{3}{2} \kappa C^*_v (\partial^v C_e \partial_\mu c^{e\mu})
    \nonumber\\[3pt]
  &\qquad
    + C^*_{ab} C'^{ab} - C^*_{ab} C'^{ba}
    - \frac{\kappa}{2} C^*_{ab} (\bar\psi^b \gamma^a C)
    + \frac{\kappa}{2} C^*_{ab} (\bar\psi^a \gamma^b C)
    \nonumber\\[3pt]
  &\qquad
    + \frac{\kappa}{2} C^*_{ab} \partial^b C^\lambda c^a{}_\lambda
    - \frac{\kappa}{2} C^*_{ab} \partial^a C^\lambda c^b{}_\lambda
    - \frac{\kappa}{2} C^*_{ab} \partial_e C^a c^{eb}
    + \frac{\kappa}{2} C^*_{ab} \partial_e C^b c^{ea}
    \nonumber\\[3pt]
  &\qquad
    + \kappa C^*_{ab} C'^a{}_e c^{eb}
    - \kappa C^*_{ab} C'^b{}_e c^{ea}
    + \frac{5}{32} \kappa^2 B \gamma_d \bar B \,(\bar C \gamma^d C)
    + u_I^*u^I + \kappa c u_I^*u^I + \frac{1}{2} \kappa^2 c^2 u_I^*u^I
    \nonumber\\[3pt]
  &\qquad    
    - \frac{1}{2} \kappa^2 {c^a}_b {c^b}_a  u_I^*u^I  + \mathcal{O}(\kappa^3)
  \Bigr) .
\end{align}

\section{The bosonic effective actions in orders of $\hbar$ and $\kappa$}
\label{Bosonic effective action}
For a general operator \(X[{c^a}_\mu,\psi,\bar C,C]\), its vacuum expectation
value is
\begin{equation}
  \bigl\langle X[{c^a}_\mu,\psi,\bar C,C] \bigr\rangle
  =
  \int \mathcal{D}{c^a}_\mu\,\mathcal{D}\psi\,\mathcal{D}\bar C\,\mathcal{D}C\,
  X[{c^a}_\mu,\psi,\bar C,C]\,
  e^{\frac{i}{\hbar} S[{c^a}_\mu,\psi,\bar C,C]} \; .
\end{equation}
If the operator \(X[{c^a}_\mu]\) depends only on the vielbein perturbation field
\(c^a{}_\mu\), we may first integrate out the remaining fields at fixed
\(c^a{}_\mu\):
\begin{align}
  &\int \mathcal{D}{c^a}_\mu\,\mathcal{D}\psi\,\mathcal{D}\bar C\,\mathcal{D}C\,
     X[{c^a}_\mu]\,
     e^{\frac{i}{\hbar} S[{c^a}_\mu,\psi,\bar C,C]}
  \nonumber\\[3pt]
  &\qquad
   = \int \mathcal{D}{c^a}_\mu\,
     \Bigl(
       X[{c^a}_\mu]\, e^{\frac{i}{\hbar} S_{\text{vielbein}}[{c^a}_\mu]}
       \int \mathcal{D}\psi\,\mathcal{D}\bar C\,\mathcal{D}C\,
       e^{\frac{i}{\hbar} S_{\text{r}}[{c^a}_\mu,\psi,\bar C,C]}
     \Bigr) ,
\end{align}
where we define the remaining action denoted by $S_r$ of the gravitino and ghost fields:  
\begin{equation}
  S_{\text{r}}[{c^a}_\mu,\psi,\bar C,C]
  = S_{\text{gravitino}}[{c^a}_\mu,\psi_\mu]
  + S_{\text{ghost}}[{c^a}_\mu,\psi_\mu,B,C,C^*_\nu,C^\lambda,C^*_{ab},C'_{ab},u_I^*,u^I] \; .
\end{equation}
We then obtain the bosonic effective action \(S_B[{c^a}_\mu]\) in the way
\begin{equation}
  e^{\frac{i}{\hbar} S_B[{c^a}_\mu]}
  =
  e^{\frac{i}{\hbar} S_{\text{vielbein}}[{c^a}_\mu]}
  \int \mathcal{D}\psi\,\mathcal{D}\bar C\,\mathcal{D}C\,
  e^{\frac{i}{\hbar} S_{\text{r}}[{c^a}_\mu,\psi,\bar C,C]} ,\label{3begin_eq}
\end{equation}
or equivalently
\begin{equation}
\label{S_B}
  i S_B[{c^a}_\mu]
  =
  i S_{\text{vielbein}}[{c^a}_\mu]
  +
  \hbar \ln\!\Biggl(
    \int \mathcal{D}\psi\,\mathcal{D}\bar C\,\mathcal{D}C\,
    e^{\frac{i}{\hbar} S_{\text{r}}[{c^a}_\mu,\psi,\bar C,C]}
  \Biggr) .
\end{equation}
The bosonic effective action \(S_B[{c^a}_\mu]\) admits an expansion in powers of \(\hbar\), ordered by loop number, as shown in Eq.~\eqref{bosonic_effective_action_hbar_expansion}. 

The tree-level action \(S^{(0)}\) is identical to \(S_{\rm vielbein}\) and decomposed into the following actions
\begin{align}
S^{(0,0)} &= -\frac{1}{2} \int d^4x \bigl(\partial^d c_{ab}(x)\,\partial_d\, c^{ab}(x)
    -\frac{1}{2}  \partial^d c(x)\,\partial_d\, c(x)\bigr), \\
    S^{(0,1)} &= -\frac{1}{2} \int d^4x ~\mathcal{L}_{\text{vielbein},1}, \\
    S^{(0,2)} &= -\frac{1}{2} \int d^4x ~\mathcal{L}_{\text{vielbein},2}.
\end{align}
where $\mathcal{L}_{\text{vielbein},1}$ and $\mathcal{L}_{\text{vielbein},2}$ are given by the explicit Lagrangians in Eqs.~\eqref{L_vielbein_1} and ~\eqref{L_vielbein_2} in Appendix ~\ref{Appendix A}.

The loop-level actions, \(\hbar S^{(1)} + \hbar^2 S^{(2)} + \cdots\), are obtained from the second term of Eq.~\eqref{S_B}. In particular, up to the order of \(\kappa^2\), we find
\begin{align}
  & i \hbar \kappa S^{(1,1)} + i \hbar \kappa^2 S^{(1,2)} + i \hbar^2 \kappa^2 S^{(2,2)} + \cdots
  \nonumber\\[3pt]
  &= \hbar \ln\!\Biggl(
      \int \mathcal{D}\psi\,\mathcal{D}\bar C\,\mathcal{D}C\,
      e^{\frac{i}{\hbar} S_{\text{r}}[{c^a}_\mu,\psi,\bar C,C]}
     \Biggr)
  \nonumber\\[3pt]
  &= \hbar \ln\!\Biggl(
      \int \mathcal{D}\psi\,\mathcal{D}\bar C\,\mathcal{D}C\,
      \exp\Bigl(
        \frac{i}{\hbar} S_{\text{r}}^{[0]}[\psi,\bar C,C]
        + \frac{i}{\hbar}\kappa S_{\text{r}}^{[1]}[{c^a}_\mu,\psi,\bar C,C]
        + \frac{i}{\hbar}\kappa^2 S_{\text{r}}^{[2]}[{c^a}_\mu,\psi,\bar C,C]
        + \cdots
      \Bigr)
     \Biggr)
  \nonumber\\[3pt]
  &= \hbar \ln\!\Biggl(
      \int \mathcal{D}\psi\,\mathcal{D}\bar C\,\mathcal{D}C\,
      \exp\Bigl(
        \frac{i}{\hbar} S_{\text{r}}^{[0]}[\psi,\bar C,C]\Bigr)
\times \Bigr(1+ \frac{i}{\hbar}\kappa S_{\text{r}}^{[1]}[{c^a}_\mu,\psi,\bar C,C]
        + \frac{i}{\hbar}\kappa^2 S_{\text{r}}^{[2]}[{c^a}_\mu,\psi,\bar C,C]
  \nonumber\\[-1pt]
  &\hspace{3.5em}
        -\frac{1}{2\hbar^2}\kappa^2 \left(S_{\text{r}}^{[1]}[{c^a}_\mu,\psi,\bar C,C]\right)^2
        + \cdots
      \Bigr)
     \Biggr)
  \nonumber\\[3pt]
  &= \hbar \ln\!\Biggl(1+ \frac{i}{\hbar}\kappa \bigl \langle S_{\text{r}}^{[1]}[{c^a}_\mu,\psi,\bar C,C] \bigl \rangle_{r,0}
        + \frac{i}{\hbar}\kappa^2  \bigl \langle S_{\text{r}}^{[2]}[{c^a}_\mu,\psi,\bar C,C] \bigl \rangle_{r,0}
        -\frac{1}{2\hbar^2}\kappa^2 \Bigr \langle \left(S_{\text{r}}^{[1]}[{c^a}_\mu,\psi,\bar C,C]\right)^2 \Bigr \rangle_{r,0}
        + \cdots
     \Biggr)
  \nonumber\\[3pt]
  &= i \kappa \bigl \langle S_{\text{r}}^{[1]}[{c^a}_\mu,\psi,\bar C,C] \bigl \rangle_{r,0}
        + i \kappa^2  \bigl \langle S_{\text{r}}^{[2]}[{c^a}_\mu,\psi,\bar C,C] \bigl \rangle_{r,0} -\frac{1}{2\hbar}\kappa^2 \Bigr \langle \left(S_{\text{r}}^{[1]}[{c^a}_\mu,\psi,\bar C,C]\right)^2 \Bigr \rangle_{r,0}
  \nonumber\\[-1pt]
        &\quad
        + \frac{1}{2\hbar}\kappa^2 {\left\langle S_{\text{r}}^{[1]}[{c^a}_\mu,\psi,\bar C,C]\right\rangle_{r,0}}^2 +\cdots
  \nonumber\\[3pt]
  &= i \kappa \bigl \langle S_{\text{r}}^{[1]}[{c^a}_\mu,\psi,\bar C,C] \bigl \rangle_{r,0}
        + i \kappa^2  \bigl \langle S_{\text{r}}^{[2]}[{c^a}_\mu,\psi,\bar C,C] \bigl \rangle_{r,0} -\frac{1}{2\hbar}\kappa^2 \Bigr \langle \left(S_{\text{r}}^{[1]}[{c^a}_\mu,\psi,\bar C,C]\right)^2 \Bigr \rangle^{\textrm{connected}}_{r,0}+\cdots
\end{align}
where $S_r$ is expanded in powers of $\kappa$ as
\begin{align}
    S_{\mathrm r}=\sum_{n\ge0}\kappa^n S_{\mathrm r}^{[n]},
\end{align}
$\langle \cdots \rangle_{r,0}$ is expectation value in theory defined by $S_{\text{r}}^{[0]}$, namely
\begin{align}
    \langle \mathcal{O}[{c^a}_\mu,\psi,\bar C,C] \rangle_{r,0}=
  \int \mathcal{D}\psi\,\mathcal{D}\bar C\,\mathcal{D}C\,
  \mathcal{O}[{c^a}_\mu,\psi,\bar C,C]\,
  e^{\frac{i}{\hbar} S_{\text{r}}^{[0]}[\psi,\bar C,C]},
\end{align}
and the superscript $connected$ denotes the connected part, defined by
\begin{align}
\Bigl\langle \left(S_{\mathrm r}^{[1]}\right)^2 \Bigr\rangle_{r,0}^{\textrm{connected}}
\equiv
\Bigl\langle \left(S_{\mathrm r}^{[1]}\right)^2 \Bigr\rangle_{r,0}
-
\Bigl\langle S_{\mathrm r}^{[1]} \Bigr\rangle_{r,0}^{\,2}.
\end{align}
We now present the explicit loop contributions to the bosonic effective action up to order $\kappa^2$. 
To make the procedure transparent, we first display the derivation of the leading one-loop term $S^{(1,1)}$. 
Here the bosonic effective action is expanded into correlation functions with respect to $S_{\mathrm r}^{[0]}$, after which the resulting Wick contractions are evaluated in momentum space:
\begin{align}
  S^{(1,1)}&=\frac{1}{\hbar}\bigl \langle S_{\text{r}}^{[1]}[{c^a}_\mu,\psi,\bar C,C] \bigl \rangle_{r,0}
     \nonumber\\[3pt]
  &= \int d^4 x \Bigl(
     + \frac14 \bigl\langle
       \partial^a c^b{}_v(x)\,
       \bar\psi_\mu(x)\,\gamma^{\mu v p}\gamma_{ab}\psi_p(x)
     \bigr\rangle_{r,0}
+ \frac{i}{2} \bigl\langle
       c^s{}_\sigma(x)\,\epsilon^{\mu v p \sigma}\,
       \bar\psi_\mu(x)\,\gamma_* \gamma_s \partial_v \psi_p(x)
     \bigr\rangle_{r,0}
     \nonumber\\[3pt]
  &- \frac12 \bigl\langle
       \partial^a c^b{}_v(x)\,B(x)\,\gamma^v \gamma_{ab} C(x)
     \bigr\rangle_{r,0}
     + 2 \bigl\langle
       C^*_v(x)\,\partial_\mu C^\lambda(x)\,\partial_\lambda c^{\mu v}(x)
     \bigr\rangle_{r,0}
     \nonumber\\[3pt]
  &+ 2 \bigl\langle
       C^*_v(x)\,C^\lambda(x)\,\partial_\mu \partial_\lambda c^{\mu v}(x)
     \bigr\rangle_{r,0}
     - \bigl\langle
       C^*_v(x)\,\partial^v C^\lambda(x)\,\partial_\lambda c(x)
     \bigr\rangle_{r,0}
     \nonumber\\[3pt]
  &- \bigl\langle
       C^*_v(x)\,C^\lambda(x)\,\partial^v \partial_\lambda c(x)
     \bigr\rangle_{r,0}
     - \bigl\langle
       C^*_v(x)\,\partial^v \partial_\rho C^\lambda(x)\,c^\rho{}_\lambda(x)
     \bigr\rangle_{r,0}
     \nonumber\\[3pt]
  &- \bigl\langle
       C^*_v(x)\,\partial_\rho C^\lambda(x)\,\partial^v c^\rho{}_\lambda(x)
     \bigr\rangle_{r,0}
     - \frac12 \bigl\langle
       C^*_v(x)\,\partial_\mu \partial_e C^\mu(x)\,c^{ev}(x)
     \bigr\rangle_{r,0}
     \nonumber\\[3pt]
  &- \frac12 \bigl\langle
       C^*_v(x)\,\partial_e C^\mu(x)\,\partial_\mu c^{ev}(x)
     \bigr\rangle_{r,0}
     + \frac32 \bigl\langle
       C^*_v(x)\,\partial_\mu \partial^\mu C_e(x)\,c^{ev}(x)
     \bigr\rangle_{r,0}
     \nonumber\\[3pt]
  &+ \frac32 \bigl\langle
       C^*_v(x)\,\partial^\mu C_e(x)\,\partial_\mu c^{ev}(x)
     \bigr\rangle_{r,0}
     - \frac12 \bigl\langle
       C^*_v(x)\,\partial_\mu \partial_e C^v(x)\,c^{e\mu}(x)
     \bigr\rangle_{r,0}
     \nonumber\\[3pt]
  &- \frac12 \bigl\langle
       C^*_v(x)\,\partial_e C^v(x)\,\partial_\mu c^{e\mu}(x)
     \bigr\rangle_{r,0}
     + \frac32 \bigl\langle
       C^*_v(x)\,\partial_\mu \partial^v C_e(x)\,c^{e\mu}(x)
     \bigr\rangle_{r,0}
     \nonumber\\[3pt]
  &+ \frac32 \bigl\langle
       C^*_v(x)\,\partial^v C_e(x)\,\partial_\mu c^{e\mu}(x)
     \bigr\rangle_{r,0}
     + \bigl\langle
       C^*_{ab}(x)\,C'^a{}_e(x)\,c^{eb}(x)
     \bigr\rangle_{r,0}
     \nonumber\\[3pt]
  &- \bigl\langle
       C^*_{ab}(x)\,C'^b{}_e(x)\,c^{ea}(x)
     \bigr\rangle_{r,0}
     + \bigl\langle c u_I^*u^I \bigr\rangle_{r,0}
  \Bigr)
     \nonumber\\[3pt]
  &= \int d^4 x \Bigl(
     + \frac18 \partial^a c^b{}_v(x) tr \bigl(
       \,\gamma^{\mu v p}\gamma_{ab}\gamma_\mu\gamma_\lambda \gamma_p
     \bigr)\int \frac{d^4 k}{(2\pi)^4}\frac{k^\lambda}{k^2}
+ \frac{1}{4} c^s{}_\sigma(x) 
       \,\epsilon^{\mu v p \sigma}\,tr \bigl(
       \,\gamma_* \gamma_s \gamma_\mu \gamma_\alpha \gamma_p
     \bigr)\int \frac{d^4 k}{(2\pi)^4}\frac{k_vk^\alpha}{k^2}
     \nonumber\\[3pt]
  &+ \frac12 \partial^a c^b{}_v(x) tr\bigl(\gamma^v \gamma_{ab} \gamma_\lambda \bigr)
       \int \frac{d^4 k}{(2\pi)^4}\frac{k^\lambda}{k^2}
     + 2\partial_\lambda c^{\mu v}(x) \delta^\lambda_v\int \frac{d^4 k}{(2\pi)^4}\frac{k_\mu}{k^2}
     + 2i \partial_\mu \partial_\lambda c^{\mu v}(x)
      \delta^\lambda_v\int \frac{d^4 k}{(2\pi)^4}\frac{1}{k^2}
     \nonumber\\[3pt]
  &- \partial_\lambda c(x) \delta^\lambda_v\int \frac{d^4 k}{(2\pi)^4}\frac{k^v}{k^2}
     - i \partial^v \partial_\lambda c(x) \delta^\lambda_v\int \frac{d^4 k}{(2\pi)^4}\frac{1}{k^2}
     +i c^\rho{}_\lambda(x) \delta^\lambda_v\int \frac{d^4 k}{(2\pi)^4}\frac{k^v k_\rho}{k^2}
     - \partial^v c^\rho{}_\lambda(x) \delta^\lambda_v\int \frac{d^4 k}{(2\pi)^4}\frac{k_\rho}{k^2}     
     \nonumber\\[3pt]
  &+ \frac12 i c^{ev}(x) \delta^\mu_v\int \frac{d^4 k}{(2\pi)^4}\frac{k^\mu k_e}{k^2}
     - \frac12 \partial_\mu c^{ev}(x) \delta^\mu_v\int \frac{d^4 k}{(2\pi)^4}\frac{k_e}{k^2}
     - \frac32 i c^{ev}(x) \eta_{ev}\delta(0)
     + \frac32 \partial_\mu c^{ev}(x) \eta_{ev}\int \frac{d^4 k}{(2\pi)^4}\frac{k^\mu}{k^2} 
     \nonumber\\[3pt]
  &+ \frac12 i c^{e\mu}(x) \delta^v_v\int \frac{d^4 k}{(2\pi)^4}\frac{k_\mu k_e}{k^2}
     - \frac12 \partial_\mu c^{e\mu}(x) \delta^v_v\int \frac{d^4 k}{(2\pi)^4}\frac{k_e}{k^2}
     - \frac32 i c^{e\mu}(x) \eta_{ev}\int \frac{d^4 k}{(2\pi)^4}\frac{k_\mu k^v}{k^2}  
     \nonumber\\[3pt]
  &+ \frac32 \partial_\mu c^{e\mu}(x) \eta_{ev}\int \frac{d^4 k}{(2\pi)^4}\frac{k^v}{k^2}
     -\frac14 i c^{eb}(x) \bigl(\delta^a_a\eta_{eb}-\delta^a_b\delta_{ea}\bigr) \delta(0)
     +\frac14 i c^{ea}(x) \bigl(\delta^b_a\delta_{eb}-\delta^b_b\delta_{ea}\bigr) \delta(0)
     +ic(x)\delta^I_I\delta(0)
  \Bigr)
     \nonumber\\[3pt]
  &= \int d^4 x \Bigl(
     - \partial^a c^b{}_v(x) \bigl(\delta^v_a\eta_{b\lambda}-\delta^v_b\eta_{a\lambda} \bigr)\int \frac{d^4 k}{(2\pi)^4}\frac{k^\lambda}{k^2}
+2 i c^s{}_\sigma(x) 
       \,\bigl(
       \delta_s^{\nu}\delta_\alpha^{\sigma}-\delta_\alpha^{\nu}\delta_s^{\sigma}
     \bigr)\int \frac{d^4 k}{(2\pi)^4}\frac{k_v k^\alpha}{k^2}
     \nonumber\\[3pt]
  &+ 2 \partial^a c^b{}_v(x) \bigl(\delta^v_a\eta_{b\lambda}-\delta^v_b\eta_{a\lambda} \bigr)
       \int \frac{d^4 k}{(2\pi)^4}\frac{k^\lambda}{k^2}
     + 2\partial_v c^{\mu v}(x) \int \frac{d^4 k}{(2\pi)^4}\frac{k_\mu}{k^2}
     + 2i \partial_\mu \partial_v c^{\mu v}(x)
      \int \frac{d^4 k}{(2\pi)^4}\frac{1}{k^2}
     \nonumber\\[3pt]
  &- \partial_v c(x) \int \frac{d^4 k}{(2\pi)^4}\frac{k^v}{k^2}
     - i \partial^v \partial_v c(x) \int \frac{d^4 k}{(2\pi)^4}\frac{1}{k^2}
     +i c^\rho{}_v(x) \int \frac{d^4 k}{(2\pi)^4}\frac{k^v k_\rho}{k^2}
     - \partial^v c^\rho{}_v(x) \int \frac{d^4 k}{(2\pi)^4}\frac{k_\rho}{k^2}    
     \nonumber\\[3pt]
  &+ \frac12 i c^{ev}(x) \int \frac{d^4 k}{(2\pi)^4}\frac{k_v k_e}{k^2}
     - \frac12 \partial_v c^{ev} \int \frac{d^4 k}{(2\pi)^4}\frac{k_e}{k^2}
     - \frac32 i c(x) \delta(0)
     + \frac32 \partial_\mu c \int \frac{d^4 k}{(2\pi)^4}\frac{k^\mu}{k^2} 
     \nonumber\\[3pt]
  &+ 2 i c^{e\mu}(x) \int \frac{d^4 k}{(2\pi)^4}\frac{k_\mu k_e}{k^2}
     - 2 \partial_\mu c^{e\mu}(x) \int \frac{d^4 k}{(2\pi)^4}\frac{k_e}{k^2}
     - \frac32 i {c_v}^\mu(x) \int \frac{d^4 k}{(2\pi)^4}\frac{k_\mu k^v}{k^2}  
     \nonumber\\[3pt]
  &+ \frac32 \partial_\mu {c_v}^\mu(x) \int \frac{d^4 k}{(2\pi)^4}\frac{k^v}{k^2}
     -\frac34 i c(x) \delta(0)
     -\frac34 i c(x) \delta(0)
     +4 i c(x) \delta(0)
  \Bigr)
     \nonumber\\[3pt]
  &= \int d^4 x \Bigl(
     - \partial^a c^b{}_a(x) \int \frac{d^4 k}{(2\pi)^4}\frac{k_b}{k^2}
     + \partial^a c(x) \int \frac{d^4 k}{(2\pi)^4}\frac{k_a}{k^2}
+2 i c^v{}_\sigma(x) \int \frac{d^4 k}{(2\pi)^4}\frac{k_vk^\sigma}{k^2}
     + 2 \partial^v c^b{}_v(x) 
       \int \frac{d^4 k}{(2\pi)^4}\frac{k_b}{k^2}
     \nonumber\\[3pt]
  &- 2 \partial^a c(x) 
       \int \frac{d^4 k}{(2\pi)^4}\frac{k_a}{k^2}+ 2\partial_v c^{\mu v}(x) \int \frac{d^4 k}{(2\pi)^4}\frac{k_\mu}{k^2}
     + 2i \partial_\mu \partial_v c^{\mu v}(x)
      \int \frac{d^4 k}{(2\pi)^4}\frac{1}{k^2}
     - \partial_v c(x) \int \frac{d^4 k}{(2\pi)^4}\frac{k^v}{k^2}  
     \nonumber\\[3pt]
  &- i \partial^v \partial_v c(x) \int \frac{d^4 k}{(2\pi)^4}\frac{1}{k^2}+i c^\rho{}_v(x) \int \frac{d^4 k}{(2\pi)^4}\frac{k^v k_\rho}{k^2}
     - \partial^v c^\rho{}_v(x) \int \frac{d^4 k}{(2\pi)^4}\frac{k_\rho}{k^2}
     + \frac12 i c^{ev}(x) \int \frac{d^4 k}{(2\pi)^4}\frac{k_v k_e}{k^2}
     \nonumber\\[3pt]
  & - \frac12 \partial_v c^{ev} \int \frac{d^4 k}{(2\pi)^4}\frac{k_e}{k^2} 
     + \frac32 \partial_\mu c \int \frac{d^4 k}{(2\pi)^4}\frac{k^\mu}{k^2}
     + 2 i c^{e\mu}(x) \int \frac{d^4 k}{(2\pi)^4}\frac{k_\mu k_e}{k^2}
     - 2 \partial_\mu c^{e\mu}(x) \int \frac{d^4 k}{(2\pi)^4}\frac{k_e}{k^2}  
     \nonumber\\[3pt]
  &- \frac32 i {c_v}^\mu(x) \int \frac{d^4 k}{(2\pi)^4}\frac{k_\mu k^v}{k^2}
  + \frac32 \partial_\mu {c_v}^\mu(x) \int \frac{d^4 k}{(2\pi)^4}\frac{k^v}{k^2}
     -2 i c(x) \delta(0)
     - \frac32 i c(x) \delta(0)
     -\frac34 i c(x) \delta(0)
     -\frac34 i c(x) \delta(0)
     \nonumber\\[3pt]
    &+4 i c(x) \delta(0)
  \Bigr)
     \nonumber\\[3pt]
&= \int d^4 x\,
     \Bigl(+ \partial^a c^b{}_a (x) \int \frac{d^4 k}{(2\pi)^4}\frac{k_b}{k^2}
  - {\frac{1}{2}} \partial^a c (x) \int \frac{d^4 k}{(2\pi)^4}\frac{k_a}{k^2}
  + 2i \partial_\mu \partial_\nu c^{\mu\nu} (x) \int \frac{d^4 k}{(2\pi)^4}\frac{1}{k^2}
     \nonumber\\[3pt]
  &- i \partial^v \partial_v c(x) \int \frac{d^4 k}{(2\pi)^4}\frac{1}{k^2}
  + 4i c^{ev} (x) \int \frac{d^4 k}{(2\pi)^4}\frac{k_v k_e}{k^2}
  - i c (x) \delta(0) \Bigr)
     \nonumber\\[3pt]
  &= \int d^4 x\,
     \Bigl(
       +4 i\, c^b{}_a(x)\,\partial^b \partial_a G(0)
       - i\, c(x)\,\delta(0)
     \Bigr),\label{EA_S_one_one}
\end{align}
where the omitted terms are total derivatives in \(x\) and therefore vanish upon spacetime integration.

The next one-loop term, $S^{(1,2)}$, is obtained by the same steps: one expands the corresponding correlators, performs the contractions in the free theory defined by $S_{\mathrm r}^{[0]}$, and then reduces the resulting expressions using momentum-space integrals. The final result is

\begin{align}
&S^{(1,2)}
  =    i \bigl \langle S_{\text{r}}^{[2]}[{c^a}_\mu,\psi,\bar C,C] \bigl \rangle_{r,0} \Bigr|_{\hbar} -\frac{1}{2\hbar} \Bigr \langle \left(S_{\text{r}}^{[1]}[{c^a}_\mu,\psi,\bar C,C]\right)^2 \Bigr \rangle^{\textrm{connected}}_{r,0} \Bigr|_{\hbar^2}
  \nonumber\\[0.8em]
  &= \int d^4 x\,
     \Bigl(
       \frac{1}{8}i\,c(x)c(x)\,\delta(0)
       -\frac{7}{8}i\,c_{o}{}^{\nu}(x)\,c_{\nu}{}^{o}(x)\,\delta(0)
       +3i\,c^{e\nu}(x)\,\partial_{\pi}\partial_{\nu}c^{\pi}{}_{e}(x)\,G(0)
  \nonumber\\[-1pt]
  &-\frac{3}{2}i\,c^{e\nu}(x)\,\partial_{e}\partial_{\nu}c(x)\,G(0)
       -2i\,c^{e\mu}(x)\,c^{f}{}_{e}(x)\,\partial_{\mu}\partial_{f}G(0)
       -\frac{1}{4}i\,c(x)\,c^{f\pi}(x)\,\partial_{\pi}\partial_{f}G(0)
     \Bigr)
  \nonumber\\[0.8em]
  &+ \int d^4 x\,d^4 y\,
     \Bigl(
       2i\,\partial_{\mu}\partial_{\lambda}c^{\mu\nu}(x)\,G(x-y)G(x-y)\,\partial_{\pi}\partial_{\nu}c^{\pi\lambda}(y)       -2i\,\partial_{\mu}\partial_{\lambda}c^{\mu\nu}(x)\,G(x-y)G(x-y)\,\partial^{\lambda}\partial_{\nu}c(y)
  \nonumber\\[-1pt]
  &+\frac{1}{2}i\,\partial^{\nu}\partial_{o}c(x)\,G(x-y)G(x-y)\,\partial^{o}\partial_{\nu}c(y)
     \Bigr)
  \nonumber\\[0.8em]
  &+ \int d^4 x\,d^4 y\,
     \Bigl(
       +i\,\partial_{o}c(x)\,\partial^{\nu}G(x-y)\,G(x-y)\,\partial^{o}\partial_{\nu}c(y)
       -2i\,\partial_{\lambda}c(x)\,\partial^{\nu}G(x-y)\,G(x-y)\,\partial_{\pi}\partial_{\nu}c^{\pi\lambda}(y)
  \nonumber\\[-1pt]
  &+\frac{1}{2}i\,\partial_{\mu}c^{e\mu}(x)\,\partial_{e}G(x-y)\,G(x-y)\,\partial^{\nu}\partial_{\nu}c(y)
       -\frac{1}{2}i\,\partial_{\lambda}c^{\mu\nu}(x)\,\partial_{\mu}G(x-y)\,G(x-y)\,\partial^{\lambda}\partial_{\nu}c(y)
  \nonumber\\[-1pt]
  &-\frac{3}{2}i\,\partial_{\mu}c^{e\nu}(x)\,\partial^{\mu}G(x-y)\,G(x-y)\,\partial_{e}\partial_{\nu}c(y)
       -\frac{3}{2}i\,\partial_{\mu}c^{e\mu}(x)\,\partial^{\nu}G(x-y)\,G(x-y)\,\partial_{e}\partial_{\nu}c(y)
  \nonumber\\[-1pt]
  &-3i\,\partial_{\mu}\partial_{\lambda}c^{\mu\nu}(x)\,\partial_{\pi}G(x-y)\,G(x-y)\,\partial_{\nu}c^{\pi\lambda}(y)
       +2i\,\partial_{\mu}\partial_{\lambda}c^{\mu\nu}(x)\,\partial_{\sigma}G(x-y)\,G(x-y)\,\partial^{\lambda}c^{\sigma}{}_{\nu}(y)
  \nonumber\\[-1pt]
  &-3i\,\partial_{\mu}\partial_{\lambda}c^{\mu\nu}(x)\,\partial^{\pi}G(x-y)\,G(x-y)\,\partial_{\pi}c_{\nu}{}^{\lambda}(y)
       +i\,\partial_{\mu}\partial_{\nu}c^{\mu\nu}(x)\,\partial_{f}G(x-y)\,G(x-y)\,\partial_{\pi}c^{f\pi}(y)
  \nonumber\\[-1pt]
  &-3i\,\partial_{\mu}\partial_{\lambda}c^{\mu\nu}(x)\,\partial^{\lambda}G(x-y)\,G(x-y)\,\partial_{\pi}c_{\nu}{}^{\pi}(y))
     \Bigr)
  \nonumber\\[0.8em]
  &+ \int d^4 x\,d^4 y\,
     \Bigl(
-\frac{3}{4}i{\partial}^{\alpha }c\left(x\right){\partial}^{\phi }G(x-y){\partial }_{\phi}G(x-y){\partial}_{\alpha}c\left(y\right)
+\frac{3}{4}i{\partial}_{\mu}c^{e\mu}\left(x\right){\partial}_{e}G\left(x-y\right){\partial}^{\pi}G(x-y){\partial}_{\pi}c\left(y\right)
  \nonumber\\[-1pt]
&+3i{\partial }_{\mu }c^{ev}\left(x\right){\partial }^{\mu }G\left(x-y\right){\partial }_{e}G(x-y){\partial }_{v}c\left(y\right)-i{\partial }^{v}{c^{\rho }}_{\lambda }\left(x\right){\partial }_{\rho }G\left(x-y\right){\partial }^{\lambda }G(x-y){\partial }_{v}c\left(y\right)
  \nonumber\\[-1pt]
&+\frac{3}{2}i{\partial }^{\alpha }c_{\alpha \omega }\left(x\right){\partial }^{\phi }G(x-y){\partial }_{\phi }G(x-y){\partial }^{\omega }c\left(y\right)-\frac{5}{8}i{\partial }_{\mu }{c_{o }}^{v}\left(x\right){\partial }^{\mu }G\left(x-y\right){\partial }^{\pi }G(x-y){\partial }_{\pi }{c_{v}}^{o}\left(y\right)
  \nonumber\\[-1pt]
&-\frac{1}{8}i{\partial }_{\mu }{c_{o }}^{\mu }\left(x\right){\partial }^{v}G\left(x-y\right){\partial }^{o }G(x-y){\partial }_{\pi }{c_{v}}^{\pi }\left(y\right)-\frac{3}{4}i{\partial }^{\alpha }c_{\tau \alpha }\left(x\right){\partial }^{\phi }G(x-y){\partial }_{\phi }G(x-y){\partial }^{\omega }{c^{\tau }}_{\omega }\left(y\right)
  \nonumber\\[-1pt]
&-\frac{1}{4}i{\partial }^{\alpha }c^{\beta v}\left(x\right){\partial }^{\phi }G(x-y){\partial }_{\phi }G(x-y){\partial }_{\alpha }c_{\beta v}\left(y\right)+\frac{1}{4}i{\partial }^{\alpha }c^{\beta v}\left(x\right){\partial }^{\phi }G(x-y){\partial }_{\phi }G(x-y){\partial }_{\beta }c_{\alpha v}\left(y\right)
  \nonumber\\[-1pt]
&+\frac{5}{2}i{\partial }^{\alpha }c_{\phi \varepsilon }\left(x\right){\partial }^{\phi }G(x-y){\partial }^{\pi }G(x-y){\partial }_{\alpha }{c_{\pi }}^{\varepsilon }\left(y\right)-\frac{7}{4}i{\partial }_{\phi }c^{\beta v}\left(x\right){\partial }^{\phi }G(x-y){\partial }^{\pi }G(x-y){\partial }_{\beta }c_{\pi v}\left(y\right)
  \nonumber\\[-1pt]
&-\frac{17}{8}i{\partial }^{\alpha }c_{\phi \omega }\left(x\right){\partial }^{\phi }G(x-y){\partial }^{\pi }G(x-y){\partial }^{\omega }c_{\alpha \pi }\left(y\right)-\frac{9}{2}i{\partial }_{\mu }{c_{o}}^{\mu }\left(x\right){\partial }^{v}G\left(x-y\right){\partial }_{f}G(x-y){\partial }_{v}c^{fo }\left(y\right)
  \nonumber\\[-1pt]
&+\frac{3}{2}i{\partial }_{\mu }{c_{o}}^{\mu }\left(x\right){\partial }^{v}G\left(x-y\right){\partial }_{\sigma }G(x-y){\partial }^{o }{c^{\sigma }}_{v}\left(y\right)
     \Bigr)
  \nonumber\\[0.8em]
  &+ \int d^4 x\,d^4 y\,
     \Bigl(
       -i\,c^{e\nu}(x)\,\partial_{\mu}\partial_{e}G(x-y)\,G(x-y)\,\partial_{\pi}\partial_{\nu}c^{\pi\mu}(y)
       -i\,c^{e\mu}(x)\,\partial_{\mu}\partial_{e}G(x-y)\,G(x-y)\,\partial_{\pi}\partial_{\nu}c^{\pi\nu}(y)
  \nonumber\\[-1pt]
  &+\frac{1}{2}i\,c^{e\mu}(x)\,\partial_{\mu}\partial_{e}G(x-y)\,G(x-y)\,\partial^{\nu}\partial_{\nu}c(y)
       +i\,c^{e\mu}(x)\,\partial_{\mu}\partial^{\nu}G(x-y)\,G(x-y)\,\partial_{\pi}\partial_{\nu}c^{\pi}{}_{e}(y)
     \Bigr)
  \nonumber\\[0.8em]
  &+ \int d^4 x\,d^4 y\,
     \Bigl(-\frac{3}{2}i{\partial }_{v}c^{ev}\left(x\right){\partial }_{\mu }{\partial }_{e}G\left(x-y\right){\partial }^{\mu }G\left(x-y\right)c\left(y\right)+\frac{7}{4}i{\partial }_{\pi }c^{e\mu }\left(x\right){\partial }_{\mu }{\partial }_{e}G\left(x-y\right){\partial }^{\pi }G\left(x-y\right)c\left(y\right)
  \nonumber\\[-1pt]
  &-\frac{5}{2}i{\partial }_{v}c^{e\mu }\left(x\right){\partial }_{\mu }{\partial }^{v}G\left(x-y\right){\partial }_{e}G\left(x-y\right)c\left(y\right)+\frac{7}{4}i{\partial }_{\mu }{c_{o }}^{\mu }\left(x\right){\partial }^{v}G\left(x-y\right){\partial }_{\pi }{\partial }^{o }G\left(x-y\right){c_{v}}^{\pi }\left(y\right)
  \nonumber\\[-1pt]
  &+\frac{3}{4}i{\partial }_{\mu }{c_{o }}^{\mu }\left(x\right){\partial }^{v}G\left(x-y\right){\partial }_{v}{\partial }_{f}G\left(x-y\right)c^{fo }\left(y\right)-\frac{1}{4}i{\partial }_{\mu }c^{ev}\left(x\right){\partial }_{e}G\left(x-y\right){\partial }_{v}{\partial }_{f}G\left(x-y\right)c^{f\mu }\left(y\right)
  \nonumber\\[-1pt]
  &+\frac{7}{4}i{\partial }_{v}{c_{o }}^{\mu }\left(x\right){\partial }_{\mu }{\partial }^{v}G\left(x-y\right){\partial }_{f}G\left(x-y\right)c^{fo }\left(y\right)-i{\partial }^{\alpha }c^{\omega \varepsilon }\left(x\right){\partial }_{\alpha }G(x-y){\partial }_{\varepsilon }{\partial }_{\tau }G(x-y){c^{\tau }}_{\omega }\left(y\right)
  \nonumber\\[-1pt]
  &-i{\partial }^{\alpha }c^{\beta \varepsilon }\left(x\right){\partial }_{\tau }G(x-y){\partial }_{\varepsilon }{\partial }_{\beta }G(x-y){c^{\tau }}_{\alpha }\left(y\right)-2i{\partial }^{\alpha }{c^{\beta }}_{\alpha }\left(x\right){\partial }_{\beta }G(x-y){\partial }_{\tau }{\partial }^{\pi }G(x-y){c^{\tau }}_{\pi }\left(y\right)
  \nonumber\\[-1pt]
  &+i{\partial }^{\alpha }c\left(x\right){\partial }^{\phi }G(x-y){\partial }_{\phi }{\partial }_{\alpha }G(x-y)c\left(y\right)+i{\partial }_{\phi }{c^{s}}_{\sigma }\left(x\right){\partial }_{v}{\partial }^{\phi }G(x-y){\partial }^{\sigma }G(x-y){c^{v}}_{s}\left(y\right)
     \Bigr)
  \nonumber\\[0.8em]
  &+ \int d^4 x\,d^4 y\,
     \Bigl(
       +\frac{3}{4}i\,c^{\rho}{}_{\lambda}(x)\,\partial^{\nu}\partial_{\rho}G(x-y)\,\partial_{\nu}\partial_{f}G(x-y)\,c^{f\lambda}(y)
       +\frac{3}{4}i\,c^{e\nu}(x)\,\partial_{\mu}\partial_{e}G(x-y)\,\partial_{\nu}\partial_{f}G(x-y)\,c^{f\mu}(y).
\end{align}

The two-loop contribution at the same order in $\kappa$ is generated by the quadratic term in the logarithmic expansion. Since the intermediate expressions are substantially longer, we only record the final result:
\begin{align}
  S^{(2,2)}
  &=  -i \bigl \langle S_{\text{r}}^{[2]}[{c^a}_\mu,\psi,\bar C,C] \bigl \rangle_{r,0} \Bigr|_{\hbar^2} +\frac{1}{2\hbar}  \Bigr \langle \left(S_{\text{r}}^{[1]}[{c^a}_\mu,\psi,\bar C,C]\right)^2 \Bigr \rangle^{\textrm{connected}}_{r,0} \Bigr|_{\hbar^3}
\\ \nonumber
  &= \int d^4 x\,d^4 y\,
     \Bigl(
       -4\, G(x-y)\,
           \partial^\alpha \partial_t \partial^t G(x-y)\,
           \partial_\alpha G(x-y)
       + 3\, G(x-y)\,
           \partial^\alpha \partial_t G(x-y)\,
           \partial_\alpha \partial^t G(x-y)
  \\[-1pt] \nonumber
  &\hspace{8em}
        -6\, \partial^\mu G(x-y)\, \partial^\alpha G(x-y)\,
             \partial_\mu \partial_\alpha G(x-y)- 2 G(x-y) \partial^\alpha \partial_\alpha G(x-y) \partial_t \partial^t G(x-y)
     \Bigr) .   
\end{align}

\section{Specified defining conditions for the Nicolai map $T_\kappa c$}
\label{Appendix C}
In this appendix, we present the universal conditions consisting of 18 equations for the first-type variables obtained in Section~\ref{section4}. In other words, we give the universal conditions on the coefficients of the diagrammatic factors in the order-\(\kappa\) Nicolai map after taking into account the constraints up to order \(\kappa^2\). Ideally, we would also like to present Eq.~\eqref{Linear_eq_Mj} in full, since it gives the complete set of conditions, up to order \(\kappa^2\), on the coefficients of the diagrammatic factors in the Nicolai map of $4D$ $\mathcal{N}=1$ pure supergravity. However, Eq.~\eqref{Linear_eq_Mj} consists of 736 equations, and even a single one of them already takes the following form:
\begin{align}
&-4M^{(0,1)}_{1,1}M^{(0,1)}_{5,5}
+4M^{(0,1)}_{1,1}M^{(0,1)}_{6,5}
-4M^{(0,1)}_{1,1}M^{(0,1)}_{9,5}
-2M^{(0,1)}_{5,1}M^{(0,1)}_{5,5}
+2M^{(0,1)}_{5,1}M^{(0,1)}_{6,5}
-2M^{(0,1)}_{5,1}M^{(0,1)}_{9,5}
\nonumber\\
&=\,2M^{(0,2),B}_{119,31}
-2M^{(0,2),B}_{120,31}
+2M^{(0,2),B}_{125,31}
+4M^{(0,2),B}_{13,1}
+2M^{(0,2),B}_{14,1}
+8M^{(0,2),B}_{151,1}
\nonumber\\
&\quad
+4M^{(0,2),B}_{152,1}
+4M^{(0,2),B}_{153,1}
+2M^{(0,2),B}_{154,1}
+4M^{(0,2),B}_{157,1}
+4M^{(0,2),B}_{163,1}
+2M^{(0,2),B}_{164,1}
\nonumber\\
&\quad
+2M^{(0,2),B}_{169,1}
+2M^{(0,2),B}_{169,31}
-2M^{(0,2),B}_{170,31}
+2M^{(0,2),B}_{175,31}
-4M^{(0,2),B}_{176,1}
-4M^{(0,2),B}_{177,1}
\nonumber\\
&\quad
-4M^{(0,2),B}_{182,1}
+2M^{(0,2),B}_{194,31}
-2M^{(0,2),B}_{195,31}
+2M^{(0,2),B}_{19,1}
+2M^{(0,2),B}_{19,31}
+4M^{(0,2),B}_{1,1}
\nonumber\\
&\quad
+2M^{(0,2),B}_{200,31}
-2M^{(0,2),B}_{201,1}
-2M^{(0,2),B}_{202,1}
-2M^{(0,2),B}_{207,1}
-2M^{(0,2),B}_{20,31}
-2M^{(0,2),B}_{244,31}
\nonumber\\
&\quad
+2M^{(0,2),B}_{245,31}
-2M^{(0,2),B}_{250,31}
+2M^{(0,2),B}_{25,31}
+4M^{(0,2),B}_{26,1}
+4M^{(0,2),B}_{28,1}
+2M^{(0,2),B}_{29,1}
\nonumber\\
&\quad
+4M^{(0,2),B}_{301,1}
+4M^{(0,2),B}_{302,1}
+4M^{(0,2),B}_{307,1}
+4M^{(0,2),B}_{319,31}
-4M^{(0,2),B}_{320,31}
+4M^{(0,2),B}_{325,31}
\nonumber\\
&\quad
+2M^{(0,2),B}_{326,1}
+2M^{(0,2),B}_{327,1}
+2M^{(0,2),B}_{332,1}
-4M^{(0,2),B}_{369,31}
+4M^{(0,2),B}_{370,31}
-4M^{(0,2),B}_{375,31}
\nonumber\\
&\quad
+4M^{(0,2),B}_{38,1}
+2M^{(0,2),B}_{39,1}
+4M^{(0,2),B}_{3,1}
+2M^{(0,2),B}_{44,1}
+2M^{(0,2),B}_{451,1}
+2M^{(0,2),B}_{452,1}
\nonumber\\
&\quad
+2M^{(0,2),B}_{457,1}
+4M^{(0,2),B}_{469,31}
-4M^{(0,2),B}_{494,31}
+2M^{(0,2),B}_{4,1}
+8M^{(0,2),B}_{619,31}
-2M^{(0,2),B}_{69,31}
\nonumber\\
&\quad
+2M^{(0,2),B}_{70,31}
-2M^{(0,2),B}_{75,31}.
\nonumber
\end{align}
We first present the 16 equations, out of the total 18, that correspond to Eq.~\eqref{1st_class_eqs}. Among these, the 14 equations derived from Eq.~\eqref{Condition_zero_one} are as follows:
\begin{align}
\label{begin_eq}
2
=&
-4M^{(0,1)}_{1,1}
-2M^{(0,1)}_{1,5}
-2M^{(0,1)}_{2,1}
-2M^{(0,1)}_{3,5}
-2M^{(0,1)}_{5,1}
-2M^{(0,1)}_{5,5}
+2M^{(0,1)}_{6,5}
-4M^{(0,1)}_{9,5},
\\[4pt]
0
=&
-4M^{(0,1)}_{1,1}
-2M^{(0,1)}_{2,1}
+2M^{(0,1)}_{2,5}
-2M^{(0,1)}_{3,5}
-2M^{(0,1)}_{6,1}
+2M^{(0,1)}_{6,5}
-4M^{(0,1)}_{9,5},
\\[4pt]
-2
=&
-4M^{(0,1)}_{1,2}
+M^{(0,1)}_{1,3}
+M^{(0,1)}_{1,5}
+2M^{(0,1)}_{1,9}
-2M^{(0,1)}_{2,2}
+M^{(0,1)}_{2,3}
+2M^{(0,1)}_{2,9}
+M^{(0,1)}_{3,5}
+M^{(0,1)}_{4,3}
-2M^{(0,1)}_{5,2}
\nonumber\\
&+2M^{(0,1)}_{5,3}
+M^{(0,1)}_{5,5}
+4M^{(0,1)}_{5,9}
-M^{(0,1)}_{6,3}
-M^{(0,1)}_{6,5}
-2M^{(0,1)}_{6,9}
-M^{(0,1)}_{8,3}
+M^{(0,1)}_{9,3}
+2M^{(0,1)}_{9,5}
+2M^{(0,1)}_{9,9},
\\[4pt]
0
=&
-4M^{(0,1)}_{1,2}
-2M^{(0,1)}_{2,2}
+M^{(0,1)}_{2,3}
-M^{(0,1)}_{2,5}
+2M^{(0,1)}_{2,9}
-M^{(0,1)}_{3,3}
+M^{(0,1)}_{3,5}
-2M^{(0,1)}_{3,9}
+M^{(0,1)}_{4,3}
+2M^{(0,1)}_{5,3}
\nonumber\\
&+4M^{(0,1)}_{5,9}
-2M^{(0,1)}_{6,2}
-M^{(0,1)}_{6,3}
-M^{(0,1)}_{6,5}
-2M^{(0,1)}_{6,9}
-M^{(0,1)}_{7,3}
-M^{(0,1)}_{8,3}
+2M^{(0,1)}_{9,5},
\\[4pt]
-2
=&
2M^{(0,1)}_{1,3}
+2M^{(0,1)}_{2,3}
+2M^{(0,1)}_{4,3}
-2M^{(0,1)}_{5,10}
+2M^{(0,1)}_{5,3}
-M^{(0,1)}_{5,4}
-M^{(0,1)}_{5,7}
+2M^{(0,1)}_{6,10}
+M^{(0,1)}_{6,4}
+M^{(0,1)}_{6,7}
\nonumber\\
&+2M^{(0,1)}_{8,10}
+M^{(0,1)}_{8,4}
+M^{(0,1)}_{8,7}
-M^{(0,1)}_{9,4},
\\[4pt]
0
=&
2M^{(0,1)}_{1,3}
-2M^{(0,1)}_{2,10}
+4M^{(0,1)}_{2,3}
-M^{(0,1)}_{2,4}
-M^{(0,1)}_{2,7}
+2M^{(0,1)}_{3,10}
-2M^{(0,1)}_{3,3}
+M^{(0,1)}_{3,4}
+M^{(0,1)}_{3,7}
+M^{(0,1)}_{4,4}
\nonumber\\
&-2M^{(0,1)}_{5,10}
+2M^{(0,1)}_{5,3}
-M^{(0,1)}_{5,4}
-M^{(0,1)}_{5,7}
+4M^{(0,1)}_{6,10}
-2M^{(0,1)}_{6,3}
+2M^{(0,1)}_{6,4}
+2M^{(0,1)}_{6,7}
+2M^{(0,1)}_{7,3}
-M^{(0,1)}_{7,4}
\nonumber\\
&+2M^{(0,1)}_{8,3}
-M^{(0,1)}_{9,4},
\\[4pt]
3
=&
2M^{(0,1)}_{1,10}
+M^{(0,1)}_{1,4}
+M^{(0,1)}_{1,7}
+2M^{(0,1)}_{2,3}
+2M^{(0,1)}_{3,10}
-2M^{(0,1)}_{3,3}
+M^{(0,1)}_{3,4}
+M^{(0,1)}_{3,7}
+M^{(0,1)}_{4,4}
+2M^{(0,1)}_{6,10}
\nonumber\\
&+M^{(0,1)}_{6,4}
+M^{(0,1)}_{6,7}
+2M^{(0,1)}_{8,3}
-2M^{(0,1)}_{9,3},
\\[4pt]
6
=&
2M^{(0,1)}_{1,4}
-2M^{(0,1)}_{2,4}
+2M^{(0,1)}_{2,7}
+4M^{(0,1)}_{3,4}
-2M^{(0,1)}_{3,7}
+2M^{(0,1)}_{4,4}
+2M^{(0,1)}_{5,7}
+2M^{(0,1)}_{6,4}
-4M^{(0,1)}_{6,7}
\nonumber\\
&-2M^{(0,1)}_{8,4}
+2M^{(0,1)}_{9,4},
\\[4pt]
0
=&
-2M^{(0,1)}_{1,7}
-2M^{(0,1)}_{2,4}
+2M^{(0,1)}_{3,4}
-2M^{(0,1)}_{3,7}
-2M^{(0,1)}_{5,4}
+2M^{(0,1)}_{6,4}
-2M^{(0,1)}_{6,7},
\\[4pt]
2
=&
2M^{(0,1)}_{1,4}
+2M^{(0,1)}_{3,4}
+2M^{(0,1)}_{5,7}
-2M^{(0,1)}_{6,7}
+2M^{(0,1)}_{7,4}
-2M^{(0,1)}_{8,7}
+2M^{(0,1)}_{9,4},
\\[4pt]
1
=&
2M^{(0,1)}_{1,6}
+2M^{(0,1)}_{2,6}
-M^{(0,1)}_{5,1}
-2M^{(0,1)}_{5,11}
+2M^{(0,1)}_{5,6}
-M^{(0,1)}_{5,8}
+M^{(0,1)}_{6,1}
+2M^{(0,1)}_{6,11}
+M^{(0,1)}_{6,8},
\\[4pt]
-1
=&
M^{(0,1)}_{1,1}
+2M^{(0,1)}_{1,11}
-M^{(0,1)}_{1,6}
+M^{(0,1)}_{1,8}
+\frac12 M^{(0,1)}_{2,1}
+M^{(0,1)}_{2,11}
+\frac12 M^{(0,1)}_{2,8}
-M^{(0,1)}_{3,6}
+\frac12 M^{(0,1)}_{5,1}
\nonumber\\
&+M^{(0,1)}_{5,11}
-M^{(0,1)}_{5,6}
+\frac12 M^{(0,1)}_{5,8}
+M^{(0,1)}_{6,6}
-2M^{(0,1)}_{9,6},
\\[4pt]
-2
=&
4M^{(0,1)}_{1,8}
+2M^{(0,1)}_{2,8}
+4M^{(0,1)}_{5,8}
-2M^{(0,1)}_{6,8},
\\[4pt]
-1
=&
-4M^{(0,1)}_{1,12}
-2M^{(0,1)}_{1,2}
-2M^{(0,1)}_{1,6}
-2M^{(0,1)}_{2,12}
-M^{(0,1)}_{2,2}
-M^{(0,1)}_{2,6}
-M^{(0,1)}_{3,6}
-4M^{(0,1)}_{5,12}
-2M^{(0,1)}_{5,2}
\nonumber\\
&-2M^{(0,1)}_{5,6}
+2M^{(0,1)}_{6,12}
+M^{(0,1)}_{6,2}
+M^{(0,1)}_{6,6}
-2M^{(0,1)}_{9,6}.
\end{align}
The remaining 2 equations derived from Eq.\eqref{Condition_one_one} are as follows:
\begin{align}
\label{S^(1,1)_condtion_1}
4
=&\,
-2M^{(1,1)}_{2,1}
-2M^{(0,1)}_{1,1}
-2M^{(0,1)}_{1,10}
-M^{(0,1)}_{1,3}
-3M^{(0,1)}_{1,4}
-11M^{(0,1)}_{1,5}
-5M^{(0,1)}_{1,7}
-4M^{(0,1)}_{1,9}
-M^{(0,1)}_{2,1}
-M^{(0,1)}_{2,10}
\nonumber\\
&-\frac{5}{2}M^{(0,1)}_{2,4}
-10M^{(0,1)}_{2,5}
-\frac{5}{2}M^{(0,1)}_{2,7}
-M^{(0,1)}_{3,10}
-M^{(0,1)}_{3,3}
-\frac{1}{2}M^{(0,1)}_{3,4}
-M^{(0,1)}_{3,5}
-\frac{5}{2}M^{(0,1)}_{3,7}
-4M^{(0,1)}_{3,9}
-\frac{5}{2}M^{(0,1)}_{4,4}
\nonumber\\
&-M^{(0,1)}_{5,1}
-M^{(0,1)}_{5,10}
-\frac{5}{2}M^{(0,1)}_{5,4}
-10M^{(0,1)}_{5,5}
-\frac{5}{2}M^{(0,1)}_{5,7}
-M^{(0,1)}_{7,3}
-\frac{1}{2}M^{(0,1)}_{7,4}
-M^{(0,1)}_{9,3}
-\frac{1}{2}M^{(0,1)}_{9,4}
\nonumber\\
&-M^{(0,1)}_{9,5}
-4M^{(0,1)}_{9,9},
\end{align}
\begin{align}
\label{S^(1,1)_condtion_2}
-1
=&\,
M^{(1,1)}_{2,1}
+2M^{(1,1)}_{2,2}
-2M^{(0,1)}_{1,11}
-8M^{(0,1)}_{1,12}
-2M^{(0,1)}_{1,2}
-\frac{5}{2}M^{(0,1)}_{1,3}
-\frac{1}{2}M^{(0,1)}_{1,4}
-11M^{(0,1)}_{1,6}
-5M^{(0,1)}_{1,8}
-M^{(0,1)}_{1,9}
\nonumber\\
&-M^{(0,1)}_{2,11}
-4M^{(0,1)}_{2,12}
-M^{(0,1)}_{2,2}
-\frac{5}{2}M^{(0,1)}_{2,3}
-10M^{(0,1)}_{2,6}
-\frac{5}{2}M^{(0,1)}_{2,8}
-M^{(0,1)}_{2,9}
-\frac{1}{2}M^{(0,1)}_{3,4}
-M^{(0,1)}_{3,6}
-\frac{5}{2}M^{(0,1)}_{4,3}
\nonumber\\
&-M^{(0,1)}_{5,11}
-4M^{(0,1)}_{5,12}
-M^{(0,1)}_{5,2}
-\frac{5}{2}M^{(0,1)}_{5,3}
-10M^{(0,1)}_{5,6}
-\frac{5}{2}M^{(0,1)}_{5,8}
-M^{(0,1)}_{5,9}
-\frac{1}{2}M^{(0,1)}_{7,4}
-\frac{1}{2}M^{(0,1)}_{9,4}
-M^{(0,1)}_{9,6}.
\end{align}
Finally, we present the 2 equations corresponding to Eq.~\eqref{additional equations}:
\begin{align}
0
=&\,
-\left(M^{(1,1)}_{2,1}\right)^{2} + 8 M^{(1,1)}_{2,1} M^{(0,1)}_{5,1} - 4 M^{(1,1)}_{2,1} M^{(0,1)}_{5,2} - M^{(1,1)}_{2,1} M^{(0,1)}_{5,3} + 2 M^{(1,1)}_{2,1} M^{(0,1)}_{5,4} + M^{(1,1)}_{2,1} M^{(0,1)}_{5,5}
\nonumber\\
& - 8 M^{(1,1)}_{2,1} M^{(0,1)}_{6,1} + 4 M^{(1,1)}_{2,1} M^{(0,1)}_{6,2} + M^{(1,1)}_{2,1} M^{(0,1)}_{6,3} - 2 M^{(1,1)}_{2,1} M^{(0,1)}_{6,4} - M^{(1,1)}_{2,1} M^{(0,1)}_{6,5} + M^{(1,1)}_{2,1} M^{(0,1)}_{8,3}
\nonumber\\
& - 2 M^{(1,1)}_{2,1} M^{(0,1)}_{8,4} - M^{(1,1)}_{2,1} M^{(0,1)}_{9,3} + 2 M^{(1,1)}_{2,1} M^{(0,1)}_{9,4} + M^{(1,1)}_{2,1} M^{(0,1)}_{9,5} - 16 \left(M^{(0,1)}_{5,1}\right)^{2} + 16 M^{(0,1)}_{5,1} M^{(0,1)}_{5,2}
\nonumber\\
& + 4 M^{(0,1)}_{5,1} M^{(0,1)}_{5,3} - 8 M^{(0,1)}_{5,1} M^{(0,1)}_{5,4} - 4 M^{(0,1)}_{5,1} M^{(0,1)}_{5,5} + 32 M^{(0,1)}_{5,1} M^{(0,1)}_{6,1} - 16 M^{(0,1)}_{5,1} M^{(0,1)}_{6,2} - 4 M^{(0,1)}_{5,1} M^{(0,1)}_{6,3}
\nonumber\\
& + 8 M^{(0,1)}_{5,1} M^{(0,1)}_{6,4} + 4 M^{(0,1)}_{5,1} M^{(0,1)}_{6,5} - 4 M^{(0,1)}_{5,1} M^{(0,1)}_{8,3} + 8 M^{(0,1)}_{5,1} M^{(0,1)}_{8,4} + 4 M^{(0,1)}_{5,1} M^{(0,1)}_{9,3} - 8 M^{(0,1)}_{5,1} M^{(0,1)}_{9,4}
\nonumber\\
& - 4 M^{(0,1)}_{5,1} M^{(0,1)}_{9,5} - 4 \left(M^{(0,1)}_{5,2}\right)^{2} - 2 M^{(0,1)}_{5,2} M^{(0,1)}_{5,3} + 4 M^{(0,1)}_{5,2} M^{(0,1)}_{5,4} + 2 M^{(0,1)}_{5,2} M^{(0,1)}_{5,5} - 16 M^{(0,1)}_{5,2} M^{(0,1)}_{6,1}
\nonumber\\
& + 8 M^{(0,1)}_{5,2} M^{(0,1)}_{6,2} + 2 M^{(0,1)}_{5,2} M^{(0,1)}_{6,3} - 4 M^{(0,1)}_{5,2} M^{(0,1)}_{6,4} - 2 M^{(0,1)}_{5,2} M^{(0,1)}_{6,5} + 2 M^{(0,1)}_{5,2} M^{(0,1)}_{8,3} - 4 M^{(0,1)}_{5,2} M^{(0,1)}_{8,4}
\nonumber\\
& - 2 M^{(0,1)}_{5,2} M^{(0,1)}_{9,3} + 4 M^{(0,1)}_{5,2} M^{(0,1)}_{9,4} + 2 M^{(0,1)}_{5,2} M^{(0,1)}_{9,5} - \frac{1}{4} \left(M^{(0,1)}_{5,3}\right)^{2} + M^{(0,1)}_{5,3} M^{(0,1)}_{5,4} + \frac{1}{2} M^{(0,1)}_{5,3} M^{(0,1)}_{5,5}
\nonumber\\
& - 4 M^{(0,1)}_{5,3} M^{(0,1)}_{6,1} + 2 M^{(0,1)}_{5,3} M^{(0,1)}_{6,2} + \frac{1}{2} M^{(0,1)}_{5,3} M^{(0,1)}_{6,3} - M^{(0,1)}_{5,3} M^{(0,1)}_{6,4} - \frac{1}{2} M^{(0,1)}_{5,3} M^{(0,1)}_{6,5} + \frac{1}{2} M^{(0,1)}_{5,3} M^{(0,1)}_{8,3}
\nonumber\\
& - M^{(0,1)}_{5,3} M^{(0,1)}_{8,4} - \frac{1}{2} M^{(0,1)}_{5,3} M^{(0,1)}_{9,3} + M^{(0,1)}_{5,3} M^{(0,1)}_{9,4} + \frac{1}{2} M^{(0,1)}_{5,3} M^{(0,1)}_{9,5} - \left(M^{(0,1)}_{5,4}\right)^{2} - M^{(0,1)}_{5,4} M^{(0,1)}_{5,5}
\nonumber\\
& + 8 M^{(0,1)}_{5,4} M^{(0,1)}_{6,1} - 4 M^{(0,1)}_{5,4} M^{(0,1)}_{6,2} - M^{(0,1)}_{5,4} M^{(0,1)}_{6,3} + 2 M^{(0,1)}_{5,4} M^{(0,1)}_{6,4} + M^{(0,1)}_{5,4} M^{(0,1)}_{6,5} - M^{(0,1)}_{5,4} M^{(0,1)}_{8,3}
\nonumber\\
& + 2 M^{(0,1)}_{5,4} M^{(0,1)}_{8,4} + M^{(0,1)}_{5,4} M^{(0,1)}_{9,3} - 2 M^{(0,1)}_{5,4} M^{(0,1)}_{9,4} - M^{(0,1)}_{5,4} M^{(0,1)}_{9,5} - \frac{1}{4} \left(M^{(0,1)}_{5,5}\right)^{2} + 4 M^{(0,1)}_{5,5} M^{(0,1)}_{6,1}
\nonumber\\
& - 2 M^{(0,1)}_{5,5} M^{(0,1)}_{6,2} - \frac{1}{2} M^{(0,1)}_{5,5} M^{(0,1)}_{6,3} + M^{(0,1)}_{5,5} M^{(0,1)}_{6,4} + \frac{1}{2} M^{(0,1)}_{5,5} M^{(0,1)}_{6,5} - \frac{1}{2} M^{(0,1)}_{5,5} M^{(0,1)}_{8,3} + M^{(0,1)}_{5,5} M^{(0,1)}_{8,4}
\nonumber\\
& + \frac{1}{2} M^{(0,1)}_{5,5} M^{(0,1)}_{9,3} - M^{(0,1)}_{5,5} M^{(0,1)}_{9,4} - \frac{1}{2} M^{(0,1)}_{5,5} M^{(0,1)}_{9,5} - 16 \left(M^{(0,1)}_{6,1}\right)^{2} + 16 M^{(0,1)}_{6,1} M^{(0,1)}_{6,2} + 4 M^{(0,1)}_{6,1} M^{(0,1)}_{6,3}
\nonumber\\
& - 8 M^{(0,1)}_{6,1} M^{(0,1)}_{6,4} - 4 M^{(0,1)}_{6,1} M^{(0,1)}_{6,5} + 4 M^{(0,1)}_{6,1} M^{(0,1)}_{8,3} - 8 M^{(0,1)}_{6,1} M^{(0,1)}_{8,4} - 4 M^{(0,1)}_{6,1} M^{(0,1)}_{9,3} + 8 M^{(0,1)}_{6,1} M^{(0,1)}_{9,4}
\nonumber\\
& + 4 M^{(0,1)}_{6,1} M^{(0,1)}_{9,5} - 4 \left(M^{(0,1)}_{6,2}\right)^{2} - 2 M^{(0,1)}_{6,2} M^{(0,1)}_{6,3} + 4 M^{(0,1)}_{6,2} M^{(0,1)}_{6,4} + 2 M^{(0,1)}_{6,2} M^{(0,1)}_{6,5} - 2 M^{(0,1)}_{6,2} M^{(0,1)}_{8,3}
\nonumber\\
& + 4 M^{(0,1)}_{6,2} M^{(0,1)}_{8,4} + 2 M^{(0,1)}_{6,2} M^{(0,1)}_{9,3} - 4 M^{(0,1)}_{6,2} M^{(0,1)}_{9,4} - 2 M^{(0,1)}_{6,2} M^{(0,1)}_{9,5} - \frac{1}{4} \left(M^{(0,1)}_{6,3}\right)^{2} + M^{(0,1)}_{6,3} M^{(0,1)}_{6,4}
\nonumber\\
& + \frac{1}{2} M^{(0,1)}_{6,3} M^{(0,1)}_{6,5} - \frac{1}{2} M^{(0,1)}_{6,3} M^{(0,1)}_{8,3} + M^{(0,1)}_{6,3} M^{(0,1)}_{8,4} + \frac{1}{2} M^{(0,1)}_{6,3} M^{(0,1)}_{9,3} - M^{(0,1)}_{6,3} M^{(0,1)}_{9,4} - \frac{1}{2} M^{(0,1)}_{6,3} M^{(0,1)}_{9,5}
\nonumber\\
& - \left(M^{(0,1)}_{6,4}\right)^{2} - M^{(0,1)}_{6,4} M^{(0,1)}_{6,5} + M^{(0,1)}_{6,4} M^{(0,1)}_{8,3} - 2 M^{(0,1)}_{6,4} M^{(0,1)}_{8,4} - M^{(0,1)}_{6,4} M^{(0,1)}_{9,3} + 2 M^{(0,1)}_{6,4} M^{(0,1)}_{9,4}
\nonumber\\
& + M^{(0,1)}_{6,4} M^{(0,1)}_{9,5} - \frac{1}{4} \left(M^{(0,1)}_{6,5}\right)^{2} + \frac{1}{2} M^{(0,1)}_{6,5} M^{(0,1)}_{8,3} - M^{(0,1)}_{6,5} M^{(0,1)}_{8,4} - \frac{1}{2} M^{(0,1)}_{6,5} M^{(0,1)}_{9,3} + M^{(0,1)}_{6,5} M^{(0,1)}_{9,4}
\nonumber\\
& + \frac{1}{2} M^{(0,1)}_{6,5} M^{(0,1)}_{9,5} - \frac{1}{4} \left(M^{(0,1)}_{8,3}\right)^{2} + M^{(0,1)}_{8,3} M^{(0,1)}_{8,4} + \frac{1}{2} M^{(0,1)}_{8,3} M^{(0,1)}_{9,3} - M^{(0,1)}_{8,3} M^{(0,1)}_{9,4} - \frac{1}{2} M^{(0,1)}_{8,3} M^{(0,1)}_{9,5}
\nonumber\\
& - \left(M^{(0,1)}_{8,4}\right)^{2} - M^{(0,1)}_{8,4} M^{(0,1)}_{9,3} + 2 M^{(0,1)}_{8,4} M^{(0,1)}_{9,4} + M^{(0,1)}_{8,4} M^{(0,1)}_{9,5} - \frac{1}{4} \left(M^{(0,1)}_{9,3}\right)^{2} + M^{(0,1)}_{9,3} M^{(0,1)}_{9,4}
\nonumber\\
& + \frac{1}{2} M^{(0,1)}_{9,3} M^{(0,1)}_{9,5} - \left(M^{(0,1)}_{9,4}\right)^{2} - M^{(0,1)}_{9,4} M^{(0,1)}_{9,5} - \frac{1}{4} \left(M^{(0,1)}_{9,5}\right)^{2}, \label{last_eq1}
\end{align}

The nonlinear system ~\eqref{begin_eq}--~\eqref{last_eq} was solved using the \texttt{nonlinsolve} function in \textsc{SymPy}, yielding the following relations.
\begin{align}
M^{(0,1)}_{1,2}
&=
-\frac{1}{2}M^{(0,1)}_{2,2}
+\frac{1}{4}M^{(0,1)}_{2,3}
-\frac{1}{4}M^{(0,1)}_{2,5}
+\frac{1}{2}M^{(0,1)}_{2,9}
-\frac{1}{4}M^{(0,1)}_{3,3}
+\frac{1}{4}M^{(0,1)}_{3,5}
-\frac{1}{2}M^{(0,1)}_{3,9}
+\frac{1}{4}M^{(0,1)}_{4,3}
\nonumber\\
&\quad
+\frac{1}{2}M^{(0,1)}_{5,3}
+M^{(0,1)}_{5,9}
-\frac{1}{2}M^{(0,1)}_{6,2}
-\frac{1}{4}M^{(0,1)}_{6,3}
-\frac{1}{4}M^{(0,1)}_{6,5}
-\frac{1}{2}M^{(0,1)}_{6,9}
-\frac{1}{4}M^{(0,1)}_{7,3}
-\frac{1}{4}M^{(0,1)}_{8,3}
\nonumber\\
&\quad
+\frac{1}{2}M^{(0,1)}_{9,5},
\label{2begin_eq}\\
M^{(0,1)}_{1,10}
&=
-M^{(0,1)}_{2,3}
+\frac{1}{2}M^{(0,1)}_{2,7}
-M^{(0,1)}_{3,10}
+M^{(0,1)}_{3,3}
-\frac{1}{2}M^{(0,1)}_{3,7}
+\frac{1}{2}M^{(0,1)}_{5,4}
+\frac{1}{2}M^{(0,1)}_{5,7}
-M^{(0,1)}_{6,10}
\nonumber\\
&\quad
-\frac{1}{2}M^{(0,1)}_{6,4}
-M^{(0,1)}_{6,7}
-M^{(0,1)}_{8,3}
-\frac{1}{2}M^{(0,1)}_{8,4}
+M^{(0,1)}_{9,3}
+\frac{1}{2}M^{(0,1)}_{9,4},
\\
M^{(0,1)}_{1,12}
&=
-\frac{1}{2}M^{(0,1)}_{2,12}
-\frac{1}{8}M^{(0,1)}_{2,3}
+\frac{1}{8}M^{(0,1)}_{2,5}
+\frac{1}{4}M^{(0,1)}_{2,6}
-\frac{1}{4}M^{(0,1)}_{2,9}
+\frac{1}{8}M^{(0,1)}_{3,3}
-\frac{1}{8}M^{(0,1)}_{3,5}
-\frac{1}{4}M^{(0,1)}_{3,6}
\nonumber\\
&\quad
+\frac{1}{4}M^{(0,1)}_{3,9}
-\frac{1}{8}M^{(0,1)}_{4,3}
-\frac{1}{4}M^{(0,1)}_{5,1}
-\frac{1}{2}M^{(0,1)}_{5,11}
-M^{(0,1)}_{5,12}
-\frac{1}{2}M^{(0,1)}_{5,2}
-\frac{1}{4}M^{(0,1)}_{5,3}
-\frac{1}{4}M^{(0,1)}_{5,8}
\nonumber\\
&\quad
-\frac{1}{2}M^{(0,1)}_{5,9}
+\frac{1}{4}M^{(0,1)}_{6,1}
+\frac{1}{2}M^{(0,1)}_{6,11}
+\frac{1}{2}M^{(0,1)}_{6,12}
+\frac{1}{2}M^{(0,1)}_{6,2}
+\frac{1}{8}M^{(0,1)}_{6,3}
+\frac{1}{8}M^{(0,1)}_{6,5}
+\frac{1}{4}M^{(0,1)}_{6,6}
\nonumber\\
&\quad
+\frac{1}{4}M^{(0,1)}_{6,8}
+\frac{1}{4}M^{(0,1)}_{6,9}
+\frac{1}{8}M^{(0,1)}_{7,3}
+\frac{1}{8}M^{(0,1)}_{8,3}
-\frac{1}{4}M^{(0,1)}_{9,5}
-\frac{1}{2}M^{(0,1)}_{9,6},
\\
M^{(0,1)}_{1,6}
&=
-M^{(0,1)}_{2,6}
+\frac{1}{2}M^{(0,1)}_{5,1}
+M^{(0,1)}_{5,11}
-M^{(0,1)}_{5,6}
+\frac{1}{2}M^{(0,1)}_{5,8}
-\frac{1}{2}M^{(0,1)}_{6,1}
-M^{(0,1)}_{6,11}
-\frac{1}{2}M^{(0,1)}_{6,8}
\nonumber\\
&\quad
+\frac{1}{2},
\\
M^{(0,1)}_{2,4}
&=
M^{(0,1)}_{2,7}
+M^{(0,1)}_{3,4}
-M^{(0,1)}_{3,7}
+M^{(0,1)}_{4,4}
+M^{(0,1)}_{6,4}
-M^{(0,1)}_{6,7}
-M^{(0,1)}_{7,4}
-M^{(0,1)}_{8,4}
\nonumber\\
&\quad
+M^{(0,1)}_{8,7}
-2,
\\
M^{(0,1)}_{1,1}
&=
-\frac{1}{2}M^{(0,1)}_{2,1}
+\frac{1}{2}M^{(0,1)}_{2,5}
-\frac{1}{2}M^{(0,1)}_{3,5}
-\frac{1}{2}M^{(0,1)}_{6,1}
+\frac{1}{2}M^{(0,1)}_{6,5}
-M^{(0,1)}_{9,5},
\\
M^{(1,1)}_{2,2}
&=
M^{(0,1)}_{5,10}
+4M^{(0,1)}_{5,11}
-2M^{(0,1)}_{5,12}
-\frac{1}{4}M^{(0,1)}_{5,5}
-\frac{1}{2}M^{(0,1)}_{5,6}
+\frac{1}{4}M^{(0,1)}_{5,7}
+M^{(0,1)}_{5,8}
-\frac{1}{2}M^{(0,1)}_{5,9}
\nonumber\\
&\quad
-M^{(0,1)}_{6,10}
-4M^{(0,1)}_{6,11}
+2M^{(0,1)}_{6,12}
+\frac{1}{4}M^{(0,1)}_{6,5}
+\frac{1}{2}M^{(0,1)}_{6,6}
-\frac{1}{4}M^{(0,1)}_{6,7}
-M^{(0,1)}_{6,8}
+\frac{1}{2}M^{(0,1)}_{6,9}
\nonumber\\
&\quad
-M^{(0,1)}_{8,10}
-\frac{1}{4}M^{(0,1)}_{8,7}
-\frac{1}{4}M^{(0,1)}_{9,5}
-\frac{1}{2}M^{(0,1)}_{9,6}
-\frac{1}{2}M^{(0,1)}_{9,9},
\\
M^{(0,1)}_{1,9}
&=
\frac{1}{2}M^{(0,1)}_{2,3}
-\frac{1}{2}M^{(0,1)}_{3,3}
-M^{(0,1)}_{3,9}
+\frac{1}{2}M^{(0,1)}_{4,3}
+\frac{1}{2}M^{(0,1)}_{5,1}
-\frac{1}{2}M^{(0,1)}_{5,10}
+M^{(0,1)}_{5,2}
+\frac{1}{2}M^{(0,1)}_{5,3}
\nonumber\\
&\quad
-\frac{1}{4}M^{(0,1)}_{5,4}
-\frac{1}{4}M^{(0,1)}_{5,7}
-\frac{1}{2}M^{(0,1)}_{6,1}
+\frac{1}{2}M^{(0,1)}_{6,10}
-M^{(0,1)}_{6,2}
+\frac{1}{4}M^{(0,1)}_{6,4}
+\frac{1}{4}M^{(0,1)}_{6,7}
-\frac{1}{2}M^{(0,1)}_{7,3}
\nonumber\\
&\quad
+\frac{1}{2}M^{(0,1)}_{8,10}
+\frac{1}{4}M^{(0,1)}_{8,4}
+\frac{1}{4}M^{(0,1)}_{8,7}
-\frac{1}{2}M^{(0,1)}_{9,3}
-\frac{1}{4}M^{(0,1)}_{9,4}
-M^{(0,1)}_{9,9},
\\
M^{(0,1)}_{1,7}
&=
-M^{(0,1)}_{2,7}
-M^{(0,1)}_{4,4}
-M^{(0,1)}_{5,4}
+M^{(0,1)}_{7,4}
+M^{(0,1)}_{8,4}
-M^{(0,1)}_{8,7}
+2,
\\
M^{(0,1)}_{1,3}
&=
-M^{(0,1)}_{2,3}
-M^{(0,1)}_{4,3}
+M^{(0,1)}_{5,10}
-M^{(0,1)}_{5,3}
+\frac{1}{2}M^{(0,1)}_{5,4}
+\frac{1}{2}M^{(0,1)}_{5,7}
-M^{(0,1)}_{6,10}
-\frac{1}{2}M^{(0,1)}_{6,4}
\nonumber\\
&\quad
-\frac{1}{2}M^{(0,1)}_{6,7}
-M^{(0,1)}_{8,10}
-\frac{1}{2}M^{(0,1)}_{8,4}
-\frac{1}{2}M^{(0,1)}_{8,7}
+\frac{1}{2}M^{(0,1)}_{9,4}
-1,
\\
M^{(0,1)}_{2,10}
&=
M^{(0,1)}_{2,3}
-M^{(0,1)}_{2,7}
+M^{(0,1)}_{3,10}
-M^{(0,1)}_{3,3}
+M^{(0,1)}_{3,7}
-M^{(0,1)}_{4,3}
+M^{(0,1)}_{6,10}
-M^{(0,1)}_{6,3}
\nonumber\\
&\quad
+M^{(0,1)}_{6,7}
+M^{(0,1)}_{7,3}
-M^{(0,1)}_{8,10}
+M^{(0,1)}_{8,3}
-M^{(0,1)}_{8,7},
\\
M^{(1,1)}_{2,1}
&=
4M^{(0,1)}_{5,1}
-2M^{(0,1)}_{5,2}
-\frac{1}{2}M^{(0,1)}_{5,3}
+M^{(0,1)}_{5,4}
+\frac{1}{2}M^{(0,1)}_{5,5}
-4M^{(0,1)}_{6,1}
+2M^{(0,1)}_{6,2}
+\frac{1}{2}M^{(0,1)}_{6,3}
\nonumber\\
&\quad
-M^{(0,1)}_{6,4}
-\frac{1}{2}M^{(0,1)}_{6,5}
+\frac{1}{2}M^{(0,1)}_{8,3}
-M^{(0,1)}_{8,4}
-\frac{1}{2}M^{(0,1)}_{9,3}
+M^{(0,1)}_{9,4}
+\frac{1}{2}M^{(0,1)}_{9,5},
\\
M^{(0,1)}_{1,8}
&=
-\frac{1}{2}M^{(0,1)}_{2,8}
-M^{(0,1)}_{5,8}
+\frac{1}{2}M^{(0,1)}_{6,8}
-\frac{1}{2},
\\
M^{(0,1)}_{1,4}
&=
-M^{(0,1)}_{3,4}
-M^{(0,1)}_{5,7}
+M^{(0,1)}_{6,7}
-M^{(0,1)}_{7,4}
+M^{(0,1)}_{8,7}
-M^{(0,1)}_{9,4}
+1,
\\
M^{(0,1)}_{1,5}
&=
-M^{(0,1)}_{2,5}
-M^{(0,1)}_{5,1}
-M^{(0,1)}_{5,5}
+M^{(0,1)}_{6,1}
-1,
\\
M^{(0,1)}_{1,11}
&=
-\frac{1}{2}M^{(0,1)}_{2,11}
-\frac{1}{4}M^{(0,1)}_{2,5}
-\frac{1}{2}M^{(0,1)}_{2,6}
+\frac{1}{4}M^{(0,1)}_{3,5}
+\frac{1}{2}M^{(0,1)}_{3,6}
+\frac{1}{2}M^{(0,1)}_{5,8}
-\frac{1}{2}M^{(0,1)}_{6,11}
-\frac{1}{4}M^{(0,1)}_{6,5}
\nonumber\\
&\quad
-\frac{1}{2}M^{(0,1)}_{6,6}
-\frac{1}{2}M^{(0,1)}_{6,8}
+\frac{1}{2}M^{(0,1)}_{9,5}
+M^{(0,1)}_{9,6}.
\label{2last_eq}
\end{align}

\section{The Nicolai map of $4D$ $\mathcal{N}=1$ pure supergravity to the second order in $\kappa$}
\label{Appendix D}
In this appendix, we spell out the representative solutions of $T^{(0,2)}c$ and $T^{(1,2)}c$ to the Nicolai map
constraints through $\mathcal O(\kappa^2)$. 

The solution of $T^{(0,2)}c$ is given by
{\footnotesize

}
The solution of $T^{(1,2)}c$ is found as follows:
{\footnotesize
\begin{align}
&T^{(1,2)}c=  -2i\Bigl( +1/2 \int d^4 y d^4 z \partial^{\mu} \partial^{\rho} \partial^{\sigma} G(x-y) \partial^{\nu} c_{\rho\sigma}(y) G(y-z) G(0)
-1 \int d^4 y d^4 z \partial^{\mu} \partial^{\rho} G(x-y) \partial^{\nu} \partial^{\sigma} c_{\rho\sigma}(y) G(y-z) G(0) \nonumber\\ \nonumber
&-11/2 \int d^4 y d^4 z G(x-y) c^{\rho\sigma}(y) \partial^{\mu} \partial^{\nu} G(y-z) \partial_{\rho} \partial_{\sigma} G(0)
+4 \int d^4 y d^4 z G(x-y) \partial^{\rho} c_{\rho}^{\sigma}(y) \partial^{\mu} G(y-z) \partial^{\nu} \partial_{\sigma} G(0) \\ \nonumber
&+1 \int d^4 y d^4 z \partial^{\mu} G(x-y) \partial^{\nu} \partial^{\rho} \partial^{\sigma} c_{\rho\sigma}(y) G(y-z) G(0)
-1/4 \int d^4 y d^4 z \partial^{\mu} \partial^{\nu} G(x-y) c^{\rho\sigma}(y) G(y-z) \partial_{\rho} \partial_{\sigma} G(0) \\ \nonumber
&-1/2 \int d^4 y d^4 z \partial^{\mu} \partial^{\nu} \partial^{\rho} \partial^{\sigma} G(x-y) c_{\rho\sigma}(y) G(y-z) G(0)
-1/2 \int d^4 y d^4 z \partial^{\mu} G(x-y) \partial^{\rho} \partial^{\sigma} c_{\rho\sigma}(y) \partial^{\nu} G(y-z) G(0) \\ \nonumber
&+1/4 \int d^4 y d^4 z \partial^{\rho} \partial^{\sigma} G(x-y) c_{\rho\sigma}(y) G(y-z) \partial^{\mu} \partial^{\nu} G(0)
-7/4 \int d^4 y d^4 z \partial^{\mu} \partial^{\rho} G(x-y) c_{\rho}^{\sigma}(y) G(y-z) \partial^{\nu} \partial_{\sigma} G(0) \\ \nonumber
&+51/4 \int d^4 y d^4 z \partial^{\rho} G(x-y) c_{\rho}^{\sigma}(y) \partial^{\mu} G(y-z) \partial^{\nu} \partial_{\sigma} G(0)
-1/4 \eta^{\mu\nu} \int d^4 y d^4 z \partial^{\rho} \partial^{\sigma} \partial^{\lambda} G(x-y) \partial_{\rho} c_{\sigma\lambda}(y) G(y-z) G(0) \\ \nonumber
&-7/4 \int d^4 y d^4 z \partial^{\mu} \partial^{\nu} G(x-y) c^{\sigma}_{\sigma}(y) G(y-z) \partial^{\rho} \partial_{\rho} G(0)
-7/2 \int d^4 y d^4 z \partial^{\sigma} G(x-y) c^{\nu}_{\sigma}(y) \partial^{\mu} G(y-z) \partial^{\rho} \partial_{\rho} G(0) \\ \nonumber
&-29/16 \eta^{\mu\nu} \int d^4 y d^4 z \partial^{\sigma} G(x-y) c_{\sigma}^{\lambda}(y) \partial^{\rho} G(y-z) \partial_{\rho} \partial_{\lambda} G(0)
+63/4 \int d^4 y d^4 z \partial^{\mu} \partial^{\rho} G(x-y) c^{\nu\sigma}(y) G(y-z) \partial_{\rho} \partial_{\sigma} G(0) \\ \nonumber
&-1/4 \eta^{\mu\nu} \int d^4 y d^4 z \partial^{\rho} G(x-y) \partial_{\rho} \partial^{\sigma} \partial^{\lambda} c_{\sigma\lambda}(y) G(y-z) G(0)
+1/4 \int d^4 y d^4 z \partial^{\mu} \partial^{\nu} \partial^{\rho} \partial_{\rho} G(x-y) c^{\sigma}_{\sigma}(y) G(y-z) G(0) \\ \nonumber
&-3 \int d^4 y d^4 z G(x-y) c^{\sigma}_{\sigma}(y) \partial^{\mu} \partial^{\nu} G(y-z) \partial^{\rho} \partial_{\rho} G(0)
-31/2 \int d^4 y d^4 z G(x-y) c^{\nu\sigma}(y) \partial^{\mu} \partial^{\rho} G(y-z) \partial_{\rho} \partial_{\sigma} G(0) \\ \nonumber
&-17/16 \int d^4 y d^4 z \partial^{\rho} G(x-y) c^{\sigma}_{\sigma}(y) \partial^{\mu} G(y-z) \partial^{\nu} \partial_{\rho} G(0)
-266/5 \int d^4 y d^4 z \partial^{\rho} \partial_{\rho} G(x-y) c^{\nu\sigma}(y) G(y-z) \partial^{\mu} \partial_{\sigma} G(0) \\ \nonumber
&+266/5 \int d^4 y d^4 z \partial^{\rho} G(x-y) \partial_{\rho} c^{\nu\sigma}(y) G(y-z) \partial^{\mu} \partial_{\sigma} G(0)
+13 \int d^4 y d^4 z \partial^{\rho} \partial^{\sigma} G(x-y) c^{\nu}_{\sigma}(y) G(y-z) \partial^{\mu} \partial_{\rho} G(0) \\ \nonumber
&+1/4 \eta^{\mu\nu} \int d^4 y d^4 z \partial^{\rho} \partial_{\rho} \partial^{\sigma} \partial^{\lambda} G(x-y) c_{\sigma\lambda}(y) G(y-z) G(0)
-1/2 \int d^4 y d^4 z \partial^{\mu} \partial^{\rho} G(x-y) c^{\sigma}_{\sigma}(y) G(y-z) \partial^{\nu} \partial_{\rho} G(0) \\ \nonumber
&-1/4 \int d^4 y d^4 z \partial^{\mu} \partial^{\sigma} G(x-y) c^{\nu}_{\sigma}(y) G(y-z) \partial^{\rho} \partial_{\rho} G(0)
-3/8 \eta^{\mu\nu} \int d^4 y d^4 z \partial^{\sigma} \partial^{\lambda} G(x-y) c_{\sigma\lambda}(y) G(y-z) \partial^{\rho} \partial_{\rho} G(0) \\ \nonumber
&-9/16 \int d^4 y d^4 z G(x-y) \partial^{\rho} c^{\sigma}_{\sigma}(y) \partial^{\mu} G(y-z) \partial^{\nu} \partial_{\rho} G(0)
-31/2 \int d^4 y d^4 z \partial^{\rho} G(x-y) \partial^{\sigma} c^{\nu}_{\sigma}(y) G(y-z) \partial^{\mu} \partial_{\rho} G(0) \\ \nonumber
&+1/4 \int d^4 y d^4 z G(x-y) \partial^{\sigma} c^{\nu}_{\sigma}(y) \partial^{\mu} G(y-z) \partial^{\rho} \partial_{\rho} G(0)
+133/10
 \eta^{\mu\nu} \int d^4 y d^4 z \partial^{\rho} \partial_{\rho} G(x-y) c^{\sigma\lambda}(y) G(y-z) \partial_{\sigma} \partial_{\lambda} G(0) \\ \nonumber
&+3 \int d^4 y d^4 z G(x-y) c^{\mu\nu}(y) \partial^{\rho} \partial^{\sigma} G(y-z) \partial_{\rho} \partial_{\sigma} G(0)
+1/2 \eta^{\mu\nu} \int d^4 y d^4 z \partial^{\rho} \partial^{\sigma} G(x-y) \partial_{\rho} \partial^{\lambda} c_{\sigma\lambda}(y) G(y-z) G(0) \\ \nonumber
&-1/8 \eta^{\mu\nu} \int d^4 y d^4 z \partial^{\rho} \partial_{\rho} \partial^{\sigma} \partial_{\sigma} G(x-y) c^{\lambda}_{\lambda}(y) G(y-z) G(0)
-457/80 \eta^{\mu\nu} \int d^4 y d^4 z \partial^{\rho} \partial_{\rho} G(x-y) c^{\lambda}_{\lambda}(y) G(y-z) \partial^{\sigma} \partial_{\sigma} G(0) \\ \nonumber
&+2 \eta^{\mu\nu} \int d^4 y d^4 z \partial^{\rho} \partial^{\sigma} G(x-y) c_{\sigma}^{\lambda}(y) G(y-z) \partial_{\rho} \partial_{\lambda} G(0)
-219/16 \eta^{\mu\nu} \int d^4 y d^4 z G(x-y) \partial^{\sigma} c_{\sigma}^{\lambda}(y) \partial^{\rho} G(y-z) \partial_{\rho} \partial_{\lambda} G(0) \\ \nonumber
&-78\int d^4 y d^4 z \partial^{\rho} \partial^{\sigma} G(x-y) \partial^{\mu} G(y-z) G(y-z) \partial^{\nu} c_{\rho\sigma}(z)
-15/2 \int d^4 y d^4 z G(x-y) \partial^{\mu} \partial^{\rho} \partial^{\sigma} G(y-z) \partial^{\nu} G(y-z) c_{\rho\sigma}(z) \\ \nonumber
&-10 \int d^4 y d^4 z \partial^{\rho} G(x-y) \partial^{\mu} \partial^{\sigma} G(y-z) G(y-z) \partial^{\nu} c_{\rho\sigma}(z)
+3/8 \int d^4 y d^4 z \partial^{\rho} G(x-y) \partial^{\mu} G(y-z) \partial^{\nu} G(y-z) \partial^{\sigma} c_{\rho\sigma}(z) \\ \nonumber
&-61 \eta^{\mu\nu} \int d^4 y d^4 z \partial^{\sigma} G(x-y) \partial^{\rho} G(y-z) \partial_{\rho} G(y-z) \partial^{\lambda} c_{\sigma\lambda}(z)
-137/8 \int d^4 y d^4 z \partial^{\rho} G(x-y) \partial^{\mu} G(y-z) \partial^{\nu} G(y-z) \partial_{\rho} c^{\sigma}_{\sigma}(z) \\ \nonumber
&+73/4 \int d^4 y d^4 z \partial^{\sigma} G(x-y) \partial^{\mu} G(y-z) \partial^{\rho} G(y-z) \partial_{\rho} c^{\nu}_{\sigma}(z)
-101/4 \int d^4 y d^4 z G(x-y) \partial^{\mu} \partial^{\sigma} G(y-z) \partial^{\rho} G(y-z) \partial_{\rho} c^{\nu}_{\sigma}(z) \\ \nonumber
&+29/4 \int d^4 y d^4 z G(x-y) \partial^{\rho} \partial^{\sigma} G(y-z) \partial_{\rho} \partial_{\sigma} G(y-z) c^{\mu\nu}(z)
+81/16 \int d^4 y \partial^{\rho} \partial_{\rho} G(x-y) c^{\mu\nu}(y) G(0) \\ \nonumber
&+49/2 \eta^{\mu\nu} \int d^4 y d^4 z G(x-y) \partial^{\rho} \partial^{\sigma} \partial^{\lambda} G(y-z) \partial_{\rho} G(y-z) c_{\sigma\lambda}(z)
-59/4 \int d^4 y \partial^{\mu} \partial^{\nu} G(x-y) c^{\rho}_{\rho}(y) G(0) \\ \nonumber
&+91/4 \int d^4 y \partial^{\mu} \partial^{\rho} G(x-y) c^{\nu}_{\rho}(y) G(0)
-169/4 \int d^4 y G(x-y) c^{\rho}_{\rho}(y) \partial^{\mu} \partial^{\nu} G(0) \\ \nonumber
&-289/3 \eta^{\mu\nu} \int d^4 y \partial^{\rho} \partial^{\sigma} G(x-y) c_{\rho\sigma}(y) G(0)
+521/60 \int d^4 y G(x-y) c^{\nu\rho}(y) \partial^{\mu} \partial_{\rho} G(0) \\ \nonumber
&-23/16 \int d^4 y G(x-y) c^{\mu\nu}(y) \partial^{\rho} \partial_{\rho} G(0)
+285/8 \eta^{\mu\nu} \int d^4 y G(x-y) c^{\rho\sigma}(y) \partial_{\rho} \partial_{\sigma} G(0) \\ \nonumber
&-27/2 \int d^4 y d^4 z \partial^{\sigma} G(x-y) \partial^{\rho} c^{\nu}_{\sigma}(y) G(y-z) \partial^{\mu} \partial_{\rho} G(0)
+527/40 \int d^4 y d^4 z \partial^{\rho} \partial_{\rho} G(x-y) c^{\sigma}_{\sigma}(y) G(y-z) \partial^{\mu} \partial^{\nu} G(0) \\ 
&-173/4 \int d^4 y d^4 z G(x-y) \partial^{\mu} \partial^{\rho} \partial_{\rho} G(y-z) \partial^{\nu} G(y-z) c^{\sigma}_{\sigma}(z)\Bigr) 
\end{align}
}
\end{appendices}

\end{document}